\documentclass[journal,english]{IEEEtran}
\usepackage{epsfig}
\usepackage{times}
\usepackage{float}
\usepackage{afterpage}
\usepackage{amsmath}
\usepackage{amstext}
\usepackage{amssymb,bm}
\usepackage{latexsym}
\usepackage{color}
\usepackage{graphicx}
\usepackage{amsmath}
\usepackage{amsthm}
\usepackage{graphicx}
\usepackage[center]{caption}
\usepackage{pstricks}
\usepackage{caption}
\usepackage{subcaption}
\usepackage{booktabs}
\usepackage{multicol}
\usepackage{lipsum}
 \usepackage[T1]{fontenc}
 \usepackage[utf8]{inputenc} 

\usepackage{epstopdf}
\usepackage[small,nohug,heads=vee]{diagrams}
\allowdisplaybreaks
\makeatletter

\newtheorem{thm}{Theorem}
\newtheorem{cor}{Corollary}

\newtheorem{prop}{Proposition}
\newtheorem{rem}{Remark}
\newtheorem{example}{Example}

\theoremstyle{definition}

\providecommand{\definitionname}{Definition}

\providecommand{\tabularnewline}{\\}
\newfloat{algorithm}{tbp}{loa}
\providecommand{\algorithmname}{Algorithm}
\floatname{algorithm}{\protect\algorithmname}

\usepackage[unicode=true,
 bookmarks=true,bookmarksnumbered=true,bookmarksopen=true,bookmarksopenlevel=1,
 breaklinks=false,pdfborder={0 0 0},backref=false,colorlinks=false]
 {hyperref}
\hypersetup{pdftitle={Your Title},
 pdfauthor={Your Name},
 pdfpagelayout=OneColumn, pdfnewwindow=true, pdfstartview=XYZ, plainpages=false}
\newcommand*{\rom}[1]{\expandafter\@slowromancap\romannumeral #1@}
\usepackage{changes}
\definechangesauthor[color=red,name={Mingyue Ji}]{MJ}
\UseRawInputEncoding

\long\def\comment#1{}

\newcommand{\dv}{{\mathbf d}}

\newcommand{\pv}{{\mathbf p}}

\newcommand{\Zm}{{\mathbf Z}}

\newcommand{\Ac}{{\mathcal A}}
\newcommand{\Bc}{{\mathcal B}}
\newcommand{\Cc}{{\mathcal C}}

\newcommand{\Gc}{{\mathcal G}}
\newcommand{\Hc}{{\mathcal H}}
\newcommand{\Ic}{{\mathcal I}}
\newcommand{\Jc}{{\mathcal J}}
\newcommand{\Kc}{{\mathcal K}}
\newcommand{\Lc}{{\mathcal L}}
\newcommand{\Mc}{{\mathcal M}}

\newcommand{\Oc}{{\mathcal O}}
\newcommand{\Pc}{{\mathcal P}}
\newcommand{\Qc}{{\mathcal Q}}
\newcommand{\Rc}{{\mathcal R}}
\newcommand{\Sc}{{\mathcal S}}
\newcommand{\Tc}{{\mathcal T}}
\newcommand{\Uc}{{\mathcal U}}
\newcommand{\Wc}{{\mathcal W}}
\newcommand{\Vc}{{\mathcal V}}
\newcommand{\Xc}{{\mathcal X}}
\newcommand{\Yc}{{\mathcal Y}}

\newcommand{\rsf}{{\mathsf r}}

\newcommand{\Bsf}{{\mathsf B}}

\newcommand{\Hsf}{{\mathsf H}}

\newcommand{\Ksf}{{\mathsf K}}

\newcommand{\Msf}{{\mathsf M}}
\newcommand{\Nsf}{{\mathsf N}}

\newcommand{\Rsf}{{\mathsf R}}

\renewcommand{\arg}{{\hbox{arg}}}

\newcommand{\be}{\begin{equation}}
\newcommand{\ee}{\end{equation}}
\newcommand{\bea}{\begin{eqnarray}}
\newcommand{\eea}{\end{eqnarray}}

\begin{document}

\title{Combination Networks with End-user-caches: 
Novel Achievable 
and Converse Bounds under Uncoded Cache Placement} 


\author{
Kai~Wan,~\IEEEmembership{Member,~IEEE,}
Daniela Tuninetti,~\IEEEmembership{Fellow,~IEEE}
Mingyue~Ji,~\IEEEmembership{Member,~IEEE},
Pablo Piantanida,~\IEEEmembership{Senior~Member,~IEEE}

\thanks{The results of this paper were presented in parts at 
    the  2017 IEEE  Information Theory Workshop, Kaohsiung, Taiwan,~\cite{wan2017outer}, 
and the 55th Annual Allerton Conference (2017) on Communication, Control, and Computing, Urbana, USA,~\cite{wan2017interference},
and the 2017 Asilomar Conference on Signals, Systems, and Computers, Pacific Grove, USA,~\cite{wan2017asilomar}.}
\thanks{
K.~Wan was with Laboratoire de Signaux et Syst\`emes (L2S, UMR8506), CentraleSup{\'e}lec-CNRS-Universit{\'e} Paris-Sud, 91192 Gif-sur-Yvette, France.
He is now with  Technische Universit\"at Berlin, Berlin, Germany (email: kai.wan@tu-berlin.de). The work of K.~Wan was  supported by Labex DigiCosme (project ANR11LABEX0045DIGICOSME) operated by ANR as part of the program ``Investissement d'Avenir'' Idex ParisSaclay (ANR11IDEX000302).}
\thanks{
D.~Tuninetti is with the Electrical and Computer Engineering Department, University of Illinois at Chicago, Chicago, IL 60607 USA (e-mail: danielat@uic.edu). The work of D. Tuninetti was supported in part by the National Science Foundation under award number 1527059.} 
\thanks{
M.~Ji is with the Electrical and Computer Engineering Department, University of Utah, Salt Lake City, UT 84112, USA  (e-mail: mingyue.ji@utah.edu). The work of M.~Ji was supported by NSF 1817154  and 1824558.}
\thanks{
P.~Piantanida is with CentraleSup{\'e}lec--French National Center for Scientific Research (CNRS)--Universit{\'e} Paris-Sud, 3 Rue Joliot-Curie, F-91192 Gif-sur-Yvette, France,  and with Montreal Institute for Learning Algorithms (MILA) at Universit{\'e} de Montr{\'e}al, 2920 Chemin de la Tour, Montr{\'e}al, QC H3T 1N8, Canada  (e-mail: pablo.piantanida@centralesupelec.fr). The work of  P.~Piantanida was supported by the European Commission's Marie Sklodowska-Curie Actions (MSCA), through the Marie Sklodowska-Curie IF (H2020-MSCAIF-2017-EF-797805-STRUDEL).}
}

\maketitle

\begin{abstract}
Caching is an efficient way to reduce network traffic congestion during peak hours by storing some content at  the users' local caches.
For the shared-link network with end-user-caches, Maddah-Ali and Niesen proposed a two-phase coded caching strategy.  
In practice, users may communicate with the server through intermediate relays. This paper studies the tradeoff between the memory size $\Msf$ and the network load  $\Rsf$ for the networks where a server with $\Nsf$ files is connected to $\Hsf$ relays (without caches), which in turn are connected to $\Ksf$ users equipped with caches of   $\Msf$ files. When each user is connected to a different subset of $\rsf$ relays, i.e., $\Ksf = \binom{\Hsf}{\rsf}$, the system is referred to as a {\it combination network with end-user-caches}. 

In this work, converse bounds are derived for the practically motivated case of {\it uncoded} cache contents, that is, bits of the various files are directly   pushed into  the user caches without any coding. In this case, once the cache contents and the users' demands are known, the problem reduces to a general index coding problem.
This paper shows that relying on a well-known ``acyclic index coding converse bound'' results in converse bounds that are not tight for combination networks with end-user-caches.    A novel converse bound that leverages the network topology is proposed,  which is the tightest converse bound known to date. 
As a result of independent interest, an inequality that generalizes the well-known sub-modularity of entropy is derived.

Several novel caching schemes are proposed, based on the  Maddah-Ali and Niesen  cache placement. These schemes leverage the structure of the combination network or/and perform interference elimination at the end-users.  
The proposed schemes are proved: 
(i) to be (order) optimal for some $(\Nsf,\Msf,\Hsf,\rsf)$ parameters regimes under the constraint of uncoded cache placement, and
(ii) to outperform the state-of-the-art schemes in numerical evaluations. 
\end{abstract}

\IEEEpeerreviewmaketitle{}

\section{Introduction}
\label{sec:intro}
 
\IEEEPARstart{D}{ue} to the high variability in network traffic, the network serving period can be divided into
peak traffic hours (where traffic is high and the network performance suffers)   
and off-peak hours (where traffic is low).
Caching effectively reduces peak-hour network traffic 
by storing some content at the users' local caches during peak-off hours. 
A caching scheme includes two phases:
(i) {\it Placement phase}: 
   the server pushes content into the users' local caches without knowledge of the users' future demands.
The placement is said to be {\it uncoded} if bits are directly copied into the cache; 
(ii) {\it Delivery phase}: after each user has requested one file and according to the cache contents, the server transmits packets in order to satisfy the users' demands. 
The goal is to minimize the number of transmitted bits (referred to as {\it load}) such that any set of   demands (referred to as worst-case demands) can be satisfied. 
The fundamental limits of {\it shared-link networks with end-user-caches} were originally studied by Maddah-Ali and Niesen (MAN) in~\cite{dvbt2fundamental}, where a server with $\Nsf$ files is connected to $\Ksf$ users through a shared error-free broadcast link and each user can cache up to $\Msf$ files.

\begin{figure}
\centerline{\includegraphics[scale=0.2]{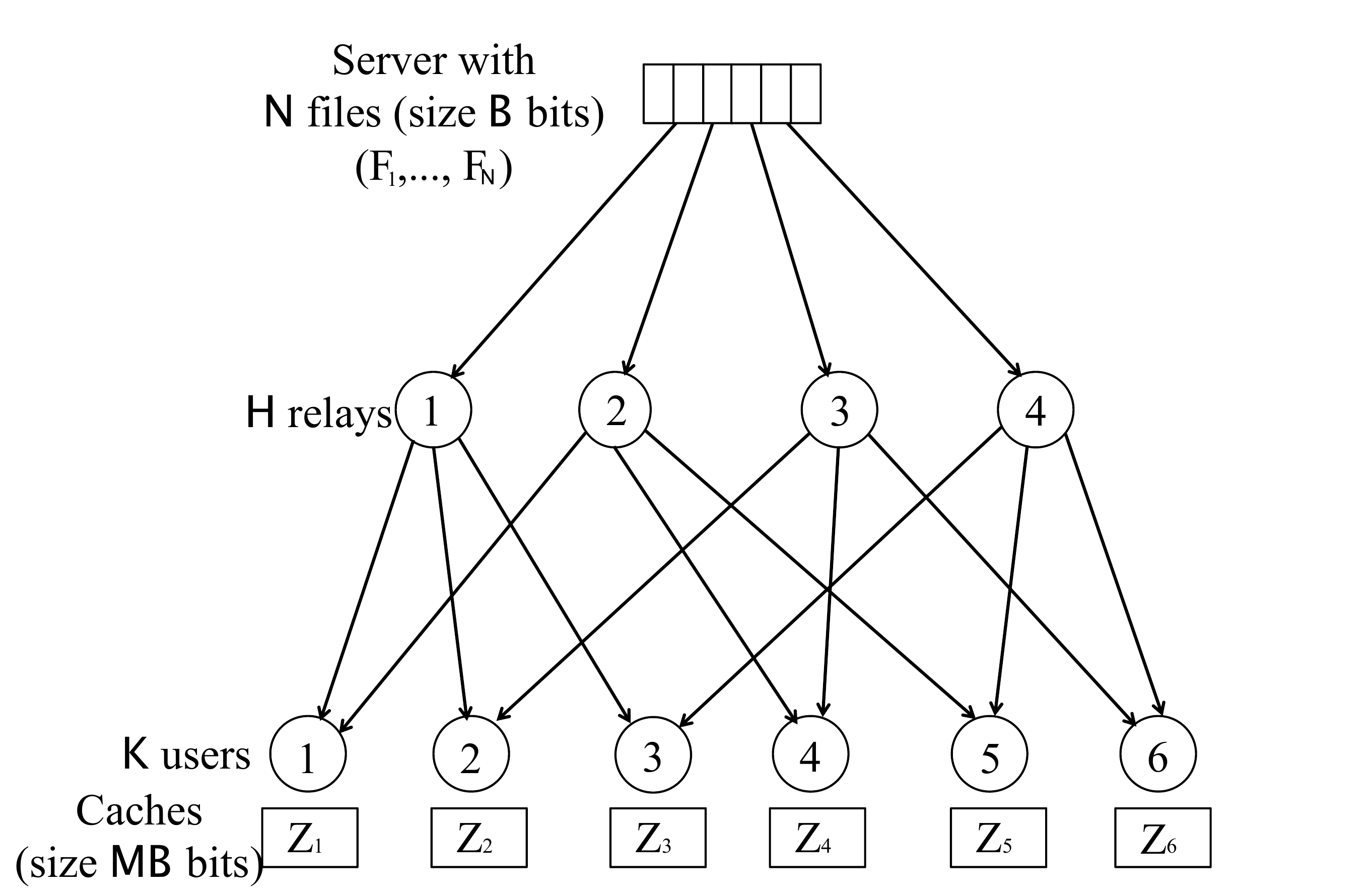}}
\caption{\small A combination network with end-user caches, with $\Hsf=4$ relays and $\Ksf=6$ users, i.e., $\rsf=2$. 
}
\label{fig: Combination_Networks}
\vspace{-5mm}
\end{figure}

Recently, a more practical case where users may communicate with the server through intermediate relays has gained attention.
As the analysis of relay networks with arbitrary topologies is challenging, we focus here on a symmetric version of this general problem known as {\it combination networks with end-user-caches}. This model was first proposed in~\cite{cachingincom,cachingJi2015}  in the context of network coding  and is shown in Fig.~\ref{fig: Combination_Networks}: 
a server with $\Nsf$ files is connected to $\Hsf$ relays (without caches), which in turn  are connected to $\Ksf = \binom{\Hsf}{\rsf}$ users, where each user is equipped with a cache of   $\Msf$ files and   is connected to a different subset of   $\rsf$ relays.
All links are assumed to be error-free and interference-free. 
The objective is to determine the optimal {\it max-link load} $\Rsf^{\star}$, defined as the smallest max-rate 
  (i.e., the maximum  number of bits sent on a link normalized by the file size, which is proportional to the overall download time)  
for the worst-case demands. {\it  The main contribution of this paper is to characterize
new (order) optimal results on the tradeoff  between the memory size $\Msf$ and the max-link load $\Rsf^{\star}$ under the constraint of uncoded cache placement. 
This is accomplished by deriving novel converse bounds and achievable schemes that leverage the network topology.}

In the rest of this section, we revise relevant past works 
    on cache-aided shared-link networks in Section~\ref{sub:shared link intro}, 
and on cached-aided combination networks in Section~\ref{sub:intro combination network}. 
Section~\ref{sub:contributions} provides the summary of our major contributions.

\subsection{Cache-aided Shared-link   Networks}
\label{sub:shared link intro}
For the shared-link model~\cite{dvbt2fundamental} Maddah-Ali and Niesen  (MAN) proposed a  coded caching scheme  that utilizes an uncoded combinatorial cache construction in the placement phase and a binary linear network code to generate multicast messages in the delivery phase. 
The achieved worst-case load of the MAN scheme satisfies
\begin{align}
 \Rsf_\textrm{\rm MAN}[t] = \frac{\Ksf-t}{1+t} = \frac{\binom{\Ksf}{t+1}}{\binom{\Ksf}{t}}
  \textrm{\rm for} \ \Msf = \Nsf \frac{t}{\Ksf} = \Nsf \frac{ \binom{\Ksf-1}{t-1} }{ \binom{\Ksf}{t} },   t\in[0:\Ksf],
\label{eq:cMANloadupper}
\end{align}
and   for $\Msf \frac{\Ksf}{\Nsf} \not= t  \in[0:\Ksf]$ one takes the lower convex envelope of the set of points $(\Msf,\Rsf)=\left(t \frac{\Nsf}{\Ksf}, \Rsf_\textrm{\rm MAN}[t]\right)$. 
When a file is requested by multiple users, Yu, Maddah-ali and Avestimehr (YMA)~\cite{exactrateuncoded} found 
that among the $\binom{\Ksf}{t+1}$ MAN multicast messages,     $\binom{\Ksf-\min(\Ksf,\Nsf)}{t+1}$ of them can be obtained as linear combinations of the others. 
This observation led to the achievable load
\begin{align}
 \Rsf_\textrm{\rm YMA}[t] = \frac{ \binom{\Ksf}{t+1} - \binom{\Ksf-\min(\Ksf,\Nsf)}{t+1} }{ \binom{\Ksf}{t} }
\ \textrm{\rm for} \ \Msf =\Nsf \frac{t}{\Ksf}, \ t\in[0:\Ksf].
\label{eq:cYMAloadupper}
\end{align}

 
In \cite[Theorem 2]{dvbt2fundamental}, Maddah-Ali and Niesen also derived a cut-set converse bound, which proved that the load of the MAN caching scheme in~\eqref{eq:cMANloadupper}  is optimal within a factor of $12$.   
An enhanced converse bound was proposed in~\cite[Theorem 2]{yufactor2TIT2018}  to prove that the load of the YMA caching scheme  in~\eqref{eq:cYMAloadupper} is order optimal within a factor   of    $2$. In other words, coded cache placement can at most half the network load compared to the best caching scheme(s) with uncoded cache placement.

As it is difficult to characterize the exact optimality, the optimality under the constraint of uncoded cache placement was originally considered in~\cite{ontheoptimality,indexcodingcaching2020}. By relating the caching problem under the constraint of uncoded placement to the index coding problem~\cite{birk1998informedsource}, the ``acyclic index coding converse bound'' in~\cite[Corollary 1]{onthecapacityindex} was exploited in~\cite{indexcodingcaching2020} to derive a converse bound on the load for the shared-link     networks under the constraint of uncoded placement. The same converse bound was also obtained   by invoking a genie-based idea in~\cite{exactrateuncoded}. By comparing this ``acyclic index coding converse bound''-based converse bound and the loads in~\eqref{eq:cMANloadupper} and~\eqref{eq:cYMAloadupper}, it was proved that the MAN caching scheme and the YMA caching scheme are exactly optimal under the constraint of uncoded placement when $\Nsf\geq \Ksf$ and $\Nsf<\Ksf$, respectively.

\subsection{Cache-aided Combination Network}
\label{sub:intro combination network} 
Past works on coded caching schemes for combination networks with end-user-caches can be divided into two classes, with uncoded or coded cache placement. 

In~\cite{cachingincom} two achievable schemes with the MAN uncoded cache placement were proposed. In the delivery phase, the first scheme uses routing while the second scheme uses Minimum Distance Separable (MDS) codes to deliver the MAN multicast messages to the users. The achieved max-link load in~\cite{cachingincom} is  
\begin{align}
\Rsf_{{\rm base}}[t] = & \min\Big\{\frac{\Ksf(1-\Msf/\Nsf)}{\Hsf}, \frac{\Ksf(1-\Msf/\Nsf)}{\rsf(1+\Ksf\Msf/\Nsf)}\Big\}, \nonumber\\& \textrm{\rm for} \ \Msf=\Nsf \frac{t}{\Ksf}, t\in[0:\Ksf].
\label{eq:Mingyue inner bound}
\end{align} 
An achievable scheme was proposed in~\cite{Li2016coded} for the case where $\rsf$ divides $\Hsf$; the idea was to split the combination network into $\Hsf$ shared-link   networks, in each of which the scheme  in~\cite{dvbt2fundamental} is used, so as to achieve the max-link load 
\begin{align}
\Rsf_{\textrm{\rm ZY}}[t^{\prime}]=\frac{\Ksf(1-\Msf/\Nsf)}{\Hsf\big(1+\frac{\Ksf\Msf\rsf}{\Hsf\Nsf} \big)}, \ \textrm{\rm for} \ \Msf=\Nsf\frac{\Hsf t^{\prime}}{\Ksf\rsf}, t^{\prime}\in [ 0: \Ksf \rsf / \Hsf ].
\label{eq:Yener load}
\end{align}
The Placement Delivery Array (PDA) scheme, originally proposed in~\cite{ontheplacementarrray} to reduce the sub-packetization of MAN caching scheme in the shared-link model, was  extended  to combination networks in~\cite{PDA2017yan} for the case where $\rsf$ divides $\Hsf$; it was shown to achieve the same load as in~\eqref{eq:Mingyue inner bound} and~\eqref{eq:Yener load} but with lower sub-packetization. For some other specific cases, improved constructions of PDA for combination networks were proposed in~\cite{pdaconmibationcont}.
%

A caching scheme with MDS coded placement was proposed in~\cite{combinationsecu2018Zewail}, which achieves the load in~\eqref{eq:Yener load} but without the constraint that $\rsf$ divides $\Hsf$. The same authors of this paper further improved on the coded placement scheme in~\cite{combinationsecu2018Zewail} by proposing an asymmetric coded cache placement in~\cite{ourcodedISIT}, and leveraging the symmetries in the topology to generate and deliver the multicast messages as in~\cite{wan2017novelmulticase}, which is not discussed here as in this paper we focus on achievable schemes and converse bounds with uncoded cache placement.

The cut-set converse bound for the shared-link model in~\cite[Theorem 2]{dvbt2fundamental} was extended to combination networks in~\cite{cachingincom}. 
 It was proved in~\cite{cachingJi2015} that the cut-set converse bound in~\cite{cachingincom} and the achieved max-link load in~\eqref{eq:Mingyue inner bound}
 are to within a factor of $12$ in the ``large memory'' regime $\Msf/\Nsf \geq 1/(2\rsf)$. 
However, the gap may be large when the ``small memory'' regime 
and this paper aims to address this open problem.\footnote{\label{foot:yener scheme} The caching scheme in~\cite{combinationsecu2018Zewail}, which achieved the load $\Rsf_{\textrm{\rm ZY}}$ in~\eqref{eq:Yener load}, generally improves on $\Rsf_{{\rm base}}$ in~\eqref{eq:Mingyue inner bound}. In addition, when the memory size is small, $\Rsf_{\textrm{\rm ZY}}$ approaches the first term in $\Rsf_{{\rm base}}$; when   the memory size is large, $\Rsf_{\textrm{\rm ZY}}$ approaches the second term in $\Rsf_{{\rm base}}$. However, $\Rsf_{\textrm{\rm ZY}}$  does not induce any order improvement compared to  $\Rsf_{{\rm base}}$.}

\subsection{Main Contributions} 
\label{sub:contributions}
Our main contributions are as follows.
\begin{enumerate}

\item {\it  Achievability:}
Based on the MAN placement, we propose four achievable schemes for combination networks with end-user-caches
to deliver the MAN multicast messages to the corresponding users:
\begin{enumerate}
\item We first propose a novel delivery scheme, named {\it Direct Independent delivery Scheme} (DIS), that exploits the fact that not all the MAN multicast messages are useful to every user.
\item For $\Msf=\Nsf/\Ksf$, we propose an {\it Interference Elimination delivery Scheme} (IES) that uses interference elimination (a form of interference alignment) to 
encode the  MAN multicast messages such that each user can cancel the interference caused by the MAN multicast messages that are not of interest.  
\item The scheme, named {\it Concatenated Inner Code delivery Scheme} (CICS), proposes a coded delivery scheme composed of two steps: (i) in the first step we directly transmit each MAN multicast message to some relay(s), which forward them to their connected users; such messages are simultaneously useful for $t=\Ksf\Msf/\Nsf$ users and
will be used as `side information' in the next step;
(ii) 
in the second step, we deliver the MAN multicast messages through a carefully designed network code that lets the remaining `unsatisfied' users recover their demanded files. In this step, a multiplicative coding gain can be achieved. 
\item By leveraging the multicasting opportunities which are ignored in the CICS, 
we finally propose a scheme, named {\it Improved Concatenated Inner Code delivery Scheme} (ICICS).
\end{enumerate}

\item {\it Converse:}
For combination networks with end-user-caches, we propose several converse bounds 
under the constraint of uncoded cache placement when $\Nsf\geq \Ksf$.\footnote{\label{foot:converse} The case $\Nsf< \Ksf$ can be treated similarly, by simply considering the users with distinct demands; this is not discussed here for sake of space; more details can be found in Remark~\ref{rem:N<K}.}

Based on the cut-set strategy in~\cite{cachingincom}, we firstly  extend the shared-link converse bound under the constraint of uncoded placement to
combination networks with end-user-caches, and propose a  converse bound based on the acyclic index coding converse bound~\cite{onthecapacityindex}.

Furthermore, 
by deriving bounds on the joint entropy of the various random variables that define the problem,
we provide a novel converse bound which tightens the acyclic index coding converse bound,
and produces the best known converse bound to date, to the best of our knowledge.

As a result of independent interest, an inequality that generalizes the well-known sub-modularity of entropy is derived, which may find applications in other network information theory problems.

\item {\it Optimality:}
By comparing the proposed achievable schemes and the proposed converse bounds under the constraint of uncoded cache placement, we obtain the (order) optimality results   summarized in Table~\ref{tab:optimality} at the top of the next page. For general $(\Nsf,\Msf,\Hsf,\rsf)$ with $\Nsf \geq \Ksf=\binom{\Hsf}{\rsf}$, the achieved max-link load in~\eqref{eq:Mingyue inner bound} was proved to be order optimal within a factor of  $12$ when $\frac{\Msf}{\Nsf}  \geq \frac{1}{2\rsf}  $ (i.e., large memory regime)~\cite{cachingJi2015}.
In this paper,  under the constraint of uncoded cache placement, we obtain the order optimality results   for $\Msf$ is small, in particular for $\frac{\Msf}{\Nsf} \leq \frac{\rsf}{\Ksf}$ where the CICS is order optimal within a factor of $2$. Hence, the order optimality under the constraint of uncoded cache placement for the regime $\frac{\rsf}{\Ksf} < \frac{\Msf}{\Nsf} < \frac{1}{2\rsf} $ remains open.


\begin{table*}
 \centering
\protect\caption{Order optimality results under the constraint of uncoded placement and $\Nsf\geq \Ksf$.}\label{tab:optimality}
\begin{tabular}{|c|c|c|}
\hline 
Delivery Scheme & Constraint of system parameters & Optimality under the constraint of uncoded cache placement\tabularnewline
\hline 
\hline 
DIS,CICS, ICICS &  $\Msf \leq \frac{\Nsf }{\Ksf} \left\lfloor \frac{\rsf}{\Hsf-\rsf} \right\rfloor $ & optimal \tabularnewline
\hline 
DIS,CICS, ICICS &   $\rsf=\Hsf-1$ & optimal \tabularnewline
\hline 
Combining CICS and~\cite{cachingJi2015}&   $\rsf=\Hsf-2$& order optimal within a factor of $12$           \tabularnewline
\hline 
CICS &   $\Msf\leq \rsf \Nsf/\Ksf$ & order optimal within  a factor of  $2$\tabularnewline
\hline 
CICS &   $\Msf=\Nsf t/\Ksf,$ $t/\rsf\to 0$ & optimal  \tabularnewline
\hline 
IES & $\Hsf\leq 2\rsf,$   $\Msf\leq \Nsf/\Ksf$ & optimal  \tabularnewline
\hline 
IES & $\Hsf >2\rsf,$   $\Msf\leq \Nsf/\Ksf$ & order optimal within a factor of $2\rsf/(2\rsf-1)\leq 4/3$  \tabularnewline
\hline 
 & $\frac{\rsf}{\Ksf} < \frac{\Msf}{\Nsf} < \frac{1}{2\rsf} $  &  open regime \tabularnewline
\hline 
\end{tabular}
\end{table*}

\item {\it Numerical Comparisons:}
Numerical results show that the proposed bounds outperform the state-of-the-art schemes. 

\end{enumerate}

\subsection{Paper Organization}
The rest of the paper is organized as follows. 
Section~\ref{sec:model} presents the system model and some  relevant past  results. 
Section~\ref{sec:main results} introduces our main results. 
Section~\ref{sec:general inner bound} provides the proofs of the proposed   novel  delivery schemes. 
Section~\ref{sec:outer bound} provides the proofs of the converse bounds. 
Section~\ref{sec:conclusion and future work} concludes the paper.
Some technical proofs are relegated in Appendix.

\subsection{Notation Convention}
\label{sub:notation}
Calligraphic symbols denote sets, 
bold symbols denote vectors, and 
sans-serif symbols denote system parameters. 
The main network parameters and notations  are given in Table~\ref{tab:notations} at the top of the next page.
We use $|\cdot|$ to represent the cardinality of a set or the length of a vector;
$X_{\Jc}:=\{X_{i}:i\in \Jc\}$;
$[a:b]:=\left\{ a,a+1,\ldots,b\right\}$ and $[n] := [1:n]$;
$\mathcal{A\setminus B}:=\left\{ x\in\Ac : x\notin\Bc\right\}$; 
$\arg \max_{x\in \Xc}f(x) := \big\{x\in \Xc:f(x)=\max_{x\in\Xc}f(x)\big\}$;
$\mathbf{p}(\Jc):=\big(p_{1}(\Jc),\ldots,p_{|\mathbf{p}(\Jc)|}(\Jc)\big)$ represents a permutation of elements of the set $\Jc$;
$\text{RLC}(m,\Sc)$ represents $m$ random linear combinations of the equal-length packets indexed by $\Sc$.  
We note that $m$ random linear combinations of $|\Sc|$ packets are linearly independent with high probability if operations are done on a large enough finite field;
the same can be obtained by using the parity-check matrix of an $(|\Sc|, |\Sc|-m)$ MDS (Maximum Distance Separable) code~\cite{detailledisit}.

\begin{table*}
\centering
\protect\caption{Notations and Acronyms}\label{tab:notations}
\begin{tabular}{|c|c|}
\hline 
\textbf{Notations} & \textbf{Semantics} \tabularnewline
\hline  
 $\Hsf$& number of relays \tabularnewline
\hline  $\rsf$ & number of relays connected to each user\tabularnewline
\hline 
 $\Ksf$& number of users \tabularnewline
\hline  $\Nsf$ & number of files \tabularnewline
\hline 
 $\Msf$&  cache size  in multiple of the file size \tabularnewline
\hline  $\Rsf$ & load \tabularnewline
\hline 
 $\Bsf$ & number of bits per file \tabularnewline
\hline   $\Hc_k$  &  set of relays connected to user $k$  \tabularnewline
\hline 
 $\Hc_{\Wc}$& set of relays connected to some user in $\Wc$ \tabularnewline
\hline $\Uc_h$ &  set of users connected to  relay $h$  \tabularnewline
\hline 
 $\Kc_{\Jc}$ & set of users whose connected relays are all in $\Jc$ \tabularnewline
\hline   $\Rc_{\Jc}$  &  set of relays connected to all the users in $\Jc$\tabularnewline
\hline 
MAN & coded caching scheme proposed by Maddah-Ali and Niesen \tabularnewline
\hline 
 YMA & coded caching scheme proposed by Yu, Maddah-ali and Avestimehr \tabularnewline
\hline DIS & Direct Independent delivery Scheme \tabularnewline
\hline 
 IES & Interference Elimination delivery Scheme \tabularnewline
\hline  CICS & Concatenated Inner Code delivery Scheme \tabularnewline
\hline 
 ICICS  &Improved Concatenated Inner Code delivery Scheme \tabularnewline
\hline  
\end{tabular}
\end{table*}

\section{System Model and Some Known Results} 
\label{sec:model}

\subsection{General System Model}
\label{sub:system model}
Consider the $(\Hsf,\rsf,\Msf,\Nsf)$ combination network with end-user-caches illustrated in Fig.~\ref{fig: Combination_Networks}.
The server has access to $\Nsf$ files denoted by $\{F_1, \cdots, F_\Nsf\}$, each composed of $\Bsf$   independent and uniformly distributed  bits.
The server is connected to $\Hsf$ relays through $\Hsf$ error-free and interference-free links. 
The relays are connected to $\Ksf = \binom{\Hsf}{\rsf}$ users  
through $\rsf \, \Ksf$ error-free and interference-free links.  Each user is  connected to a distinct subset of $\rsf$ relays. 
%
The set of users connected to the $h$-th relay is denoted by $\Uc_{h}, \ h\in[\Hsf]$. 
The set of relays connected to $k$-th user is denoted by $\Hc_{k}, k\in[\Ksf]$. For each set of users $\Wc\subseteq [\Ksf]$, let $\Hc_{\Wc}=\underset{k\in \Wc}{\cup}\Hc_{k}$.
For each set of relays $\Jc\subseteq[\Hsf]$ such that $|\Jc| \geq \rsf$, let $\Kc_{\Jc}:=\{k\in[\Ksf]:\Hc_{k}\subseteq\Jc\}$ be the set of users whose connected relays are all in $\Jc$.   For each set of users $\Wc\subseteq[\Ksf]$, define $\Rc_{\Wc}=\{h\in[\Hsf]:\Wc\subseteq\Uc_{h}\}$ as the set of relays connected to all the users in $\Wc$. 
For the network in~Fig.~\ref{fig: Combination_Networks}, we have, for example, 
$\Uc_{1}=\{1,2,3\}$, 
$\Hc_{1}=\{1,2\}$,
$\Kc_{\{1,2,3\}}=\{1,2,4\}$ and $\Rc_{\{1,2\}}=\{1\}$.

The system works in two phases.
In the placement phase, user $k\in[\Ksf]$ stores information about the $\Nsf$ files in its cache of   $\mathsf{MB}$ bits, where $\Msf \in[0,\Nsf]$.  
We denote the content in the cache of user $k\in[\Ksf]$ by $Z_{k}$ and let $\Zm:=(Z_{1},\ldots,Z_{\Ksf})$.
During the delivery phase, user $k\in[\Ksf]$ demands file $F_{d_{k}}$ where $d_k\in[\Nsf]$;
the demand vector $\dv:=(d_{1},\ldots,d_{\Ksf})$ is revealed to all nodes. 
Given $(\dv,\Zm)$, the server sends a message $X_{h}$ 
of $\Bsf \, \Rsf_{h}(\dv,\Zm,\Msf)$ bits to relay $h\in [\Hsf]$. 
Then, relay $h\in [\Hsf]$ transmits a message $X_{h\to k}$ 
of $\Bsf \, \Rsf_{h\to k}(\dv,\Zm,\Msf)$ bits to user $k \in \Uc_h$. 
User $k\in[\Ksf]$ must recover its desired file $F_{d_{k}}$ from $Z_{k}$ and $(X_{h\to k} : h\in \Hc_k)$ with high probability for some file size $\Bsf$. 


The max-link load is  defined as 
\begin{align}
\Rsf  (\Msf)
 =
\max_{\substack{h\in[\Hsf],\\ k\in\Uc_h\\ \dv\in[\Nsf]^{\Ksf}}}\max
\left\{
\Rsf_h(\dv,\Zm,\Msf),\Rsf_{h\to k}(\dv,\Zm,\Msf)
\right\}.
\end{align}
Our objective is to determine the minimum  max-link load $\Rsf^{\star} (\Msf)$ over all possible caching schemes. 
Obviously, for each relay $h$, the load on the link from the server to $h$ should not be less than the load on each link from relay $h$ to user $k\in\Uc_{h}$.  
So   in our achievable schemes,  we assume that each relay forwards all of its received packets from the server to its connected users. 

\subsection{Systems with Uncoded Cache Placement}
\label{sub:uncoded cache placement}
The minimum max-link load under the constraint of uncoded cache placement is denoted by $\Rsf^{\star}_{\mathrm{u}}(\Msf)$. 
In general, $\Rsf^{\star}_{\mathrm{u}}(\Msf) \geq \Rsf^{\star}(\Msf)$. In the rest of the paper, we simplify  $\Rsf^{\star}_{\mathrm{u}}(\Msf)$ and $\Rsf^{\star}(\Msf)$ by $\Rsf^{\star}_{\mathrm{u}}$ and $\Rsf^{\star}$, respectively. 

After the uncoded placement phase is concluded, each file can be effectively divided into non-overlapping subfiles depending on which user stores which bit.
Let 
\begin{align}
\Tc_{\Zm} := \big\{ F_{i,\Wc}: i\in [\Nsf], \ \Wc\subseteq[\Ksf] \big\}, 
\end{align}
where $F_{i,\Wc}$ is the set of bits of the file $F_{i}$ uniquely stored  by the users in $\Wc$.
After  the demand vector is revealed,  the set of requested subfiles according to the demand vector $\dv\in[\Nsf]^{\Ksf}$ is denoted by
\begin{align}
\Tc_{\dv,\Zm}:=\big\{ F_{d_{k},\Wc}:k\in [\Ksf], \ \Wc\subseteq[\Ksf], \ k\notin\Wc \big\} \subseteq \Tc_{\Zm}.
\label{eq:Tc{dv,Zm}}
\end{align}

In the delivery phase, the delivery of the subfiles in $\Tc_{\dv,\Zm}$ is equivalent to an index coding problem and can thus be represented by a directed graph $G_{\Tc_{\dv,\Zm}}$ (i.e., known as side information graph) as follows. Each node in the graph represents one subfile demanded by one user only (if the same file is demanded by multiple users, only one such user is considered at a time). A directed edge from node $i$ to node $i^\prime$ exists if the subfile represented by node $i$ is in the cache of the user who requests the subfile represented by node $i^\prime$. 
%
If $\Jc$ is a subset of vertices in the graph $G_{\Tc_{\dv,\Zm}}$  
where the subgraph of $G_{\Tc_{\dv,\Zm}}$ over $\Jc$ does not contain a directed cycle, 
then the ``acyclic index coding converse bound'' (as used in~\cite{indexcodingcaching2020} for the  shared-link   network) 
can be used to lower bound   the max-link load as a function of the total number of bits of the subfiles in $\Jc$ (see Proposition~\ref{prop:acyclic outer bound}).

\subsection{MAN Coded Caching Scheme  in the Shared-Link Network}
\label{sub:known results}
Next, we revisit the MAN coded caching scheme for the shared-link model, which will be used in our novel proposed caching schemes for  combination networks. In the shared-link model, all the $\Ksf$ users are directly connected to the central server through an error and interference free shared-link.

Let 
$\Msf=t\frac{\Nsf}{\Ksf}$
for some positive integer $t\in[\Ksf]$. 
In the placement phase, each file is split into $\binom{\Ksf}{t}$ non-overlapping equal size subfiles
of $\frac{\Bsf}{\binom{\Ksf}{t}}$ bits. The subfiles of $F_{i}$ are denoted by $F_{i,\Wc}$
for $\Wc\subseteq[\Ksf]$ where $|\Wc|=t$. 
User $k\in [\Ksf]$ 
fills its cache as
\begin{align}
Z_k 
= \Big( 
F_{i,\Wc}
:  k\in\Wc, 
\ \Wc\subseteq[\Ksf], 
\ |\Wc|=t,
\ i\in[\Nsf]
\Big).
\label{eq:cMAN cache function}
\end{align}
In the delivery phase, given the demand vector $\mathbf{d}$, 
the server 
transmits the multicast messages 
\begin{align}
W_{\Jc}:=\oplus_{k\in\Jc}F_{d_{k},\Jc\backslash\{k\}}, \forall \Jc\subseteq [\Ksf] : |\Jc|=t+1.
\label{eq:cMAN multicast messages}
\end{align}
Note that user $k\in\Jc$
wants $F_{d_{k},\Jc\backslash\{k\}}$ and knows $F_{d_{j},\Jc\backslash\{j\}}$ 
for all $j\in (\Jc\setminus \{k\})$, so it can recover $F_{d_{k},\Jc\backslash\{k\}}$ from the MAN multicast message and its cache content. 
Thus the load of the MAN caching scheme is given by~\eqref{eq:cMANloadupper}.

\subsection{Results from~\cite{cachingincom} for Combination Networks} 
\label{sub:Mingyue paper}

\subsubsection*{\textbf{Achievable Schemes}}
With the MAN placement in~\eqref{eq:cMAN cache function}, the authors in~\cite{cachingincom} proposed two delivery schemes for combination networks. 

In the first scheme, for each user $k$, we divide the bits of $F_{d_k}$ which are not cached by user $k$ into $\rsf$ non-overlapping and equal-length pieces, and transmit one different piece to each relay in $\Hc_k$, such that the max-link load is the first term of~\eqref{eq:Mingyue inner bound}. 

In the second scheme, the MAN multicast messages in~\eqref{eq:cMAN multicast messages} are generated. For each multicast message $W_{\Jc}$, we divide it into $\rsf$ non-overlapping and equal-length pieces, which are then encoded by an $(\Hsf,\rsf)$ MDS code. We then transmit one different MDS symbol of $W_{\Jc}$ to each relay $h\in[\Hsf]$. The max-link load is the second term of~\eqref{eq:Mingyue inner bound}.

\subsubsection*{\textbf{Cut-set Converse Bound}}
Each time  one cut including $\lceil \alpha \rsf\rceil\in[\rsf:\Hsf]$ relays and $\binom{\lceil \alpha \rsf\rceil}{\rsf}$ users  is considered.  
The total load from the server to the  $\lceil \alpha \rsf\rceil$ relays should not be less than the converse bound of the 
load in a shared-link system  with $\Nsf$ files and $\binom{\lceil \alpha \rsf\rceil}{\rsf}$ users. Then the max-link load is lower bounded by this total load divided by the number 
of relays in the cut.
By using this strategy, the cut-set converse bound in~\cite{dvbt2fundamental} was extended to combination networks in~\cite{cachingincom} to show that the optimal   max-link  load $\Rsf^{\star}(\Msf)$ must satisfy
\begin{align}
& \Rsf^{\star}(\Msf) \geq  
\max_{\alpha \in [1,\frac{\Hsf}{\rsf}]}   \frac{1}{\lceil \alpha \rsf\rceil}
\max_{l \in \left[ \min\left\{\Nsf,{\lceil \alpha \rsf\rceil \choose \rsf}\right\} \right]} \left(l - \frac{l}{\lfloor\frac{\Nsf}{l}\rfloor}\Msf\right). \label{eq: Mingyue converse}
\end{align}

\section{Main Results on  Combination Networks  with End-Cache-Users}
\label{sec:main results}
In this section, we introduce our main results on  combination networks with end-cache-users, including  novel achievable schemes in Section~\ref{sub:results on inner bounds}, novel converse bounds in Section~\ref{sub:results on outer bounds}, and finally optimality results in Section~\ref{sub:optimality results}.

\subsection{Novel Achievable Schemes}
\label{sub:results on inner bounds}
In this paper we propose four novel delivery schemes for combination networks with end-user-caches by using  
the MAN placement in~\eqref{eq:cMAN cache function} and  multicast message generation in~\eqref{eq:cMAN multicast messages}.
The novelty of our schemes is on the delivery of the MAN multicast messages.  Recall that $\Rc_{\Jc}$ is the set of relays
connected to all the users in $\Jc$.  For each   $t \in [0:\Ksf]$,
we define   
\begin{subequations}
\begin{align}
& \Vc_{1}:=\{\Jc\subseteq [\Ksf]:|\Jc|=t+1,\Rc_{\Jc}\neq\emptyset\}, \label{eq:def of v 1} \\
& \Vc_{2}:=\{\Jc\subseteq [\Ksf]:|\Jc|=t+1,\Rc_{\Jc}=\emptyset\}, \label{eq:def of v 2}
\end{align}
\end{subequations}
where $\Vc_{1}$ and $\Vc_{2}$ represent the families of the subsets with cardinality $t+1$ of the set of users who have and do not have a common connected relay, respectively.

In this paper, we mainly propose four novel achievable schemes: (i) Direct Independent delivery Scheme (DIS) in Section~\ref{sub:DIS}, which improves on the second scheme in~\cite{cachingincom}  by observing that the multicast message $W_{\Jc}$ is only useful to the users in $\Jc$ and thus we need not   let all users recover it; (ii) Interference Elimination delivery Scheme (IES) in Section~\ref{sub:IES}, based on a novel topological interference alignment idea; (iii) Concatenated Inner Code delivery Scheme (CICS) in Section~\ref{sub:CICS}, which includes two steps to deliver each $W_{\Jc}$ to the users in $\Jc$; (iv)
Improved Concatenated Inner Code delivery Scheme (ICICS)  in Section~\ref{sub:ICICS}, which     leverages the ignored multicasting opportunities of the CICS. The achievable loads of the proposed scheme is given in the following theorem.\footnote{\label{foot:no load} The ICICS is expressed as an algorithm, for which we could not find a closed form solution for the load.}

\begin{thm}[Achievable schemes]
\label{thm:achievability}
For the $(\Hsf,\rsf,\Msf,\Nsf)$ combination network with end-user-caches,
the optimal tradeoff under the constraint of uncoded cache placement $(\Msf,\Rsf^{\star}_{\mathrm{u}})$  is upper bounded by the lower convex envelop of the following  points: 
\begin{itemize}
\item 
 $\left(\frac{\Nsf t}{\Ksf},  \Rsf_{\mathrm{DIS}}[t] \right)$ for each $t\in [0:\Ksf]$, where 
\begin{align}
\Rsf_{\mathrm{DIS}}[t]
:=\frac{\sum_{\mathcal{J}_1 \in \Vc_1} 1
+ \sum_{\mathcal{J}_2 \in \Vc_2} |\{h\in[\Hsf]:\Uc_h\cap\Jc_2\neq 0\}|}{\Hsf\binom{\Ksf}{t} };\label{eq: load of DIS}
\end{align}
\item 
\begin{align}
\left(\frac{\Nsf  }{\Ksf}, \Rsf_{\mathrm{IES}}\right)&= \left(\frac{\Nsf  }{\Ksf},\frac{1}{2\Hsf}\left(\Ksf-1-\binom{\Hsf-\rsf}{\rsf}\right)+ \right. \nonumber\\& \left. \frac{\binom{2\rsf-1}{\rsf-1}}{(2\rsf-1)\Ksf}\binom{\Hsf-1}{2\rsf-1}\right), \label{eq:achievable interference load}
\end{align}
  if   $2\rsf-1=p^{v}$ or $2\rsf-1 = pq$ where $p,q$ are different primes and $v$ is a positive integer;
  \item  $\left(\frac{\Nsf t}{\Ksf}, \Rsf_{\mathrm{CICS}}[t] \right)$ for each $t\in [0:\Ksf]$, where 
  \begin{subequations}
  \begin{align}
\Rsf_{\mathrm{CICS}}[t] &:= 
\sum_{\Jc\subseteq[\Ksf]:|\Jc|=t+1}\frac{1+\min_{h\in[\Hsf]}|\Jc\setminus \Uc_h|/\rsf}{\Hsf\binom{\Ksf}{t}} \\& \leq \binom{\Ksf}{t+1}\frac{1+t/\rsf}{\Hsf\binom{\Ksf}{t}}.
\end{align} 
\label{eq:load of CICS}
  \end{subequations}
\end{itemize}
\end{thm}

Comparing the loads in~\eqref{eq:Mingyue inner bound} and~\eqref{eq: load of DIS}, we have the following corollary, whose proof is also in Section~\ref{sub:DIS}.
\begin{cor}
\label{cor:strictly better than MJ}
For the $(\Hsf,\rsf,\Msf,\Nsf)$ combination network with end-user-caches, it holds that
\begin{align}
\Rsf_{\mathrm{DIS}}\leq  \Rsf_{{\rm base}} .
\end{align}
\end{cor}

\begin{rem}
\label{foot:conjecture}
As mentioned in Theorem~\ref{thm:achievability}, the IES  works for  the combination networks with   $2\rsf-1=p^{v}$ or $2\rsf-1 = pq$, where $p,q$ are different primes and $v$ is a positive integer.
The smallest value of $\rsf$ not satisfying the above condition is $23$. If $\rsf=23$ and $\Hsf=2\rsf=46$, in the network there are more than $8.23\times 10^{12}$ users, which is not practical. 
Note that by the proof of Theorem~\ref{thm:circulant matrix rank} in Appendix~\ref{sec:discussion of full rank} (upon which the above condition is built), the circulant matrix in Theorem~\ref{thm:circulant matrix rank} may not be invertible when $\rsf \geq 23$ (i.e., when $\rsf=45$);  but this does not imply that Problem 1 in Appendix~\ref{sec:discussion of full rank} does not 
have a solution for $\rsf \geq 23$. We conjecture that the IES  works in general.
\end{rem}

\subsection{Novel Converse bounds}
\label{sub:results on outer bounds}
First we use the cut-set strategy of the converse bound in~\eqref{eq: Mingyue converse} to extend the converse bound in~\cite[Theorem 2]{yufactor2TIT2018} for the shared-link model to combination networks with end-user-caches. 
\begin{thm}[Enhanced cut-set, any placement]
\label{thm:enhanced outer bound combination}
For the $(\Hsf,\rsf,\Msf,\Nsf)$ combination network with end-user-caches, it must satisfy that
\begin{align}
\Rsf^{\star}&\geq 
 \max_{x\in[\rsf:\Hsf]} \max_{s \in \left[ \min\{\Nsf,\binom{x}{\rsf}\} \right]} \max_{\alpha\in[0,1]} \nonumber\\&  \frac{ s-1+\alpha-\frac{s(s-1)-l(l-1)+2\alpha s}{2(\Nsf-l+1)}\Msf}{x}, 
\label{eq:enhanced outer bound combination}
\end{align}
 where $l\in [s]$  is the minimum value such that $\frac{s(s-1)-l(l-1)}{2}+\alpha s\leq(\Nsf-l+1)l^{2}$.
\end{thm}

\begin{IEEEproof}
Each time we consider a cut with $x$ relays. The  total 
 load  transmitted 
from the server to this $x$ relays 
 is  lower bounded by the load in a shared-link model with $\Nsf$ files, $\binom{x}{\rsf}$ users and memory of $\Msf\Bsf$ bits. So we can use this strategy to extend any converse bound for the shared-link model  to combination networks. Theorem~\ref{thm:enhanced outer bound combination} uses the enhance cut-set converse bound in~\cite[Theorem 2]{yufactor2TIT2018}.
\end{IEEEproof}

Similarly to Theorem~\ref{thm:enhanced outer bound combination},
we next extend the shared-link converse bound under the constraint of uncoded placement 
to combination networks. 
\begin{thm}[Cut-set, uncoded placement]
\label{thm:cut-set outer bound combination}
For the $(\Hsf,\rsf,\Msf,\Nsf)$ combination network with end-user-caches,   the optimal tradeoff under the constraint of uncoded cache placement $(\Msf,\Rsf^{\star}_{\mathrm{u}})$  is lower bounded by the lower convex envelop of    
\begin{align}
& \left(t\frac{\Nsf}{\binom{x}{\rsf}}, \frac{1}{x} \ \frac{ \binom{\binom{x}{\rsf}}{t+1}-\binom{\max\Big\{\binom{x}{\rsf}-\Nsf,0\Big\}}{t+1}}{\binom{\binom{x}{\rsf}}{t}}\right),  \nonumber\\
&\forall  t\in \left[0:\binom{x}{\rsf}\right],  x\in[\rsf:\Hsf].
\label{eq:cut-set outer bound combination}
\end{align}
\end{thm}

We then propose the following converse bound 
which is tighter than the cut-set bound in Theorem~\ref{thm:cut-set outer bound combination}. 
As this result follows 
straightforwardly from the work 
 for shared-link network done by some of the authors of this paper in~\cite{indexcodingcaching2020}, 
 here we consider the resulting bound as our `baseline' bound. The proof is in  Section~\ref{subset:acyclic outer bound}.
\begin{thm}[Acyclic index coding, uncoded placement]
\label{thm:acyclic outer bound} 
Consider the $(\Hsf,\rsf,\Msf,\Nsf)$ combination network with end-user-caches where $\Nsf \geq \Ksf=\binom{\Hsf}{\rsf}$.
For each subset $\Qc\subseteq[\Hsf]$ such that $|\Qc|\in [\rsf:\Hsf]$,
and each permutation $\mathbf{p}(\Kc_{\Qc})$, we have
\begin{subequations}
\begin{align}
&|\Qc|\Rsf^{\star}_{\mathrm{u}} \geq\ 
\sum_{i\in[|\Kc_{\Qc}|]}
\sum_{\medspace\medspace\medspace
\Wc\subseteq[\Ksf]\setminus\{p_{1}(\Kc_{\Qc}),\ldots,p_{i}(\Kc_{\Qc})\}
}
x_{\Wc},
\label{eq:outer bound 2 constrant 1}
\\
&x_{\Wc}:=\frac{1}{\Nsf\Bsf}\sum_{i\in[\Nsf]}|F_{i,\Wc}|, \ \forall\Wc\subseteq[\Ksf],
\label{eq:defxw}
\\
&\sum_{\Wc\subseteq[\Ksf]}x_{\Wc}=1,\label{eq:xfile size}
\\ 
&\sum_{\Wc\subseteq[\Ksf]:i\in\Wc}x_{\Wc}\leq\frac{\Msf}{\Nsf}, \ \forall i\in[\Ksf].
\label{eq:memory size}
\end{align}
\end{subequations}
\end{thm}
 
Recall that for any uncoded cache placement, $F_{i,\Wc}$ represents the set of bits uniquely cached by the users in $\Wc$. Thus $x_{\Wc} \Nsf \Bsf$  represents the total number of bits uniquely cached by the users in $\Wc$.

\begin{rem}
The converse bound in Theorem~\ref{thm:acyclic outer bound} 
can be numerically computed by means of a linear programming 
with variables $(\Rsf^{\star}_{\mathrm{u}},x_{\Wc}:\Wc\subseteq[\Ksf])$ and 
constraints in~\eqref{eq:outer bound 2 constrant 1}-\eqref{eq:memory size}. 
\end{rem}

The ``baseline'' bound in Theorem~\ref{thm:acyclic outer bound} can be improved as follows (see also examples in Sections~\ref{ex:example1} and~\ref{ex:example2}) whose detailed proof can be found in Section~\ref{subset:improved outer bound 3}. This is  a tightened  converse bound obtained by leveraging the network topology and a generalized version of the submodularity of entropy.

\begin{thm}[Improved  converse, uncoded placement]
\label{thm:improved outer bound 3}
Consider the $(\Hsf,\rsf,\Msf,\Nsf)$ combination network with end-user-caches where $\Nsf \geq \Ksf=\binom{\Hsf}{\rsf}$.
For each integer $b\in [\rsf:\Hsf]$, 
each set of relays $\Qc\subseteq[\Hsf]$ with $|\Qc|=b$, 
each integer $a\in [\left\lfloor b/\rsf\right\rfloor ]$, 
each disjoint partition $\Qc=\Qc_{1}\cup \ldots\cup\Qc_{a}$ where $|\Qc_{i}|\geq \rsf$ and $i\in [a]$, 
and each combination of permutations $\mathbf{p}(\Kc_{\Qc_{1}}),\ldots,\mathbf{p}(\Kc_{\Qc_{a}}),\mathbf{p}\big(\Kc_{\Qc}\setminus (\Kc_{\Qc_{1}}\cup\cdots\cup\Kc_{\Qc_{a}})\big)$,
  the following must hold
  \begin{subequations}
\begin{align}
 |\Qc|\Rsf^{\star}_{\mathrm{u}} 
&\geq \sum_{i\in[a]}\sum_{\medspace\medspace j\in[|\Kc_{\Qc_{i}}|]}\sum_{\medspace\medspace\Wc\subseteq[\Ksf]\setminus\cup_{k\in[j]}\{ p_{k}(\Kc_{\Qc_{i}})\}}%
x_{\Wc} \nonumber\\& +\sum_{j\in\big[\big|\Kc_{\Qc}\setminus\Vc\big|\big]}\sum_{\medspace\medspace\Wc\subseteq([\Ksf]\setminus\Vc)\setminus\cup_{k\in[j]}\{p_{k}(\Kc_{\Qc}\setminus\Vc)\}}%
\negmedspace\negmedspace\negmedspace \negmedspace x_{\Wc}+y_{\Qc},\label{eq:combine outer bounds xw} 
\end{align}
satisfying~\eqref{eq:xfile size} and~\eqref{eq:memory size}, where $\Vc:=\cup_{i\in[a]}\Kc_{\Qc_{i}}$,
      the variables $x_{\Wc}$ are as in~\eqref{eq:defxw}, and 
for each permutation $\pv([\Ksf])$ we have,
\begin{align}
 & \sum_{\Qc:|\Qc|=b}%
y_{\Qc}\geq  %
\sum_{i\in[\Ksf]}\sum_{\medspace\medspace\Wc\subseteq[\Ksf]\setminus\cup_{j\in[i]}\{p_{j}([\Ksf])\}} %
\negmedspace\negmedspace\negmedspace \negmedspace \negmedspace \negmedspace   c\big(\{p_{i}([\Ksf])\}\cup\Wc,b\big)x_{\Wc},\label{eq:prop 3 const 2}\\
& c(\Wc_{1},l):= \max\Big\{  \binom{\Hsf-1}{l-1} -  \nonumber\\&  |\{\Qc \subseteq [\Hsf] : |\Qc| =  l,\Kc_{\Qc}\nsubseteq
[\Ksf]\setminus\Wc_{1}\}|,0 \Big\}.\label{eq:def of c}
\end{align}      
 \end{subequations} 
\end{thm}

Intuitively, compared to the acyclic index coding converse bound in Theorem~\ref{thm:acyclic outer bound}, the converse bound in Theorem~\ref{thm:improved outer bound 3} is tighter because of the following two improvements:
\begin{enumerate}

\item 
As discussed in Remark~\ref{rem:why better 1}, the sum 
\begin{align*}
&\sum_{i\in[a]}\sum_{\medspace\medspace j\in[|\Kc_{\Qc_{i}}|]}\sum_{\medspace\medspace\Wc\subseteq[\Ksf]\setminus\cup_{k\in[j]}\{ p_{k}(\Kc_{\Qc_{i}})\}}%
x_{\Wc} \\& +\sum_{j\in\big[\big|\Kc_{\Qc}\setminus\Vc\big|\big]}\sum_{\medspace\medspace\Wc\subseteq([\Ksf]\setminus\Vc)\setminus\cup_{k\in[j]}\{p_{k}(\Kc_{\Qc}\setminus\Vc)\}}%
\negmedspace\negmedspace\negmedspace \negmedspace x_{\Wc}
\end{align*} 
in~\eqref{eq:combine outer bounds xw} may contain the cycles in the directed graph that represents the equivalent index coding problem, i.e., it may contain more terms than the RHS of~\eqref{eq:outer bound 2 constrant 1}.

\item 
In~\eqref{eq:combine outer bounds xw}, there is an additional non-negative term $y_{\Qc}$, which further tightens the   acyclic index coding converse bound in~\eqref{eq:outer bound 2 constrant 1}. Intuitively, 
Theorem~\ref{thm:acyclic outer bound} uses the cut-set strategy, that is,  each time we select a set of relays $\Qc$ and   only consider the users in $\Kc_{\Qc}$ (i.e., whose connected relays are all in $\Qc$). The 
  total load from the server  to these relays is lower bounded by the load for the shared-link caching model containing $|\Kc_{\Qc}|$ users. However, there are some coded messages transmitted to the relays in $\Qc$ which are only useful to the users in $[\Ksf] \setminus \Qc$. The non-negative term $y_{\Qc}$ characterizes the  entropy of these messages.

\end{enumerate}

\begin{rem}
The converse bound in Theorem~\ref{thm:improved outer bound 3} 
can be numerically computed by means of a linear programming with
variables $(\Rsf^{\star}_{\mathrm{u}},x_{\Wc}:\Wc\subseteq[\Ksf],y_{\Qc}:\Qc\subseteq [\Hsf])$ and 
constraints in~\eqref{eq:combine outer bounds xw},~\eqref{eq:prop 3 const 2},~\eqref{eq:xfile size} and~\eqref{eq:memory size}. 
Notice that the computation complexity orders of Theorems~\ref{thm:acyclic outer bound} ans~\ref{thm:improved outer bound 3} are   no more than that of the linear programming including 
$\Oc(2^{\Ksf})$ variables and $\Oc( \Hsf  \Ksf !)$ constraints.
 
\end{rem}

\subsection{Optimality Results}
\label{sub:optimality results}
We first compare the cut-set converse bound 
in Theorem~\ref{thm:cut-set outer bound combination} and the converse bound in Theorem~\ref{thm:improved outer bound 3} to the aforementioned proposed schemes, in order to derive the following optimality results under the constraint of uncoded cache placement. The detailed proof can be found in Appendix~\ref{sec:proof of thm exact opt}.
\begin{thm}[Exact Optimality Under Uncoded Placement]
\label{thm:exact optimallity}
For the $(\Hsf,\rsf,\Msf,\Nsf)$ combination network with end-user-caches where $\Nsf \geq \Ksf=\binom{\Hsf}{\rsf}$,  
\begin{enumerate}

\item Case $\Msf \leq \frac{\Nsf }{\Ksf} \left\lceil  \frac{\rsf}{\Hsf-\rsf}-1 \right\rceil$.
The optimal tradeoff under the constraint of uncoded cache placement $(\Msf,\Rsf^{\star}_{\mathrm{u}})$  is the   lower convex envelop of the following corner points 
\begin{align}
\left( \frac{ \Nsf t}{\Ksf},  \frac{\Ksf-t}{\Hsf(t+1)} \right) \ \textrm{\rm for} \ t \in \left[0:\left\lceil  \frac{\rsf}{\Hsf-\rsf}-1 \right\rceil\right],
\label{eq:optimality case 1}
\end{align}
which is achieved by the DIS, the CICS   and the ICICS.

\item Case $\rsf=\Hsf-1$.  The optimal tradeoff under the constraint of uncoded cache placement $(\Msf,\Rsf^{\star}_{\mathrm{u}})$  is the   lower convex envelop of the following corner points 
\begin{subequations}
\begin{align}
&\left( \frac{ \Nsf t}{\Ksf}, \frac{\Ksf-t}{\Hsf(t+1)} \right)  \ \textrm{\rm for} \ t \in[0:\Ksf-2], \label{eq:optimality case 2-1}\\
&\left( \frac{ \Nsf t}{\Ksf},\frac{\Ksf-t}{\Hsf(\Ksf-1)} \right)\ \textrm{\rm for} \ t \in[\Ksf-1:\Ksf], \label{eq:optimality case 2-2}
\end{align}
\end{subequations}
which is achieved by the DIS, the CICS   and  the ICICS. 

\item Case $\Msf\leq\frac{\Nsf}{\Ksf}$. The optimal max-link load under the constraint of uncoded cache placement is
\begin{subequations}
\begin{align}
\Rsf^{\star}_{\mathrm{u}} & =\frac{\Ksf}{\Hsf}-\frac{\Ksf+1}{2\Hsf}\frac{\Ksf\Msf}{\Nsf}
   \ \textrm{\rm  when }\Hsf<2\rsf, \label{eq:H<2r}\\
\Rsf^{\star}_{\mathrm{u}} & =\frac{\Ksf(\Hsf-1)-(\frac{\Ksf\Hsf+\Hsf-\Ksf}{2}-1)\frac{\Ksf\Msf}{\Nsf}}{\Hsf(\Hsf-1)}
   \ \textrm{\rm  when }\Hsf=2\rsf,\label{eq:H=2r}
\end{align}
\end{subequations}
which is achieved by the IES. 

\end{enumerate}
\end{thm}


We then compare the cut-set converse bound in Theorem~\ref{thm:cut-set outer bound combination} with the achieved max-link loads of the CICS    in~\eqref{eq:load of CICS} and of the IES    in~\eqref{eq:achievable interference load} to derive the following order optimality results, whose detailed proof is in Appendix~\ref{sec:proof of thm order opt}.
\begin{thm}[Order Optimality Under Uncoded Placement]
\label{thm:order optimality}
For the $(\Hsf,\rsf,\Msf,\Nsf)$ combination network with end-user-caches where $\Nsf \geq \Ksf=\binom{\Hsf}{\rsf}$,   
\begin{enumerate}
\item Case $\Msf=\frac{\Nsf t}{\Ksf}$ where $t\in [0,\Ksf]$. The multiplicative gap between 
 the memory sharing of the DIS  and of the scheme in~\cite{cachingincom}  and the converse bound under the constraint of uncoded cache placement in Theorem~\ref{thm:cut-set outer bound combination} is within a factor of  $\min\{\Hsf/\rsf,\left\lceil t \right\rceil +1\}$.  
\item Case $\Msf=\frac{t\Nsf}{\Ksf}$ where $t\in [0,\Ksf]$. The multiplicative gap between 
 the CICS  and the converse bound  under the constraint of uncoded cache placement in Theorem~\ref{thm:cut-set outer bound combination} is within a factor of $1+\left\lceil t \right\rceil/\rsf$.
\item Case $\Msf\leq \frac{\rsf\Nsf}{\Ksf}$. The CICS  is order optimal under the constraint of uncoded cache placement within a factor of $2$. 
\item Case $\Hsf> 2\rsf$ and $\Msf\leq\frac{\Nsf}{\Ksf}$. The IES  is order optimal under the constraint of uncoded cache placement within a factor of $\frac{2\rsf}{2\rsf-1}\leq \frac{4}{3}$.
\end{enumerate}
\end{thm}

\begin{rem}
By using the technique  in~\cite{yufactor2TIT2018}, in the regimes identified by Theorem~\ref{thm:order optimality}, the mentioned schemes are optimal   within the reported factor multiplied by $2$   when no constraint is imposed on the placement phase. 
\end{rem}

\begin{rem}\label{rem:any M, r>H-2, N>K}
We remark that for the small cache size regime given by $\Msf\leq \frac{\rsf\Nsf}{\Ksf}$, the proposed CICS is optimal within a factor 
no more than two under the constraint of uncoded placement.
By combining the order optimality result in~\cite{cachingJi2015} (i.e., the order optimality within a factor of $12$ when $\frac{\Msf}{\Nsf}   \geq \frac{1}{2\rsf}  $) 
with the one 
  in Theorem~\ref{thm:order optimality}, 
for general $(\Hsf,\rsf,\Msf,\Nsf)$ with $\Nsf \geq \Ksf$,   the order optimality under the constraint of uncoded cache placement for the regime   $\frac{\rsf}{\Ksf} < \frac{\Msf}{\Nsf} < \frac{1}{2\rsf}$   remains open.   
\end{rem}

\begin{rem}\label{rem:decen}
 Our proposed delivery schemes can be extended to ``decentralized systems'' where users are not allowed to coordinate in the placement phase. 
With the placement phase  in [21], the delivery is divided into $\Ksf+1$ rounds and in the $i$-round where $i\in [0:\Ksf]$ we can use the proposed achievable schemes in this paper with parameter $t=i$. 
   \end{rem}

\subsection{Numerical Evaluations} 
\label{sub:centralized numerical}
\begin{figure}
\centerline{\includegraphics[scale=0.4]{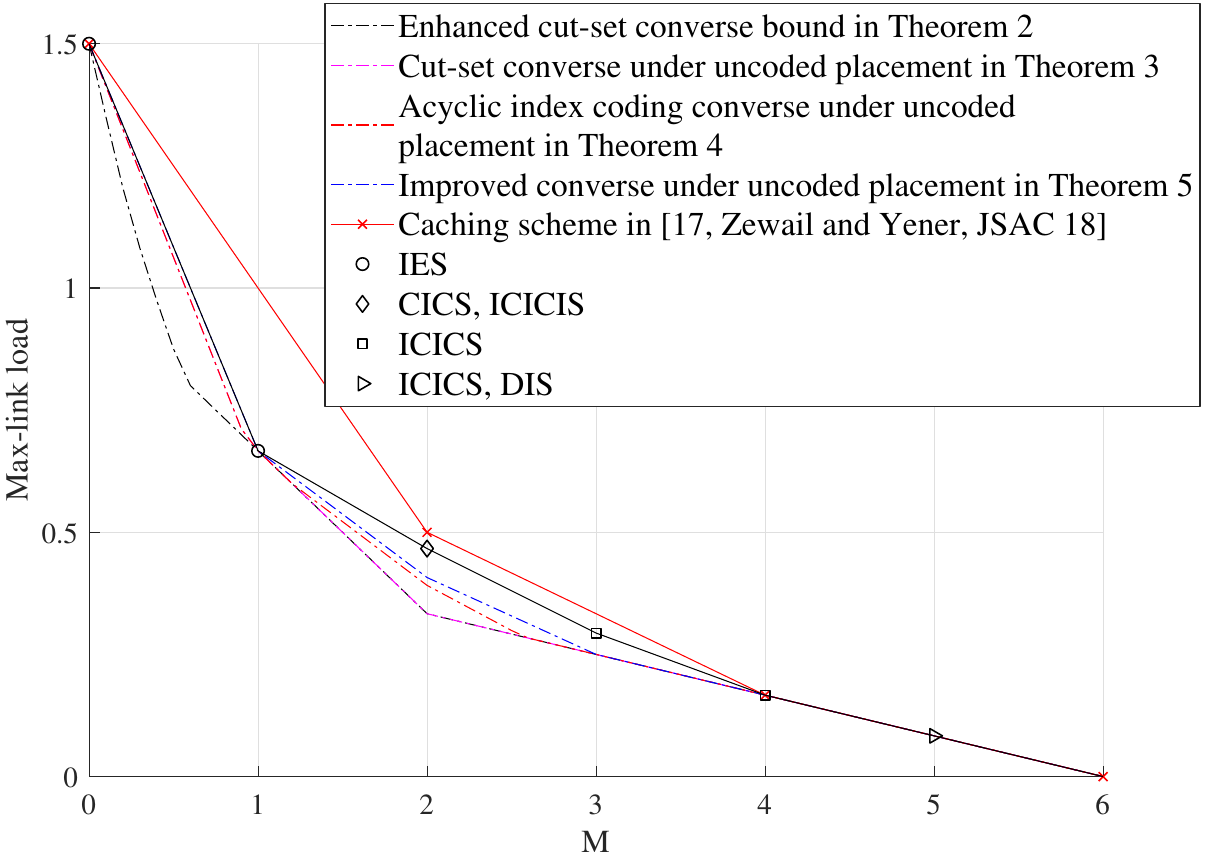}}
\caption{\small Max-link loads for the combination networks with end-user-caches where $\Hsf=4$,  $\rsf=2$, and $\Ksf=\Nsf=6$.}
\label{fig: Combination_Networks_h4r2}
\vspace{-5mm}
\end{figure}

\begin{figure}
\centering{}
\includegraphics[scale=0.4]{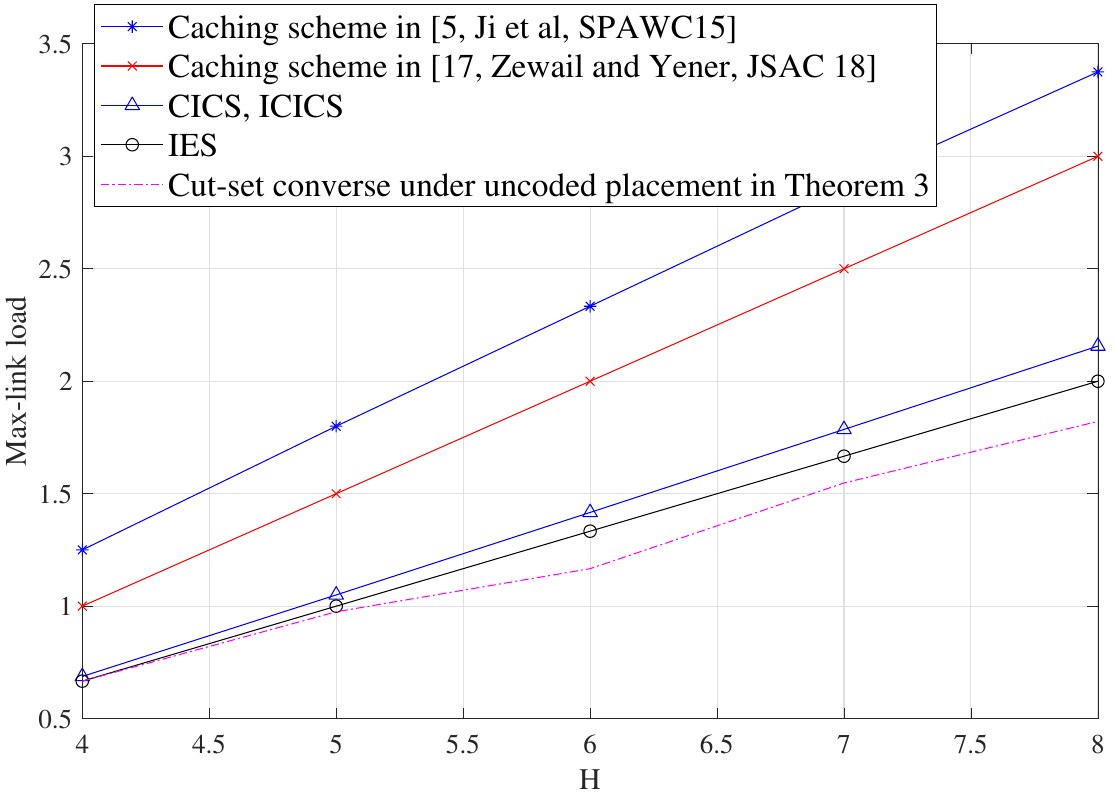}
\caption{Max-link loads for the combination networks with end-user-caches where  $\Hsf\in [4:8]$, $\rsf=2$, $\Nsf=\Ksf=\binom{\Hsf}{\rsf}$, and $\Msf=1$.}
\label{fig:numerical 2}
\end{figure}

Finally,  in Table~\ref{tab:h4r2example} at the top of the next page and Fig.~\ref{fig: Combination_Networks_h4r2} we provide numerical results 
for $\Hsf=4$, $\rsf=2$ and $\Nsf=\Ksf=\binom{\Hsf}{\rsf}=6$, to illustrate the proposed achievable bounds and converse bounds compared to the state-of-the-art. It can be seen that 
our results improve on the literature. In addition, when $\Msf \in [0,1]$, the proposed IES   coincides with the proposed converse  bound in Theorem~\ref{thm:improved outer bound 3}. 
For the case $\Hsf\in [4:8]$, $\Ksf=\Nsf$, $\rsf=2$ and $\Msf=1$, Fig.~\ref{fig:numerical 2}  shows that IES outperforms all of the other schemes.

\begin{table*}
 \centering
\protect\caption{Max-link loads for the combination networks with end-user-caches where $ \Hsf=4$, $\rsf=2$, and $\Ksf=\Nsf=6 $. 
  Best achievable and converse bounds for each case are highlighted in bold phase font. 
}\label{tab:h4r2example}
\begin{tabular}{|c|c|c|c|c|c|c|c|c|}
\hline 
  & $\Msf=0.5$ & $\Msf=1$ & $\Msf=1.5$ & $\Msf=2$ & $\Msf=2.5$ & $\Msf=3$ & $\Msf=4$ & $\Msf=5$ \tabularnewline
\hline 
\hline 
$\Rsf_{{\rm base}}$ in~\eqref{eq:Mingyue inner bound} & $ 1.375$ & $1.25$& $ 0.959$ & $0.667$ & $ 0.521$& $0.375$ & $0.2$ & {\bf 0.083} \tabularnewline
\hline 
$\Rsf_{\textrm{\rm ZY}}$ in~\eqref{eq:Yener load} & $1.25$ & $1$ & $0.75$& $0.5$ & $  0.417$& $0.333$ & {\bf 0.167} & {\bf 0.083} \tabularnewline
\hline 
$\Rsf_{\textrm{\rm DIS}}$ in~\eqref{eq: load of DIS} & $1.125$ & $0.75$ & $ 0.658$ & $0.567$& $0.471 $ & $0.375$ & $0.2$ & {\bf 0.083}\tabularnewline
\hline 
$\Rsf_{\textrm{\rm IES}}$ in~\eqref{eq:achievable interference load} & {\bf  1.084 } & {\bf 0.667} & & & &  &  &   \tabularnewline
\hline 
$\Rsf_{\textrm{\rm CICS}}$ in~\eqref{eq:load of CICS} & $ 1.094$ & $0.688$ & {\bf 0.578} & {\bf 0.467} & $ 0.384$& $0.3$ & $0.2$ & $0.104$ \tabularnewline
\hline 
$\Rsf_{\textrm{\rm ICICS}}$ by Algorithm 1 &$1.094$ & $0.688$ & {\bf 0.578}& {\bf 0.467} & {\bf   0.378 } & {\bf 0.288} & {\bf 0.167} & {\bf 0.083}\tabularnewline
\hline 
\hline 
 Cut-set converse bound in~\eqref{eq: Mingyue converse} & $0.75$ & $0.5$ & $0.375$& $0.333$ & $ 0.292$& {\bf 0.25} & {\bf 0.167} & {\bf 0.083} \tabularnewline
\hline 
Enhanced cut-set converse bound in~\eqref{eq:enhanced outer bound combination} & $0.875$ & {\bf 0.667} & $0.5$ & $0.333$& $0.292$ & {\bf 0.25} & {\bf 0.167} & {\bf 0.083} \tabularnewline
\hline 
Cut-set converse bound under & & & & & & & & \tabularnewline
 uncoded placement in~\eqref{eq:cut-set outer bound combination}& $1.063$ & {\bf 0.667} & $0.5$& $0.333$ & $0.292$ & {\bf 0.25} & {\bf 0.167} & {\bf 0.083} \tabularnewline
\hline 
Acyclic index coding  converse bound under & & & & & & & & \tabularnewline
uncoded placement in~\eqref{eq:outer bound 2 constrant 1}  & $1.063$ & {\bf 0.667} & $0.522$ & $0.391$ & $0.299$ &{\bf 0.25} & {\bf 0.167} & {\bf 0.083} \tabularnewline
\hline 
Improved  converse bound under & & & & & & & & \tabularnewline
uncoded placement in~\eqref{eq:combine outer bounds xw}  &  {\bf 1.083 } & {\bf 0.667} & {\bf 0.539} & {\bf 0.412} & {\bf 0.330} & {\bf 0.25} & {\bf 0.167} &{\bf 0.083}\tabularnewline
\hline 
\end{tabular}
\end{table*}

\section{Novel Achievable Delivery Schemes} 
\label{sec:general inner bound}
In this section, we introduce our proposed delivery schemes corresponding to the achievable bounds in Section~\ref{sub:results on inner bounds}.
The placement phase is the same as the MAN placement described in Section~\ref{sub:known results}. Hence, we focus on each memory $\Msf =\frac{\Nsf t}{\Ksf}$ where $t \in [0:\Ksf]$.
 
\subsection{Direct Independent delivery Scheme (DIS)}
\label{sub:DIS}
Recall that $\Rc_{\Jc}$ is   the  set of relays connected to all the users in $\Jc$, and that the definitions of $\Vc_{1}$ and $\Vc_{2}$ are given  in~\eqref{eq:def of v 1} and~\eqref{eq:def of v 2}, respectively.
The DIS   is based on the observation that each MAN multicast message $W_{\Jc}$ is only useful to the users in $\Jc$;
therefore, if $\Rc_{\Jc}\neq\emptyset$, i.e., there exists at least one relay that is connected to all the users in $\Jc$, it is enough to transmit $W_{\Jc}$ only to the relays in $\Rc_{\Jc}$; this observation motivates the next steps.

{\it Step~1:} For each 
$\Jc\in \Vc_{1}$, we  divide  $W_{\Jc}$ into $|\Rc_{\Jc}|$ non-overlapping pieces with equal length and  directly transmit  each different piece to the relays in  $\Rc_{\Jc}$. 

{\it Step~2:} For each $\Jc\in  \Vc_{2}$, we  divide  $W_{\Jc}$ into $\rsf$ non-overlapping and equal-length pieces, which are then encoded by an $(|\{h\in[\Hsf]:\Uc_{h}\cap\Jc\neq\emptyset\}|,\rsf)$ MDS code. We then transmit one different MDS symbol to each   relay in $ \{h\in[\Hsf]:\Uc_{h}\cap\Jc\neq\emptyset\}$.
Each user in $\Jc$ is connected to $\rsf$ relays, and thus can receive $\rsf$ MDS symbols of $W_{\Jc}$ to recover $W_{\Jc}$.

By Step~1 and Step~2, each user $k\in [\Ksf]$ can recover all the  multicast messages $W_{\Jc}$ where $k\in \Jc$ and then recover the subfile $F_{d_{k},\Jc\setminus \{k\}}$. The resulting load is in~\eqref{eq: load of DIS}.

It is 
  not difficult to  prove that the DIS    is better than the second scheme in~\cite{cachingincom}, which transmits $|\Wc_{\Jc}|/\rsf$ bits to each relay $h\in[\Hsf]$. The first scheme in~\cite{cachingincom} is equivalent to using $|\Jc||\Wc_{\Jc}|$ bits to transmit $\Wc_{\Jc}$. However, in the DIS, we use at most $|\Jc||\Wc_{\Jc}|$ bits to transmit $\Wc_{\Jc}$; only when for each pair of  users $k_1, k_2\in \Jc$ we have $\Hc_{k_1}\cap \Hc_{k_2}=\emptyset$, we use  $|\Jc||\Wc_{\Jc}|$ bits. Hence, the max-link load of the DIS    is 
 no larger  than the one in~\cite{cachingincom}, as we state in Corollary~\ref{cor:strictly better than MJ}.

\subsection{Interference Elimination delivery Scheme (IES) for the Case $t=1$}
\label{sub:IES}
The IES  is 
for $\Msf=\Nsf/\Ksf$ (i.e., $t=1$) only.
Notice that when  $\Hsf<2\rsf$ and $\Msf=\Nsf/\Ksf$,  $\Vc_{2}$ contains the sets $\Jc$ where no relay is connected to all the users in $\Jc$ with $|\Jc|=t+1=2$. So we have $\Vc_{2}=\emptyset$. 

\paragraph*{Delivery of multicast messages in $\Vc_{1}$} 
For each $\Jc\in \Vc_{1}$ with $\Vc_{1}$ in~\eqref{eq:def of v 1}, we divide $W_{\Jc}$ into $|\Rc_{\Jc}|$ non-overlapping and equal-length pieces, i.e., $W_{\Jc}=\{W_{\Jc,h}:h\in \Rc_{\Jc}\}$; we transmit $W_{\Jc,h}$ to each relay $h\in \Rc_{\Jc}$. Note that each relay $h\in \Rc_{\Jc}$ is connected to all of the users in $\Jc$. Hence, if relay $h$ forwards $W_{\Jc,h}$ to the users in $\Jc$, each user in $\Jc$ can recover $W_{\Jc}$. The  link load to deliver the coded multicast messages in $\Vc_{1}$ to each relay is $\frac{1}{2\Hsf}\left(\Ksf-1-\binom{\Hsf-\rsf}{\rsf}\right)$.\footnote{\label{foot:IES v1}Note that when $t=1$, for each $k\in [\Ksf],$ the number of sets in $\Vc_1$ which contains $k$ is $\Ksf-1-\frac{\Hsf-\rsf}{\rsf}$. Hence, by the symmetry, the number of sets in $\Vc_1$ is $\Ksf\frac{\Ksf-1-\binom{\Hsf-\rsf}{\rsf}}{2}$. Since each multicast message contains $\frac{1}{\binom{\Ksf}{t}}=\frac{1}{\Ksf}$ bits and the link load  to deliver  the coded multicast messages in $\Vc_{1}$ to each relay is the same, we can compute that this link load is equal to  $\frac{1}{2\Hsf}\left(\Ksf-1-\binom{\Hsf-\rsf}{\rsf}\right)$.}

The key idea of the IES  is using an interference elimination scheme to transmit the messages $W_{\Jc}$ where $\Jc\in \Vc_{2}$. 
We examine three examples to highlight the key idea.

\begin{example}[$\Hsf=2\rsf,\rsf=2$]
\label{ex:example of IE H4r2}
\rm
Consider the combination network  with end-user-caches in Fig.~\ref{fig: Combination_Networks} where $\Hsf=4$, $\rsf=2$, $\Msf=1$ and $\Ksf=\Nsf=6$.
Assume that $\mathbf{d}=(1:6)$, so that
\begin{align*}
\Vc_{1}=&\big\{\{1,2\},\{1,3\},\{1,4\},\{1,5\},\{2,3\},\{2,4\},\{2,6\},\\ & \{3,5\},\{3,6\}, \{4,5\}, \{4,6\}, \{5,6\}\big\},\\
 \Vc_{2} =&\{\{1,6\},\{2,5\},\{3,4\}\}.
\end{align*} 

For the messages in $\Vc_{1}$,
the server transmits
\begin{align*}
&W_{\{1,2\}},W_{\{1,3\}},W_{\{2,3\}}\ \textrm{\rm  to relay~1},\\
&W_{\{1,4\}},W_{\{1,5\}},W_{\{4,5\}}\ \textrm{\rm  to relay~2},\\
&W_{\{2,4\}},W_{\{2,6\}},W_{\{4,6\}}\ \textrm{\rm  to relay~3},\\
&W_{\{3,5\}},W_{\{3,6\}},W_{\{5,6\}}\ \textrm{\rm  to relay~4}.  
\end{align*}
We then use an interference elimination scheme to transmit the messages $W_{\Jc}$ where $\Jc\in \Vc_{2}$ and $\Vc_{2}=\{\{1,6\},\{2,5\},\{3,4\}\}$.  Consider user~1, who is connected to relays~1 and~2, and only needs to recover $W_{\{1,6\}}$ (i.e., $W_{\{2,5\}}$ and $W_{\{3,4\}}$ are interference). 
In order to perform the IES  we transmit (with operations on some finite field\footnote{
We would like to perform the operations ``$+$'' and ``$-$'' on some finite field of sufficiently large size. 
In other words, with an abuse of notation, we use $W_{\Jc}$ to denote both the binary multicast messages as well as its representation on some larger field. 
In this example we need a field size larger than $3$; { on GF(3) the ``$-1$'' becomes ``$+2$''} so that 
$X_1 = W_{\{1,6\}}+ W_{\{2,5\}}+ W_{\{3,4\}}$, 
$X_2 = W_{\{1,6\}}+2W_{\{2,5\}}+2W_{\{3,4\}}$, 
and thus
$X_1+X_2= (2W_{\{1,6\}}+3W_{\{2,5\}}+3W_{\{3,4\}})_{\textrm{\rm  mod } 3} = 2W_{\{1,6\}}$, etc. 
})
\begin{align*}
&X_1 :=+W_{\{1,6\}}+W_{\{2,5\}}+W_{\{3,4\}}\ \textrm{\rm  to relay~1},\\
&X_2 :=+W_{\{1,6\}}-W_{\{2,5\}}-W_{\{3,4\}}\ \textrm{\rm  to relay~2},\\
&X_3 :=-W_{\{1,6\}}+W_{\{2,5\}}-W_{\{3,4\}}\ \textrm{\rm  to relay~3},\\
&X_4 :=-W_{\{1,6\}}-W_{\{2,5\}}+W_{\{3,4\}}\ \textrm{\rm  to relay~4}, 
\end{align*}
and the relays then forward what they received to the users.
Then, user~1 computes $X_1+X_2=2 W_{\{1,6\}}$
and recovers $W_{\{1,6\}}$.  
Similarly, user~2 computes $X_1+X_3=2 W_{\{2,5\}}$, and so on.
{ With this, all users recover the missing subfiles   of the demanded file}.
As the length of $W_{\Jc}$, equal to $\Bsf/6$, goes to infinity as $\Bsf \to \infty$, we can divide each $W_{\Jc}$ where $\Jc \in  \Vc_{2}$ into $P$ sub-packets with length $\Bsf/(6P)$ such that we can do operations among multicast messages on a finite field of size $3$. Notice that each file is composed of  $6P$ sub-packets. 

The link load to transmit the multicast messages in $\Vc_{2}$ to each relay is $P/(6P)=1/6$, while the link load to transmit   the multicast messages in $\Vc_1$ to each relay is $1/2$.
Hence,  
the max-link load of this scheme is $2/3 = 1/2 + 1/6$, coinciding with the  cut-set converse bound under the constraint of uncoded placement in~\eqref{eq:cut-set outer bound combination}, while that of the CICS  (ICICS) and the schemes in~\cite{cachingincom,combinationsecu2018Zewail} are $0.6875$,  $5/4$, and $1$, respectively. 
\hfill$\square$
\end{example}

In the next example, we generalize the IES  to transmit $W_{\Jc}$ where $\Jc \in  \Vc_{2}$ to the case $\Hsf>2\rsf$ and $\rsf=2$.  
\begin{example}[$\Hsf>2\rsf,\rsf=2$]
\label{ex:example of IE h5r2}
\rm
Consider the combination networks with end-user-caches where $\Hsf=5$, $\rsf=2$, $\Msf=1$, and $\Ksf=\Nsf=10$.
Assume that $\mathbf{d}=(1:10)$, so that
\begin{align*}
\Uc_{1}&=[1:4],\\  
\Uc_{2}&=\{1,5,6,7\},\\ 
 \Uc_{3}&=\{2,5,8,9\},\\
\Uc_{4}&=\{3,6,8,10\},\\ 
\Uc_{5}&=\{4,7,9,10\},\\
\Vc_{2}&=\big\{\{1,8\},\{1,9\},\{1,10\},\{2,6\},\{2,7\},\{2,10\},\{3,5\},\\ 
&\{3,7\},\{3,9\},\{4,5\},\{4,6\},\{4,8\},\{5,10\},\{6,9\},\{7,8\}\big\}.
\end{align*}
The   link load to transmit the multicast messages in $\Vc_{1}$ to each relay is $3/5$.
 For $W_{\Jc}$ where $\Jc \in  \Vc_{2}$, we expand on the IES  introduced in Example~\ref{ex:example of IE H4r2}.

For each subset of relays $\Bc\subseteq[\Hsf]$ with cardinality $|\Bc|=2\rsf=4$, 
the set of users connected to $\rsf=2$ of the chosen relays is denoted as $\Pc_{\Bc}$ and we also let
\begin{align}
\Tc_{\Bc} :=\{ \Jc\in  \Vc_{2}: \Jc\subseteq \Pc_{\Bc}\},
\label{eq:def of TB}
\end{align}
   be the set of multicast messages in $\Vc_{2}$ which are  useful to some users in $\Pc_{\Bc}$ and not useful to the users not in $\Pc_{\Bc}$.
We then use the scheme in Example~\ref{ex:example of IE H4r2} to transmit the codewords for the multicast messages in $\Tc_{\Bc}$ to the relays in $\Bc$.  

For example, for $\Bc=\{1,2,3,4\}$ we have $\Pc_{\{1,2,3,4\}}=\{1,2,3,5,6,8\}$ and $\Tc_{\{1,2,3,4\}}=\big\{\{1,8\},\{2,6\},\{3,5\}\big\}$; we transmit 
\begin{align*}
&+W_{\{1,8\}}+W_{\{2,6\}}+W_{\{3,5\}}\ \textrm{\rm  to relay~1},\\
&+W_{\{1,8\}}-W_{\{2,6\}}-W_{\{3,5\}}\ \textrm{\rm  to relay~2},\\
&-W_{\{1,8\}}+W_{\{2,6\}}-W_{\{3,5\}}\ \textrm{\rm  to relay~3},\\
&-W_{\{1,8\}}-W_{\{2,6\}}+W_{\{3,5\}}\ \textrm{\rm  to relay~4}. 
\end{align*}
Hence, each multicast message  $W_{\Jc}$ where $\Jc\in \Tc_{\{1,2,3,4\}}$ can be recovered by the users in $\Jc$. We proceed similarly to transmit the codewords for the multicast messages in $\Tc_{\Bc}$ to the relays $\Bc$ such that each user in $\Jc$ can recover $W_{\Jc}$ where $\Jc \in \Tc_{\Bc}$. 

The link load to transmit the multicast messages in $\Vc_{2}$ to each relay is $2/5$. Hence, the max-link load of this scheme is $3/5+2/5=1$, while that of the  CICS (ICICS) and the schemes  in~\cite{cachingincom,combinationsecu2018Zewail} are $1.05$, $2.25$,  and $1.5$, respectively. 
\hfill$\square$
\end{example}

In the final example, we generalize the IES  to transmit $W_{\Jc}$ where $\Jc \in  \Vc_{2}$ 
to any $\rsf\geq 2$.
\begin{example}[$\Hsf\geq 2\rsf,\rsf>2$]
\label{ex:example of IE h6r3}
\rm
Consider the combination networks with end-user-caches where $\Hsf=6$, $\rsf=3$, $\Ksf=\Nsf=20$, and $\Msf=1$.
Assume that $\mathbf{d}=(1:20)$ so that
\begin{align*}
  &\Uc_{1}=\{1,2,3,4,5,6,7,8,9,10\},
\\&\Uc_{2}=\{1,2,3,4,11,12,13,14,15,16\}, 
\\&\Uc_{3}=\{1,5,6,7,11,12,13,17,18,19\},
\\&\Uc_{4}=\{2,5,8,9,11,14,15,17,18,20\},
\\&\Uc_{5}=\{3,6,8,10,12,14,16,17,19,20\},
\\&\Uc_{6}=\{4,7,9,10,13,15,16,18,19,20\},
\\&\Vc_{2}=\big\{\{1,20\},\{2,19\},\{3,18\},\{4,17\},\{5,16\}, \{6,15\}, \nonumber\\& \{7,14\},\{8,13\},\{9,12\},\{10,11\}\big\}.
\end{align*}
We focus on the transmission of the multicast messages in $\Vc_{2}$.
If $\Hsf=2\rsf$ { (the case $\Hsf>2\rsf$ can be dealt as in Example~\ref{ex:example of IE h5r2})}, there are $\binom{2\rsf}{\rsf}/2=\binom{2\rsf-1}{\rsf-1}=10$ elements in $\Vc_{2}$. 
We partition $\Vc_{2}$ into $\binom{2\rsf-1}{\rsf-1}/(2\rsf-1)=2$ groups where each group contains $2\rsf-1=5$ elements  and some constraint should be satisfied (to be clarified soon). The existence of such a partition is discussed in Appendix~\ref{sec:discussion of full rank}.
In this example the groups could be chosen as 
\begin{subequations}
\begin{align}
\Gc_{1}&=\big\{\{1,20\},\{2,19\},\{3,18\},\{5,16\},\{ 7,14\}\big\},
\\
\Gc_{2}&=\big\{\{4,17\},\{6,15\},\{8,13\},\{9,12\},\{10,11\}\big\}. 
\end{align}
\end{subequations}
We focus on group $\Gc_{1}$; { the same applies to $\Gc_{2}$}. 
The 
vector of multicast messages in $\Gc_{1}$ is indicated by $\mathbb{M}_{1}=[W_{\{1,20\}};W_{\{2,19\}};W_{\{3,18\}};W_{\{5,16\}};W_{\{7,14\}}]$. 
{ We now design a linear code for $\mathbb{M}_{1}$ on some finite field;} 
we denote the coding matrix by $\mathbb{A}_{1}$,
the element on $i^{\textrm{\rm th}}$ row $j^{\textrm{\rm th}}$ column of $\mathbb{A}_{1}$ by $a_{1,i,j}$, and
the $i^{\textrm{\rm th}}$ row of $\mathbb{A}_{1}$ by $\mathbb{A}_{1,i}$;
$\mathbb{A}_{1,i}\times\mathbb{M}_{1}$ represents the codeword for $\mathbb{M}_{1}$ transmitted to relay $i\in [6]$. 

Recall that each multicast message  $W_{\Jc}$ where  $\Jc \in \Gc_{1}$ is useful to the users in $\Jc$ and is interference to the users in $\cup_{\Jc_{1}\in\Gc_{1}:\Jc_{1}\neq\Jc}\Jc_{1}$. Our objective is to eliminate the interference caused by $W_{\Jc}$ to those users who do not need it. 
To achieve our objective, we construct the linear code as follows.

\begin{subequations}
We focus on 
$W_{\{1,20\}}$ (assumed to be the 
 first element in the vector $\mathbb{M}_{1}$). 
%
Firstly,  on some finite field, we let 
\begin{align}
\sum_{i\in[6]}a_{1,i,1}=0.
\label{eq:ex3 total sum const}
\end{align}
 The reason for the choice in~\eqref{eq:ex3 total sum const} will become clear later.
Then, for each  $\Jc_{1}\in \Gc_{1} \backslash \{ \{1,20\} \}$ 
we eliminate the interference caused by $W_{\{1,20\}}$ to users in $\Jc_{1}=\{k_1,k_2\}$ (where $k_{1}$ is the user connected to relay $6$) by letting\footnote{\label{choose user}
{
For each pair of users $\Jc_1=\{k_1,k_2\}$, we can choose any one of them (denoted by $k^\prime$) and write the equation $\sum_{i\in\mathcal{H}_{k^\prime}}a_{1,i,1}=0$ which can also lead to $\sum_{i\notin\mathcal{H}_{k^\prime}}a_{1,i,1}=0$ from~\eqref{eq:ex3 total sum const}. Thus,  the interference caused by $W_{\{1,20\}}$ to these two users is eliminated. Here we choose the user connected to relay $6$, i.e., $k^\prime=k_1$.}
}
\begin{align}
\sum_{i\in\mathcal{H}_{k_{1}}}a_{1,i,1}=0.
\label{eq:ex3 interference}
\end{align}
It can be seen that each set $\Jc\in  \Vc$ contains two users $k_1$ and $k_2$ for which $\Hc_{k_1}\cap \Hc_{k_2}=\emptyset$ and $\Hc_{k_1}\cup \Hc_{k_2}=[\Hsf]$,
therefore, from~\eqref{eq:ex3 total sum const} and~\eqref{eq:ex3 interference}, we have
\begin{align}
\sum_{i\in\mathcal{H}_{k_{2}}}a_{1,i,1}=0.
\label{eq:ex3 interference another one}
\end{align} 
Hence, if users $k_1$ and $k_2$ sum their received codewords 
from their connected relays, they can eliminate the interference caused by $W_{\{1,20\}}$. 
 This construction is repeated for all $\Jc_{1}\in \Gc_{1} \backslash \{ \{1,20\} \}$.
Last, as user $20$ requires $W_{\{1,20\}}$, we let
\begin{align}
\sum_{i\in\mathcal{H}_{20}}a_{1,i,1}=s : s\not= 0.
\label{eq:ex3 useful}
\end{align} 
As $\Hc_{1}\cap \Hc_{20}=\emptyset$ and $\Hc_{1}\cup \Hc_{20}=[\Hsf]$, from~\eqref{eq:ex3 total sum const} and~\eqref{eq:ex3 useful}, it can be seen that 
\begin{align}
\sum_{i\in\mathcal{H}_{1}}a_{1,i,1}=-s\not= 0.
\label{eq:ex3 useful another one}
\end{align} 
Hence, if user~1 and~20 sum their received codewords 
from their connected relays, they can recover $W_{\{1,20\}}$. 
\label{eq:all for code construction}
\end{subequations}

To summarize, by collecting all the constraint in~\eqref{eq:all for code construction}
we obtain the following system of equations to solve
\begin{align}
\begin{bmatrix}
1 & 1 & 1 & 1 & 1 & 1\\
0 & 0 & 1 & 0 & 1 & 1\\
0 & 0 & 1 & 1 & 0 & 1\\
0 & 1 & 0 & 0 & 1 & 1\\
1 & 0 & 1 & 0 & 0 & 1\\
0 & 0 & 0 & 1 & 1 & 1\\
\end{bmatrix}
\times
\begin{bmatrix}
a_{1,1,1}\\
a_{1,2,1}\\
a_{1,3,1}\\
a_{1,4,1}\\
a_{1,5,1}\\
a_{1,6,1}\\
\end{bmatrix}
=
\begin{bmatrix}
0\\
0\\
0\\
0\\
0\\
s\\
\end{bmatrix}.
\label{eq:ex3 equation}
\end{align}
  So the group division must satisfy the constraint that the first matrix in~\eqref{eq:ex3 equation} is full-rank.  Thus
if we let for example $s=-3$, we have $[a_{1,1,1};a_{1,2,1};a_{1,3,1}$ $;a_{1,4,1};a_{1,5,1};a_{1,6,1}]=[1;-2;-2;1;1;1]$. 
Similarly, we can get all the elements in $\mathbb{A}_{1}$. 

Hence, for all the multicast messages $W_{\Jc}$ where $\Jc \in \Gc_{1}$, 
we transmit (note that the following operations can be done on a finite field with size larger than $7$)
\begin{align*}
\begin{bmatrix}
X_{1,1} \\X_{2,1} \\X_{3,1} \\X_{4,1} \\X_{5,1} \\X_{6,1} \\
\end{bmatrix}
=
\begin{bmatrix}
+1 & +1 & -2 & +1 & -2 \\
-2 & +1 & +1 & -2 & +1 \\
-2 & -2 & +1 & +1 & +1 \\
+1 & +1 & +1 & +1 & +1 \\
+1 & -2 & -2 & +1 & +1 \\
+1 & +1 & +1 & -2 & -2 \\
\end{bmatrix}
\times
\begin{bmatrix}
W_{\{1,20\}} \\ W_{\{2,19\}} \\ W_{\{3,18\}} \\ W_{\{5,16\}} \\ W_{\{7,14\}} \\
\end{bmatrix}
.
\end{align*}
%
%
Similarly, for all the multicast messages $W_{\Jc}$ where $\Jc \in \Gc_{2}$, 
we transmit
\begin{align*}
\begin{bmatrix}
X_{1,2} \\X_{2,2} \\X_{3,2} \\X_{4,2} \\X_{5,2} \\X_{6,2} \\
\end{bmatrix}
=
\begin{bmatrix}
-2 & -2 & +1 & +1 & +1 \\
-2 & +1 & +1 & -2 & +1 \\
+1 & -2 & -2 & +1 & +1 \\
+1 & +1 & +1 & +1 & +1 \\
+1 & +1 & +1 & -2 & -2 \\
+1 & +1 & -2 & +1 & -2 \\
\end{bmatrix}
\times
\begin{bmatrix}
W_{\{4,17\}} \\ W_{\{6,15\}} \\ W_{\{8,13\}} \\ W_{\{9,12\}} \\ W_{\{10,11\}} \\
\end{bmatrix}
.
\end{align*}
%
%
Each user  $k\in\Jc$ can recover $W_{\Jc}$ where $\Jc \in \Gc_{u}, u\in\{1,1\}$, by summing the received codewords (corresponding to $\Gc_{u}$) from the relays in $\Hc_{k}$. 


We can now compute the max-link load.
The link load to transmit the multicast messages in $ \Vc_{1}$  to each relay is $3/2$. 
The link load  to transmit the multicast messages in $ \Vc_{2}$ to each relay is $1/10$.
Thus, the max-link load of this scheme is $3/2+1/10=8/5$ coinciding with the improved converse bound under the constraint of uncoded placement in Theorem~\ref{thm:improved outer bound 3},
while that of the CICS  (ICICS) and the schemes in~\cite{cachingincom,combinationsecu2018Zewail}  are $1.6111$,  $19/6 \approx 3.1667$, and $29/12\approx 2.41667$, respectively. 
\hfill$\square$
\end{example}

We now generalize the scheme to transmit the messages $W_{\Jc}$ where $\Jc\in \Vc_{2}$, described in the above examples to the general case of $t=1$ and $\Hsf\geq 2\rsf$. We use the MAN cache placement and let $W_{\Jc}=\underset{j\in\Jc}{\oplus}F_{d_{j},\Jc\setminus\{j\}}$ for each $\Jc\subseteq [\Ksf]$ where $|\Jc|=2$.  Notice that since $t=1$, $\Vc_{2}$ contains all the sets of two users $k$ and $k^{\prime}$ where $\Hc_{k}\cap \Hc_{k^{\prime}}=\emptyset$.

\paragraph{Transmission of $\Vc_{2}$ for $\Hsf=2\rsf$}
If $\Hsf=2\rsf$, we have 
$|\Vc_{2}|
=\binom{2\rsf}{\rsf}/2
=\binom{2\rsf-1}{\rsf-1}$. 
We partition $\Vc_{2}$ into 
$\binom{2\rsf-1}{\rsf-1}/(2\rsf-1)
=\binom{2\rsf-2}{\rsf-1}-\binom{2\rsf-2}{\rsf}$ 
groups, each containing $2\rsf-1$ elements (the group division is explained in Appendix~\ref{sec:discussion of full rank}). 

For each group $\Gc_{g}, \ g\in \left[\binom{2\rsf-1}{\rsf-1}/(2\rsf-1)\right]$, the  vector 
$\mathbb{M}_{g}$ of dimension $(2\rsf-1)\times 1$ lists the multicast messages $W_{\Jc}$ where  $\Jc \in \Gc_{g}$. Let $m_{g,j}$ denote the $j^{\textrm{\rm th}}$ element of $\mathbb{M}_{g}$.
We design a linear code, { on some finite field of sufficiently large size}, to transmit $\mathbb{M}_{g}$, whose coding matrix is denoted by $\mathbb{A}_{g}$.
Let the element on $i^{\textrm{\rm th}}$ row and $j^{\textrm{\rm th}}$ column of $\mathbb{A}_{g}$ be denoted by $a_{g,i,j}$, the $i^{\textrm{\rm th}}$ row by $\mathbb{A}_{g,i}$; 
the message $\mathbb{A}_{g,i}\times\mathbb{M}_{g}$ 
is transmitted to relay $i\in [\Hsf]$.

\begin{subequations} 
We construct the 
$j^{\textrm{\rm th}}$ column of $\mathbb{A}_{g}$, 
whose elements are the coefficients for the  multicast messages $m_{g,j}$, 
as follows.
First, we let
\begin{align}
\sum_{i\in[\Hsf]}a_{g,i,j}=0.
\label{eq:general sum}
\end{align} 
Then, for each $\Jc_{1}\in \Gc_{g}\setminus \{\Jc\}$, we let
\begin{align}
\sum_{i\in\mathcal{H}_{k_{1}}}a_{g,i,j}=0,
\label{eq:general interference}
\end{align}
where $k_{1}$ is the user in $\Jc_{1}$ connected to relay $\Hsf$. 
Assume that $k_2$ is the other user in $\Jc_{1}$. As $\Hc_{k_1}\cap \Hc_{k_2}=\emptyset$ and $\Hc_{k_1}\cup \Hc_{k_2}=[\Hsf]$, from~\eqref{eq:general sum} and~\eqref{eq:general interference} it can be seen that
\begin{align}
\sum_{i\in\mathcal{H}_{k_{2}}}a_{g,i,j}=0. 
\end{align}
Users $k_1$ and $k_2$ can eliminate the interference caused by $W_{\Jc}$ by summing their received codewords for $\Gc_{g}$ from their connected relays. 
Last, we let
\begin{align}
\sum_{i\in\mathcal{H}_{k_{3}}}a_{g,i,j}=s, \ \text{where } s\not=0,
\label{eq:general useful}
\end{align} 
where $k_{3}$ is the user in $\Jc$ who is connected to relay $\Hsf$.
%
Assume $k_4$ is the other user in $\Jc$, as $\Hc_{k_{3}}\cap \Hc_{k_{4}}=\emptyset$ and $\Hc_{k_{3}}\cup \Hc_{k_{4}}=[\Hsf]$, from~\eqref{eq:general sum} and~\eqref{eq:ex3 useful}, it can be seen that
\begin{align}
\sum_{i\in\mathcal{H}_{k_{4}}}a_{g,i,j}=-s.
\end{align} 
Hence, if the users in $\Jc$ sum  their received codewords for $\Gc_{g}$ from their connected relays, they can recover $W_{\Jc}$.
\label{eq:poiuytrewqasdasdfghjkl}
\end{subequations} 
 
By collecting all the equations in~\eqref{eq:poiuytrewqasdasdfghjkl}, we can write the following system of equations
\begin{align}
\mathbb{C}_{g}
\times
\begin{bmatrix} a_{g,1,j}\\ \vdots \\ a_{g,\Hsf-1,j}\\ a_{g,\Hsf,j}\end{bmatrix}
=
\begin{bmatrix}0\\ \vdots \\ 0 \\ s \end{bmatrix}, \ j\in{[2\rsf-1]}.
\label{eq:general equation}
\end{align}
In Appendix~\ref{sec:discussion of full rank}, we 
show how the proposed group division results in a matrix $\mathbb{C}_{g}$ that is full-rank. 
With this, we design the $j^{\textrm{\rm th}}$ column of $\mathbb{A}_{g}$, and then the whole matrix of $\mathbb{A}_{g}$.  

Finally, each user  $k\in \Jc$ can recover $W_{\Jc}$ for $\Jc \in \Gc_{g}$ by summing the received codewords (corresponding to $\Gc_{g}$) from the relays in $\Hc_{k}$.
Having recovered the multicast messages $W_{\Jc}$ where $k\in \Jc$, user $k$ then decodes $F_{d_{k},\Jc\setminus \{k\}}$. 


It can be   seen that the   link load to transmit the multicast messages  in $\Vc_{2}$ to each relay 
is $\frac{\binom{2\rsf-1}{\rsf-1}}{(2\rsf-1)K}.$

\paragraph{Transmission of $\Vc_{2}$ for $\Hsf>2\rsf$}
For each set of relays $\Bc$ where $|\Bc|=2\rsf$, we find the users who are connected to $\rsf$ of the chosen relays, and denote them by $\Pc_{\Bc}$. We can see that $|\Pc_{\Bc}|=\binom{2\rsf}{\rsf}$. It can  be also seen that $|\Tc_{\Bc}|=\binom{2\rsf}{\rsf}/2=\binom{2\rsf-1}{\rsf-1}$ where $\Tc_{\Bc}$ is defined in~\eqref{eq:def of TB}. So we can use the same scheme for the case $\Hsf=2\rsf$ to transmit the multicast messages in $\Tc_{\Bc}$ to each relay in $\Bc$ with load $\frac{\binom{2\rsf-1}{\rsf-1}}{(2\rsf-1)K}$, such that each user $k\in \Pc_{\Bc}$ can recover $W_{\Jc}$ where $\Jc \in \Tc_{\Bc}$ and $k\in \Jc$. Notice that as $\Vc_{2}$ contains all the sets of two users $k$ and $k^{\prime}$ where $\Hc_{k}\cap \Hc_{k^{\prime}}=\emptyset$, we have $\underset{\Bc\subseteq [\Hsf]:|\Bc|=2\rsf}{\cup}\Tc_{\Bc}= \Vc_{2}$. Hence, after considering all the sets of relay  $\Bc$ where $|\Bc|=2\rsf$, each user $k$ can recover all the  multicast messages $W_{\Jc}$ where $k\in \Jc$ and $\Jc \in  \Vc_{2}$, and then decodes $F_{d_{k},\Jc\setminus \{k\}}$.   

\paragraph*{Performance} The   max-link load  to transmit the multicast messages in $\Vc_{2}$ is $\frac{\binom{2\rsf-1}{\rsf-1}}{(2\rsf-1)\Ksf}\binom{\Hsf-1}{2\rsf-1}$.
In Appendix~\ref{sec:discussion of full rank}, we show that when $2\rsf-1=p^{v}$ or $pq$, where $p$ and $q$ are as in the  statement of Theorem~\ref{thm:achievability}, we can partition $\Vc_{2}$ in such a way that the IES  is doable. By summing the loads for the two classes of multicast messages, the achieved max-link load of the IES  is  given in~\eqref{eq:achievable interference load}.

\subsection{Concatenated Inner Code delivery Scheme (CICS)}
\label{sub:CICS}
At a high level, the CICS   contains two delivery steps.
We directly transmit each  MAN multicast message to some relays in the first step such that each $W_{\Jc}$ can be recovered by a subset of users in $\Jc$;
these messages can be seen as side information for the second step. 
In the second step, we design linear combinations of the MAN multicast messages such that  the remaining users in $\Jc$ recover $W_{\Jc}$. We illustrate this idea by means of one example first.

\begin{example}
\label{ex:ex1 of CICS}
\rm
Consider the combination networks with end-user-caches in Fig.~\ref{fig: Combination_Networks} where $\Hsf=4$, $\rsf=2$,  $\Ksf=\Nsf=6$, and 
  $\Msf=t=3$. 
Let $\mathbf{d}=(1:6)$.  For each $\Jc\subseteq [6]$ where $|\Jc|=t+1=4$, the MAN multicast message $W_{\Jc}$ in~\eqref{eq:cMAN multicast messages}  
contain $\Bsf/20$ bits, because each file was split into ${\Ksf \choose t} = 20$ equal-length parts.  
We now look in detail into the two-step delivery of the CICS.

\paragraph*{First step}  
For each 
$\Jc\subseteq [6]$ of size $|\Jc|=4$, 
we compute the set of relays each of which is connected to the largest number of users  in $\Jc$, $\Sc_{\Jc} := \arg\max_{h\in [\Hsf]}|\Uc_h\cap \Jc|$. We then partition $W_{\Jc}$ into $|\Sc_{\Jc}|$ non-overlapping and equal-length pieces, $W_{\Jc}=(W^{|\Sc_{\Jc}|}_{\Jc,h}:h\in\Sc_{\Jc})$, and transmit $W^{|\Sc_{\Jc}|}_{\Jc,h}$ to each relay $h\in\Sc_{\Jc}$.

For example, 
for $\Jc=\{1,2,3,4\}$,
relay $1$ is connected to three users in $\Jc$ (users $1$, $2$ and $3$), 
relay $2$ is connected to two users in $\Jc$ (users $1$, $4$), 
relay $3$ is connected to two users in $\Jc$ (users $2$, $4$), and 
relay $4$ is connected to one user in $\Jc$ (user $3$). 
So we have $\Sc_{\{1,2,3,4\}} = \arg\max_{h\in [\Hsf]}|\Uc_h\cap \Jc|=1$.
Therefore, we have $W_{\{1,2,3,4\}}=\{W^1_{\{1,2,3,4\},1}\}$, where $W^1_{\{1,2,3,4\},1}$ contains $\Bsf/20$ bits.

As another example, for $\Jc=\{1,2,5,6\}$,
relay $1$ is connected to two users in $\Jc$ (users $1$ and $2$), 
relay $2$ is connected to two users in $\Jc$ (users $1$ and $5$), 
relay $3$ is connected to two users in $\Jc$ (users $2$ and $6$), and 
relay $4$ is connected to two users in $\Jc$ (users $5$ and $6$). 
So we have $\Sc_{\Jc} =\arg\max_{h\in [\Hsf]}|\Uc_h\cap \Jc|=4$.
Therefore, we have $W_{\{1,2,5,6\}}=\{W^4_{\{1,2,5,6\},1}, W^4_{\{1,2,5,6\},2}, W^4_{\{1,2,5,6\},3}, W^4_{\{1,2,5,6\},4}\}$, where  $W^4_{\{1,2,5,6\},i}$, $i\in [4]$,  contains $\Bsf/80$ bits.

After considering all the sets $\Jc\subseteq [\Ksf]$ where $|\Jc|=4$, 
 we see that the server 
has sent to relay $1$ (and similarly for all other relays) the following messages
\begin{align*}
&W^1_{\{1,2,3,4\},1}, W^1_{\{1,2,3,5\},1} ,W^1_{\{1,2,3,6\},1}, 
W^4_{\{1,2,5,6\},1}, \nonumber\\&  W^4_{\{1,3,4,6\},1}, W^4_{\{2,3,4,5\},1}, 
\end{align*}
for a total of $3\frac{\Bsf}{20}+ 3\frac{\Bsf}{80}=  \frac{3\Bsf}{16}$ bits;
these messages are then transmitted by relay $1$ to the users in $\Uc_1$. 
 
\paragraph*{Second step} 
Recall that $W_{\{1,2,3,4\}}=\{W^1_{\{1,2,3,4\},1}\}$. In the first step,  $W^1_{\{1,2,3,4\},1}$ is   transmitted to relay $1$ and can be recovered by users $1,2,3$. In the second step, we aim to transmit $W^1_{\{1,2,3,4\},1}$ to user  $4$, while considering it as the side information for users $1,2,3$. 
For user $4$, we partition $W^1_{\{1,2,3,4\},1}$ into $\rsf=2$ equal-length parts and denote $W^1_{\{1,2,3,4\},1}=\{W^{1,4}_{\{1,2,3,4\},1,h} : h\in\Hc_4\}$.  
We then let user $4$ recover the first part $W^{1,4}_{\{1,2,3,4\},1,2}$ from relay $2$, and the second part $W^{1,4}_{\{1,2,3,4\},1,3}$ from relay $3$. 
As user $1$  connected to relay $2$  knows $W^{1,4}_{\{1,2,3,4\},1,2}$, we assign $W^{1,4}_{\{1,2,3,4\},1,2}$ to $\Pc^{2}_{4,\{1\}}$,
representing the set of bits needed to be recovered by user $4$ (the first entry in the subscript) from relay $2$ (the superscript) and already known by user $1$ (the second entry in the subscript) who is also connected to relay $2$. 
 As user $2$ connected to relay $3$  knows  $W^{1,4}_{\{1,2,3,4\},1,3}$, we assign $W^{1,4}_{\{1,2,3,4\},1,3}$ to $\Pc^{3}_{4,\{2\}}$, representing the set of bits needed to be recovered by user $4$ (the first entry in the subscript) from relay $3$ (the superscript) and already known by user $2$ (the second entry in the subscript) who is also connected to relay $3$.

Recall that $W_{\{1,2,5,6\}}=\{W^4_{\{1,2,5,6\},1},W^4_{\{1,2,5,6\},2},W^4_{\{1,2,5,6\},3},W^4_{\{1,2,5,6\},4}\}$. Let us focus on $W^4_{\{1,2,5,6\},1}$, 
 which is directly transmitted to relay $1$ in the first step and can be recovered by users $1,2,3$. In the second step, we aim to transmit $W^4_{\{1,2,5,6\},1}$ to users $5,6$, while considering it as the side information for users $1,2,3$. 
For user $5$, we partition $W^4_{\{1,2,5,6\},1}$ into $\rsf=2$ equal-length parts and denote $W^4_{\{1,2,5,6\},1}
=\{W^{4,5}_{\{1,2,5,6\},1,k} : k\in\Hc_5\}$ 
where $\Hc_5=\{2,4\}$. 
We then let user $5$ recover the first part $W^{4,5}_{\{1,2,5,6\},1,2}$ from relay $2$, and the second part $W^{4,5}_{\{1,2,5,6\},1,4}$ from relay $4$. 
As user $1$  connected to relay $2$  knows $W^{4,5}_{\{1,2,5,6\},1}$, we assign $W^{4,5}_{\{1,2,5,6\},1,2}$ to $\Pc^{2}_{5,\{1\}}$. As user $3$ connected to relay $4$ knows  $W^{4,5}_{\{1,2,5,6\},4}$,  we assign $W^{4,5}_{\{1,2,5,6\},1,4}$ to $\Pc^{4}_{5,\{3\}}$. 
For user $6$, we partition $W^{4}_{\{1,2,5,6\},1}$ into $\rsf=2$ equal-length parts  (i.e., $W^4_{\{1,2,5,6\},1}
=\{W^{4,6}_{\{1,2,5,6\},1,k} : k\in\Hc_6\}$), and assign $W^{4,6}_{\{1,2,5,6\},1,3}$ to $\Pc^{3}_{6,\{2\}}$, $W^{4,6}_{\{1,2,5,6\},1,4}$ to $\Pc^{4}_{6,\{3\}}$.

We perform the same procedure for the other pieces in $W_{\{1,2,5,6\}}$.

After considering all the pieces of the multicast messages $W_{\Jc}$, for relay  $1$ we have (and similarly for all other relays) 
\begin{align*}
&\Pc^{1}_{1,\{2\}}=\left\{W^{1,1}_{\{1,2,4,6\},1,1},W^{4,1}_{\{1,2,5,6\},3,1},W^{4,1}_{\{1,3,4,6\},3,1} \right\},\\ 
&\Pc^{1}_{1,\{3\}}=\left\{W^{1,1}_{\{1,3,5,6\},1,1},W^{4,1}_{\{1,2,5,6\},4,1},W^{4,1}_{\{1,3,4,6\},4,1}\right\}, \\
&\Pc^{1}_{2,\{1\}}=\left\{W^{1,2}_{\{1,2,4,5\},1,1},W^{4,2}_{\{1,2,5,6\},2,1},W^{4,2}_{\{2,3,4,5\},2,1}\right\},\\
&\Pc^{1}_{2,\{3\}}=\left\{W^{1,2}_{\{2,3,5,6\},1,1},W^{4,2}_{\{1,2,5,6\},4,1},W^{4,2}_{\{2,3,4,5\},4,1}\right\},\\
&\Pc^{1}_{3,\{1\}}=\left\{W^{1,3}_{\{1,3,4,5\},1,1},W^{4,3}_{\{1,3,4,6\},2,1},W^{4,3}_{\{2,3,4,5\},2,1}\right\},\\ 
&\Pc^{1}_{3,\{2\}}=\left\{W^{1,3}_{\{2,3,4,6\},1,1},W^{4,3}_{\{1,3,4,6\},3,1},W^{4,3}_{\{2,3,4,5\},3,1}\right\};
\end{align*}
all such $\Pc$'s contains $3\Bsf/80$ bits.
Finally  the server transmits to relay $1$ 
\begin{align*}
&\Pc^{1}_{1,\{2\}}\oplus\Pc^{1}_{2,\{1\}},\ \ \Pc^{1}_{1,\{3\}}\oplus\Pc^{1}_{3,\{1\}},\ \ \Pc^{1}_{2,\{3\}}\oplus\Pc^{1}_{3,\{2\}},
\end{align*}
  for a total of   $9\Bsf/80$ bits. 

In the end, for each set $\Jc\subseteq \Uc_h$, each relay $h\in[\Hsf]$ transmits  $\oplus_{j\in\Jc} \Pc_{d_{j},\Jc\setminus\{j\}}$  to each user $j\in \Jc$.
In conclusion,  the  max-link load of the CICS  is $\frac{15}{4\binom{6}{3}}+\frac{9}{4\binom{6}{3}}=0.3$ while the max-link loads of the schemes in~\cite{cachingincom,combinationsecu2018Zewail} are  $0.375$ and $1/3$, respectively. 
\hfill$\square$
\end{example}

The general scheme is as follows.
\paragraph*{First step} 
For each $W_{\Jc}$ in~\eqref{eq:cMAN multicast messages} where $\Jc\subseteq [\Ksf]$ and $|\Jc|=t+1$, we find
$\Sc_{\Jc} := \arg\max_{h\in[\Hsf]}|\Uc_h\cap\Jc|$ (i.e., the set of relays 
each relay in which is connected to the largest number of users in $\Jc$).  We then partition $W_{\Jc}$ into $|\Sc_{\Jc}|$ non-overlapping and equal-length pieces, and denote $W_{\Jc}=\{W^{|\Sc_{\Jc}|}_{\Jc,h}:h\in\Sc_{\Jc}\}$.  The server transmits  $W^{|\Sc_{\Jc}|}_{\Jc,h}$ to relay $h\in \Sc_{\Jc}$, and relay $h\in \Sc$ transmits $W^{|\Sc_{\Jc}|}_{\Jc,h}$ to the users in $\Uc_h$. 

\paragraph*{Second step} 
For each $W_{\Jc}$ as in the first step
the users in $\Jc\cap \Uc_h, h\in\Sc_{\Jc},$ can recover $W^{|\Sc_{\Jc}|}_{\Jc,h}$; 
thus the second step aims to transmit $W^{|\Sc_{\Jc}|}_{\Jc,h}$ to the users in $\Jc\setminus \Uc_h$. 
For each user $k\in\Jc\setminus \Uc_h$,  $W^{|\Sc_{\Jc}|}_{\Jc,h}$ is divided into $\rsf$ non-overlapping and equal-length pieces,  $W^{|\Sc_{\Jc}|}_{\Jc,h}=\{W^{|\Sc_{\Jc}|,k}_{\Jc,h,h^{\prime}}:h^{\prime}\in \Hc_{k}\}$.
We aim to let user $k\in \Jc\setminus \Uc_h$  recover  $W^{|\Sc_{\Jc}|,k}_{\Jc,h,h^{\prime}}$ from relay $h^{\prime}\in \Hc_{k}$.   
For relays $h,h^{\prime}$ and user $k$, where user $k$ is connected to relay $h^{\prime}$ but not to relay $h$,  we define
\begin{align}
\Qc^{k}_{h,h^{\prime}}:=\big\{j\in \Uc_{h}\cap \Uc_{h^{\prime}}:\Hc_{j}\subseteq \Hc_{k}\cup \{h\}\big\}
\label{eq:notpretty}
\end{align}
and assign $W^{|\Sc_{\Jc}|,k}_{\Jc,h,h^{\prime}}$ to $\Pc^{h^{\prime}}_{k,\Qc^{k}_{h,h^{\prime}}}$, representing the set of bits known by the users in $\Qc^{k}_{h,h^{\prime}}$ and to  be recovered  by user $k$ from relay $h^{\prime}$. Note $|\Qc^{k}_{h,h^{\prime}}|=\rsf-1$, as explained in Remark~\ref{footnote:Qc}. 
Finally, for each relay $h\in[\Hsf]$ and each set $\Vc\subseteq\Uc_h$ where $|\Vc|=\rsf$, the server transmits $\oplus_{k\in\Vc}\Pc^{h}_{k,\Jc\setminus \{k\}}$ to relay $h$, which forwards it to the users in $\Vc$.

\begin{rem}
\label{footnote:Qc}
The two partition steps for each $W_{\Jc}$ (e.g., $W^{|\Sc_{\Jc}|}_{\Jc,h}$ and  $W^{|\Sc_{\Jc}|,k}_{\Jc,h,h^{\prime}}$) ensure that the number of bits transmitted from the server to each relay is the same.
 So the achieved max-link load is  proportional to $1/\Hsf$.

We assign $W^{|\Sc_{\Jc}|,k}_{\Jc,h,h^{\prime}}$ to $\Pc^{h^{\prime}}_{k,\Qc^{k}_{h,h^{\prime}}}$ where $\Qc^{k}_{h,h^{\prime}}$ is defined in~\eqref{eq:notpretty};
among   relays $h,h^{\prime}$, user $k$ is only connected to relay $h^{\prime}$.  As the users in $\Qc^{k}_{h,h^{\prime}}$ are connected to relays $h,h^{\prime}$ simultaneously and the connected relays of these users are in the set $\Hc_{k}\cup \{h\}$ including $\rsf+1$ relays, 
one has $|\Qc^{k}_{h,h^{\prime}}|=\binom{\rsf+1-2}{\rsf-2}=\rsf-1$. 
By the symmetry of the combination network, for each relay $a\in \Hc_{k}\setminus \{h^{\prime}\}$,  there must exist one set $\Jc^{\prime}$ with $|\Jc^{\prime}|=t$ where
$a$ is in the set $\arg\max_{b\in[\Hsf]}|\Uc_{b}\cap\Jc^{\prime}|$, which also includes $|\Sc_{\Jc}|$ elements, and the user (assumed to be $k^{\prime}$) connected to relays in $(\Hc_k\cup \{h\})\setminus \{a\}$ is also in $\Jc^{\prime}$.   As user $k^{\prime}$ is connected to relays $h$ and $h^{\prime}$, one has  $k^{\prime}\in \Qc^{k}_{h,h^{\prime}}$.
In addition, user $k^{\prime}$ needs to recover $W^{|\Sc_{\Jc^{\prime}}|,k^{\prime}}_{\Jc^{\prime},a,h^{\prime}}$ from relay $h^{\prime}$, whose length is equal to the length of $W^{|\Sc_{\Jc}|,k}_{\Jc,h,h^{\prime}}$. 
Notice that   $W^{|\Sc_{\Jc^{\prime}}|}_{\Jc^{\prime},a}$ is directly transmitted to relay $a$, so $W^{|\Sc_{\Jc^{\prime}}|,k^{\prime}}_{\Jc^{\prime},a,h^{\prime}}$ is
known by the $\rsf-2$ users in $\Qc^{k}_{h,h^{\prime}} \setminus \{k^{\prime}\}$ and by user $k$. Thus, we can take the XOR of  $W^{|\Sc|,k}_{\Jc,h,h^{\prime}}$ and $W^{|\Sc_{\Jc^{\prime}}|,k^{\prime}}_{\Jc^{\prime},a,h^{\prime}}$, such that users $k,k^{\prime}$ can recover their desired pieces. Similarly, there are $|\Hc_{k}\setminus \{h^{\prime}\}|=\rsf-1$ relays as relay $a$. So $W^{|\Sc|,k}_{\Jc,h,h^{\prime}}$ can be XORed with the other $\rsf-1$ pieces with the same length  (each of which is demanded by one user in $\Qc^{k}_{h,h^{\prime}}$) and then be transmitted to relay $h^{\prime}$.
\end{rem}

\paragraph*{Performance}
For each MAN multicast message $W_{\Jc}$, 
the server directly transmits $W^{|\Sc_{\Jc}|}_{\Jc,h}$ to relay $h\in\Sc_{\Jc}$ in the first step  for a total of $|W_{\Jc}| = \Bsf/\binom{\Ksf}{t}$ bits. 
In the second step, for relay $h\in\Sc_{\Jc}$, $|\Jc\setminus \Uc_h|$ users should recover $W^{|\Sc_{\Jc}|}_{\Jc,h}$. For this purpose,
the server transmits $W^{|\Sc_{\Jc}|,k}_{\Jc,h,h^{\prime}}$ to each user $k\in \Jc\setminus \Uc_h$ for $h^{\prime}\in \Hc_k$  in one linear combination with other $\rsf-1$ pieces of the same length (equal to $\frac{\Bsf}{\rsf\binom{\Ksf}{t}|\Sc_{\Jc}|}$). Hence, the total link load to transmit $W_{\Jc}$ is $\frac{1}{\binom{\Ksf}{t}}+\frac{|\Sc_{\Jc}||\Jc\setminus \Uc_h|}{\rsf\binom{\Ksf}{t}|\Sc_{\Jc}|}=\frac{1+|\Jc\setminus \Uc_h|/\rsf}{\binom{\Ksf}{t}}$.
By the symmetry of network, the number of transmitted bits to each relay is the same as in~\eqref{eq:load of CICS}.

\subsection{Improved Concatenated Inner Code delivery Scheme (ICICS)}
\label{sub:ICICS}
The main idea  in the ICICS is to leverage the multicasting opportunities which are ignored in the CICS. We also consider the network in Example~\ref{ex:ex1 of CICS} to highlight the improvement provided by  the ICICS compared to the CICS.

\begin{example}
\label{ex:ex of comparison CICS ICICS}
\rm
 Consider the same network as in   Example~\ref{ex:ex1 of CICS} where   $\Hsf=4$, $\rsf=2$,  $\Ksf=\Nsf=6$, and 
  $\Msf=t=3$. We also   let $\mathbf{d}=(1:6)$.  The MAN cache placement and the MAN  multicast messages generation are used, such that each MAN multicast message  in~\eqref{eq:cMAN multicast messages}  
contains $\Bsf/20$ bits. As the CICS, the delivery of the ICICS  also contains two steps.

\paragraph*{Improved first step of the ICICS} 
This step is the same as the CICS  with the exception that,
  each MAN multicast message $W_{\Jc}$ 
is divided into $|W_{\Jc}|/P$ packets where each packet contains $P$ bits, for some large enough length $P$, which is possible since $\Bsf$ can be taken arbitrary large  and where $P$ is such that the resulting packets can be seen as elements on a finite field of   suitable size. 

\paragraph*{Improved second step of the ICICS} 
In the CICS, recall that both $W^{4,1}_{\{1,2,5,6\},4,1}$ and  $W^{4,2}_{\{1,2,5,6\},4,1}$ are from  $W^{4 }_{\{1,2,5,6\},4 }$ and should be transmitted to relay $4$.
It can be seen that the CICS  treats $W^{4,1}_{\{1,2,5,6\},4,1}$ and  $W^{4,2}_{\{1,2,5,6\},4,1}$  (demanded by user $1$ and $2$, respectively) as two independent pieces. 
More precisely, the CICS  transmits $W^{4,1}_{\{1,2,5,6\},4,1}$ in one XORed message to relay $1$ to let user $1$ recover; transmits  $W^{4,2}_{\{1,2,5,6\},4,1}$ in another XORed message to relay $1$ to let user $2$ recover.


Instead, we can leverage the following multicasting opportunity.
We put $\text{RLC}(|W^4_{\{1,2,5,6\},4}|/(2P),W^4_{\{1,2,5,6\},4})$ in $\Xc^{1}_{\{1,2\},\{3\}}$, where  $\Xc^{1}_{\{1,2\},\{3\}}$ is the set of packets needed to be recovered by users in $\{1,2\}$ (first part of the subscript) from relay~$1$ (superscript) and already known by the users in $\{3\}$ (second part of the subscript) who are also connected to relay~$1$ (superscript). The number of packets in $\Xc^{1}_{\{1,2\},\{3\}}$ is $|\Xc^{1}_{\{1,2\},\{3\}}|/P$.
We then encode the messages at relay~$1$ as
\begin{align*}
\Xc^{1}_{\{1\},\{2\}}\oplus\Xc^{1}_{\{2\},\{1\}}, \
\Xc^{1}_{\{1\},\{3\}}\oplus\Xc^{1}_{\{3\},\{1\}}, \
\Xc^{1}_{\{2\},\{3\}}\oplus\Xc^{1}_{\{3\},\{2\}},
\end{align*}
where we used the same convention as before when it comes to `summing' sets.
We also send 
$$
\text{RLC}(2|\Xc^{1}_{\{2,3\},\{1\}}|/P,\Xc^{1}_{\{1,2\},\{3\}}\cup\Xc^{1}_{\{1,3\},\{2\}}\cup\Xc^{1}_{\{2,3\},\{1\}})
$$
to relay~$1$. 
Note that the users in $\{1,2,3\}$ know $|\Xc^{1}_{\{2,3\},\{1\}}|/P$ packets in $\Xc^{1}_{\{1,2\},\{3\}}\cup\Xc^{1}_{\{1,3\},\{2\}}\cup\Xc^{1}_{\{2,3\},\{1\}}$. So if the server transmits  $2|\Xc^{1}_{\{2,3\},\{1\}}|/P$ random linear combinations of those packets to relay~$1$, 
each user in $\{1,2,3\}$ can recover all    the packets in $\Xc^{1}_{\{1,2\},\{3\}}\cup\Xc^{1}_{\{1,3\},\{2\}}\cup\Xc^{1}_{\{2,3\},\{1\}}$ with high probability provided that $\Bsf\to \infty$.

The  max-link load of the ICICS is $\left.\frac{15}{4\binom{K}{t}}+\frac{17}{8\binom{K}{t}}\right|_{\Ksf=6,t=3}=0.29375$, which is less than the max-link load  of the CICS  (equal to $0.3$); for the same set of parameters, the max-link loads of the schemes in~\cite{cachingincom,combinationsecu2018Zewail} are   $0.375$ and $1/3$, respectively.
\hfill$\square$
\end{example}

In general, the pseudo-code for this improved delivery can be found in Algorithm~1 next.\\
\rule{\textwidth/2}{0.8pt}

\vspace{-3bp}
\textbf{Algorithm 1 }Improved Concatenated Inner Code delivery Scheme (ICICS)\vspace{-6bp}\\
\rule{\textwidth/2}{0.8pt}
\begin{enumerate}
\item \textbf{input:} $F_{i,\Wc}$ where $i\in[\Nsf]$, $\Wc\subseteq[\Ksf]$ and $|\Wc|=t$;  \textbf{initialization:} $\Xc^{h}_{\Wc_1,\Wc_2}=\emptyset$ for each $h\in [\Hsf]$, $\Wc_1 \subseteq \Uc_{h}$ and $\Wc_2\subseteq \Uc_{h}\setminus \Wc_1$;
\item \textbf{for} each $\Jc\subseteq[K]$ where $|\Jc|=t+1$,
\begin{enumerate}
\item let $W_{\Jc}=\oplus_{j\in\Jc} F_{d_{j},\Jc\setminus\{j\}}$; divide $W_{\Jc}$  into $|W_{\Jc}|/P$ packets where each packet contains $P$ bits, for some large enough length $P$;
\item $\Sc_{\Jc}=\arg\max_{h\in[\Hsf]}|\Uc_{h}\cap\Jc|$;
divide all   packets in $W_{\Jc}$ into $|\Sc_{\Jc}|$ non-overlapping parts with equal length,  $W_{\Jc}=\{W^{|\Sc_{\Jc}|}_{\Jc,h}:h\in[\Hsf] \}$;
\item \textbf{for} each $h\in \Sc_{\Jc}$,
\begin{enumerate}
\item transmit $W^{|\Sc_{\Jc}|}_{\Jc,h}$ to relay $h$; 
\item \textbf{for} each $h^{\prime}\in [\Hsf]$ where $\Uc_{h^{\prime}}\cap (\Jc\setminus \Uc_{h})\neq \emptyset$, 
assign $|W^{|\Sc_{\Jc}|}_{\Jc,h}|/(\rsf P)$ random linear combinations of packets in $W^{|\Sc_{\Jc}|}_{\Jc,h}$ to $\Xc^{h^{\prime}}_{\Ac,\Bc}$ where 
$\Ac=\Uc_{h^{\prime}}\cap (\Jc\setminus \Uc_h) $ and $\Bc=\big\{j\in \Uc_{h}\cap \Uc_{h^{\prime}}:\Hc_{j}\subseteq \Hc_{\Ac}\cup \{h\}\big\}$;
\end{enumerate}
\end{enumerate}
\item \textbf{for} each $h\in [\Hsf]$, 
\begin{enumerate}
\item for each $\Jc^{\prime}\subseteq \Uc_h$, let $\Lc^{h}_{\Jc^{\prime}}=\text{RLC}(c ,\Cc)$, where $c=\underset{k\in\Jc^{\prime}}{\max}
\underset{\Wc_1,\Wc_2:\Wc_1\cup\Wc_2=\Jc^{\prime},k\notin \Wc_2}{\sum}
 |\Xc^{h}_{\Wc_1,\Wc_2}|$ and $\Cc=\underset{\Wc_1,\Wc_2:\Wc_1\cup\Wc_2=\Jc^{\prime}}{\cup}
  \Xc^{h}_{\Wc_1,\Wc_2}$;
\item we transmit $\text{RLC}(c^{\prime} ,\Cc^{\prime})$ to relay $h$ and relay $h$ then forwards $\text{RLC}(c^{\prime} ,\Cc^{\prime})$ to users in $\Uc_h$, where $c^{\prime}=
\underset{k\in \Uc_h}{\max} \underset{\Jc^{\prime}\subseteq\Uc_h:\Lc^{h}_{\Jc^{\prime}} \textrm{\rm is unknown to }k}{ \sum}
 |\Lc^{h}_{\Jc^{\prime}}|$ and $\Cc^{\prime}=\underset{\Jc^{\prime}\subseteq\Uc_h }{\cup}
  \Lc^{h}_{\Jc^{\prime}}$;
\end{enumerate}
\end{enumerate}
\rule{\textwidth/2}{0.8pt}

\paragraph*{First step} 
This step is the same as in Section~\ref{sub:CICS}  except that each coded messages $W_{\Jc}$  is divided into $|W_{\Jc}|/P$ packets where each packet contains $P$ bits, for some large enough length $P$.
\paragraph*{Second step} 
For each  $W_{\Jc}$ where $\Jc\subseteq [\Ksf]$ and $|\Jc|=t+1$, and each $h\in\Sc_{\Jc}$,  the second step is used to transmit $W^{|\Sc_{\Jc}|}_{\Jc,h}$ to the users in $\Jc\setminus \Uc_h$. In this paragraph, to simplify the notation, we let $\Ac =\Uc_{h^{\prime}}\cap (\Jc\setminus \Uc_h) $ and $\Bc =\big\{j\in \Uc_{h}\cap \Uc_{h^{\prime}}:\Hc_{j}\subseteq \Hc_{\Ac}\cup \{h\}\big\}$. For each $h^{\prime}\in [\Hsf]\setminus \{h\}$ where $\Ac\neq \emptyset$, we assign $|W^{|\Sc_{\Jc}|}_{\Jc,h}|/(\rsf P)$ random linear combinations of packets in $W^{|\Sc_{\Jc}|}_{\Jc,h}$ to $\Xc^{h^{\prime}}_{\Ac,\Bc}$, where $\Xc^{h^{\prime}}_{\Ac,\Bc}$ represents the set of  packets to be recovered by users in $\Ac$ from relay $h^{\prime}$ and already known by the users in $\Bc$ who are also connected to relay $h^{\prime}$.\footnote{\label{foot:consistence}To make the notation consistent with that of the CICS, $|W^{|\Sc_{\Jc}|}_{\Jc,h}|$ represents the number of bits in $ W^{|\Sc_{\Jc}|}_{\Jc,h} $.}  

We aim to let each user $k\in \Jc\setminus \Uc_h$  recover   $\Xc^{h^{\prime}}_{\Wc_1,\Wc_2} $ where $h^{\prime}\in\Uc_{k}$ and $k\in \Wc_1$, such that it can recover $|W^{|\Sc_{\Jc}|}_{\Jc,h}|/P$ random linear combinations of packets in $W^{|\Sc_{\Jc}|}_{\Jc,h}$ and then recover $W^{|\Sc_{\Jc}|}_{\Jc,h}$ with high probability provided that $\Bsf\to \infty$. We use a two-stage coding.

{\it Stage~1:} For each relay $h\in[\Hsf]$ and each set $\Jc^{\prime} \subseteq \Uc_h$, we encode  the packets in $\Cc$ by $\Lc^{h}_{\Jc^{\prime}}=\text{RLC}(c ,\Cc)$ where
\begin{align*}
&\Cc=\underset{\Wc_1,\Wc_2:\Wc_1\cup\Wc_2=\Jc^{\prime}}{\cup}\Xc^{h}_{\Wc_1,\Wc_2},\\
&c=\max_{k\in\Jc^{\prime}}\sum_{\Wc_1,\Wc_2:\Wc_1\cup\Wc_2=\Vc,k\notin \Wc_2}|\Xc^{h}_{\Wc_1,\Wc_2}|,
\end{align*}  
  $|\Xc^{h}_{\Wc_1,\Wc_2}|$ represents the number of packets in  $\Xc^{h}_{\Wc_1,\Wc_2}$, and
 $c $ represents the maximal number of packets in $\Cc$ not known by  each user in $\Jc^{\prime}$.
So from  $\Lc^{h}_{\Jc^{\prime}}$, each user $k\in \Jc^{\prime}$ can recover   the packets in $\Xc^{h}_{\Wc_1,\Wc_2} $ where $k\in \Wc_1$.

{\it Stage~2:} If for each set of packets $\Xc^{h}_{\Wc_1,\Wc_2}$ where $\Wc_1\cup\Wc_2=\Jc^{\prime}$ and $|\Xc^{h}_{\Wc_1,\Wc_2}|>0$, we have $k\in \Wc_2$, then user $k$ has already known $\Lc^{h}_{\Jc^{\prime}}$ from the first stage. Hence, for relay $h\in [\Hsf]$, we can encode the packets in  $\Cc^{\prime}$   by $\text{RLC}(c^{\prime} ,\Cc^{\prime})$ where 
\begin{align*}
&\Cc^{\prime}=\underset{\Jc^{\prime}\subseteq\Uc_h }{\cup} \Lc^{h}_{\Jc^{\prime}},\\
&c^{\prime}=\max_{k\in \Uc_h}\sum_{\Jc^{\prime}\subseteq\Uc_h:\Lc^{h}_{\Jc^{\prime}} \textrm{\rm is unknown to }k}|\Lc^{h}_{\Jc^{\prime}}|,
\end{align*}
where $|\Lc^{h}_{\Jc^{\prime}}|$ represents the number of packets in  $\Lc^{h}_{\Jc^{\prime}}$.
Finally, we  transmit $\text{RLC}(c^{\prime} ,\Cc^{\prime})$ to relay $h$, and relay $h$ then forwards $\text{RLC}(c^{\prime} ,\Cc^{\prime})$ to users in $\Uc_h$.
So from  $\text{RLC}(c^{\prime} ,\Cc^{\prime})$, each user in $\Uc_h$ can recover  the packets in  $\Cc^{\prime}$.

\begin{rem}
\label{rem:file split}
We now analyze the actual file split level for our proposed schemes.
In the IES, we need not to divide $W_{\Jc}$. In the DIS, we divide each $W_{\Jc}$ into $\rsf$ non-overlapping and equal-length pieces. In the CICS, we start by dividing $W_{\Jc}$ into $|\Sc_{\Jc}|$  non-overlapping and equal-length pieces and then divide each obtained piece into $\rsf$  non-overlapping and equal-length parts. Hence, $W_{\Jc}$ is divided into $|\Sc_{\Jc}|\rsf$ pieces. In the ICICS, we first  divide $W_{\Jc}$ into $|\Sc_{\Jc}|$  non-overlapping and equal-length pieces and then divide each obtained piece into at most $\Hsf-1$  non-overlapping and equal-length parts. Hence, $W_{\Jc}$ is divided into at most $|\Sc_{\Jc}|(\Hsf-1)$ pieces.
\end{rem}

\section{Proofs of the Novel Converse Bounds under the Constraint of Uncoded Cache Placement} 
\label{sec:outer bound}
In the rest of the paper, for a set of subfiles $\Sc\subseteq\Tc_{\dv,\Zm}$ where  $\Tc_{\dv,\Zm}$ is given in~\eqref{eq:Tc{dv,Zm}},
we denote by $H(\Sc)$ the joint entropy of the subfiles in $\Sc$, and 
by $H(Y|\Sc^{c})$ the entropy of a random variable $Y$ conditioned on the subfiles in $\Sc^{c}:=\Tc_{\dv,\Zm}\setminus \Sc$. 

\subsection{Preliminaries} 
\label{subsec:Useful Propositions}
We start this section by extending the  ``acyclic index coding converse bound'' for shared-link network from~\cite{indexcodingcaching2020} to combination  networks.
\begin{prop}
\label{prop:acyclic outer bound}
Consider a combination network with end-user-caches, where the cache  placement $\Zm$ is uncoded and 
  the demands in $\dv$ are distinct.
For a set of relays $\Jc\subseteq[\Hsf]$, and 
for an acyclic set of subfiles $\Sc\subseteq\Tc_{\dv,\Zm}$ in the directed graph $G_{\Tc_{\dv,\Zm}}$
that are demanded by the users in $\Kc_{\Jc}$,
the following must hold 
\begin{subequations}
\begin{align}
H(\Sc)
  &\leq H(X_{\Jc}|\Sc^{c})+\Bsf\varepsilon_{\Bsf}
\label{eq:acyclic outer bound1}\\&\leq |\Jc|\Bsf \, \Rsf^{\star}_{\mathrm{u}} + \Bsf\varepsilon_{\Bsf}.
\label{eq:acyclic outer bound}
\end{align}
\end{subequations}
\end{prop}
\begin{IEEEproof}
\begin{subequations}
The entropy of the subfiles in $\Sc$ is bounded as
\begin{align}
H(\Sc) 
 &= H(\Sc|\Sc^{c})  \\
 &= H(X_{\Jc},\Sc|\Sc^{c})
\label{eq:ineq in outer bound 1 line 2}\\
 &=H(X_{\Jc}|\Sc^{c})+H(\Sc|X_{\Jc},\Sc^{c})
\label{eq:ineq in outer bound 1 line 3}\\
 &\leq H(X_{\Jc}|\Sc^{c})+\Bsf\varepsilon_{\Bsf}
\label{eq:ineq in outer bound 1 line 4}\\
 &\leq H(X_{\Jc})+\Bsf\varepsilon_{\Bsf} \\
 &\leq |\Jc|\Rsf^{\star}_{\mathrm{u}} \Bsf+\Bsf\varepsilon_{\Bsf},
\label{eq:ineq in outer bound 1 line 6}
\end{align}
where 
in~\eqref{eq:ineq in outer bound 1 line 2} we use  the independence of the subfiles and
the fact that $X_{\Jc}$ is function of $\Tc_{\dv,\Zm}$,
in~\eqref{eq:ineq in outer bound 1 line 4} we use the fact that $\Sc$ is acyclic
and Fano's inequality (where $\lim_{\Bsf\to\infty}\varepsilon_{\Bsf}=0$), and
in~\eqref{eq:ineq in outer bound 1 line 6} we use  the definition of $\Rsf^{\star}_{\mathrm{u}}$.
\end{subequations}
\end{IEEEproof}

Proposition~\ref{prop:acyclic outer bound} may not be tight when $|\Jc|\Rsf^{\star}_{\mathrm{u}}$ in~\eqref{eq:ineq in outer bound 1 line 6} is strictly larger than $H(X_{\Jc}|\Sc^{c})$ in~\eqref{eq:ineq in outer bound 1 line 4}. In the following, we tighten the bound in Proposition~\ref{prop:acyclic outer bound}.
\begin{prop}
\label{prop:improved acyclic outer bound} 
Consider a combination network with end-user-caches, where the cache  placement $\Zm$ is uncoded and 
  the demands in $\dv$ are distinct.
For a set of relays $\Jc\subseteq[\Hsf]$,
and for two sets of subfiles $\Sc_{1},\Sc_{2}\subseteq\Tc_{\dv,\Zm}$
that are acyclic in the graph $G_{\Tc_{\dv,\Zm}}$,
where $\Sc_{1}$ 
includes some subfiles demanded by the users in $\Kc_{\Jc}$ 
and $\Sc_{2}$ 
includes some subfiles demanded by the users in $[\Ksf]\setminus\Kc_{\Jc}$ but not cached by
the users in $\Kc_{\Jc}$, 
we have
\begin{align}
&H(\Sc_{1}) + H(X_{\Jc}|\Sc_{2}^{c}) 
\leq |\Jc|\Bsf \, \Rsf^{\star}_{\mathrm{u}}
 + 2\Bsf\varepsilon_{\Bsf}.
\label{eq:improved acyclic outer bound}
\end{align}
\end{prop}
%
\begin{IEEEproof}
\begin{subequations}
The entropy of the subfiles 
can be bounded as
\begin{align} 
 H(\Sc_{1},\Sc_{2})
 &\leq 
 H\big(X_{[\Hsf]}|(\Sc_{1} \cup \Sc_{2})^{c}\big)
+\Bsf\varepsilon_{\Bsf}
\label{eq:Proof for improved acyclic outer bound line1}\\
 &= 
 H\big(X_{\Jc}|(\Sc_{1} \cup \Sc_{2})^{c}\big)
+H\big(X_{[\Hsf]\setminus\Jc}|X_{\Jc},\Sc_{2}^{c}\big) \nonumber\\& 
 +I\big(\Sc_{1};X_{[\Hsf]\setminus\Jc}|X_{\Jc},(\Sc_{1} \cup \Sc_{2})^{c}\big)
 +\Bsf\varepsilon_{\Bsf}
 \label{eq:Proof for improved acyclic outer bound line2}\\
 & \leq 
 |\Jc|\Bsf \, \Rsf^{\star}_{\mathrm{u}}
+H\big(X_{[\Hsf]\setminus\Jc}|X_{\Jc},\Sc_{2}^{c}\big) \nonumber\\& 
+H\big(\Sc_{1}|X_{\Jc},(\Sc_{1} \cup \Sc_{2})^{c}\big)
+\Bsf\varepsilon_{\Bsf}
\label{eq:Proof for improved acyclic outer bound line3}\\
 &\leq 
 |\Jc|\Bsf \, \Rsf^{\star}_{\mathrm{u}}
+H\big(X_{[\Hsf]\setminus\Jc}|X_{\Jc},\Sc_{2}^{c}\big)
+2\Bsf\varepsilon_{\Bsf}
\label{eq:Proof for improved acyclic outer bound line4}\\
 &\leq 
 |\Jc|\Bsf \, \Rsf^{\star}_{\mathrm{u}}
-H\big(X_{\Jc}|\Sc_{2}^{c}\big)
+H(\Sc_{2})
+2\Bsf\varepsilon_{\Bsf}
\label{eq:Proof for improved acyclic outer bound line5},
\end{align}
where~\eqref{eq:Proof for improved acyclic outer bound line1} is from~\eqref{eq:acyclic outer bound};~\eqref{eq:Proof for improved acyclic outer bound line4} is from Fano's inequality (where $\lim_{\Bsf\to\infty}\varepsilon_{\Bsf}=0$), and from the fact that $\Sc_{1}$ is acyclic and  $\Sc_{2}$ does not include the side information of the user requesting $\Sc_{1}$;~\eqref{eq:Proof for improved acyclic outer bound line5} is because
\begin{align*}
H(\Sc_{2}) 
& =H(\Sc_{2}|\Sc_{2}^{c}) \\
& =H\big(\Sc_{2},X_{[\Hsf]}|\Sc_{2}^{c}\big) \\
& \geq H\big(X_{[\Hsf]}|\Sc_{2}^{c}\big) \\
& =H\big(X_{\Jc}|\Sc_{2}^{c}\big)+H\big(X_{[\Hsf]\setminus \Jc}|X_{\Jc},\Sc_{2}^{c}\big).
\end{align*}
This concludes the proof.
\end{subequations}
\end{IEEEproof}
 
Finally, we generalize the well-known sub-modularity of entropy. 
\begin{prop}
\label{prop:lem for improved outer bound 2}
Let $\Yc$ be a set of random variables, 
and $\Mc$ be a set of mutually independent random variables (but not necessary independent of $\Yc$). 
If $\Yc_{1},\Yc_{2}\subseteq \Yc$ and $\Mc_{1},\Mc_{2}\subseteq \Mc$, then
the following must hold
\begin{align}
 H(\Yc_{1}|\Mc_{1})+H(\Yc_{2}|\Mc_{2}) 
 &\geq H(\Yc_{1}\cup \Yc_{2}|\Mc_{1}\cup\Mc_{2}) 
     +\nonumber\\&  H(\Yc_{1}\cap \Yc_{2}|\Mc_{1}\cap\Mc_{2}).
\label{eq:eq in lemma1}
\end{align}
\end{prop}
\begin{IEEEproof}
Without loss of generality, assume $\Mc_{1}=\{M_{0},M_{1}\}$ and $\Mc_{2}=\{M_{0},M_{2}\}$, where $M_{0},M_{1},M_{2}$
are independent random variables. We have
\begin{subequations}
\begin{align}
 & H(\Yc_{1}|M_{0},M_{1})+H(\Yc_{2}|M_{0},M_{2})\nonumber\\
 &=H(\Yc_{1}|M_{0},M_{1},M_{2})+I(\Yc_{1};M_{2}|M_{0},M_{1})\nonumber\\ & +H(\Yc_{2}|M_{0},M_{1},M_{2})+I(\Yc_{2};M_{1}|M_{0},M_{2}) \\ 
 &=H(\Yc_{1}\cup\Yc_{2}|M_{0},M_{1},M_{2})+I(\Yc_{1};\Yc_{2}|M_{0},M_{1},M_{2}) \nonumber\\& +I(\Yc_{1};M_{2}|M_{0},M_{1})+I(\Yc_{2};M_{1}|M_{0},M_{2}) \label{eq:chain rule of entropy}\\ 
 &\geq H(\Yc_{1}\cup\Yc_{2}|M_{0},M_{1},M_{2}) +  H(\Yc_{1}\cap \Yc_{2}| M_{0}),\label{eq:last eq}
\end{align}
\end{subequations}
where~\eqref{eq:last eq} follows from   
\begin{subequations}
\begin{align}
 & I(\Yc_{1};\Yc_{2}|M_{0},M_{1},M_{2})+I(\Yc_{1};M_{2}|M_{0},M_{1})+I(\Yc_{2};M_{1}|M_{0},M_{2})\nonumber\\
 &= I(\Yc_{1};\Yc_{2}|M_{0},M_{1},M_{2})+I(\Yc_{1},M_{1};M_{2}|M_{0})\nonumber\\& +I(\Yc_{2};M_{1}|M_{0},M_{2}) \label{eq:indep}\\
 & =I(\Yc_{1};\Yc_{2}|M_{0},M_{1},M_{2})+I(\Yc_{1},M_{1};\Yc_{2},M_{2}|M_{0})\nonumber\\& -I(\Yc_{1},M_{1};\Yc_{2}|M_{0},M_{2})+I(\Yc_{2};M_{1}|M_{0},M_{2})\label{eq:chain rule of mutual 1}\\
 & =I(\Yc_{1};\Yc_{2}|M_{0},M_{1},M_{2})+I(\Yc_{1},M_{1};\Yc_{2},M_{2}|M_{0})\nonumber\\& -I(\Yc_{1};\Yc_{2}|M_{0},M_{1},M_{2})\label{eq:chain rule of mutual 2}\\
 & =I(\Yc_{1},M_{1};\Yc_{2},M_{2}|M_{0}) \geq I(\Yc_{1};\Yc_{2}|M_{0})
   \geq H(\Yc_{1}\cap \Yc_{2}| M_{0}). 
\end{align}
\end{subequations}
Notice that~\eqref{eq:indep} follows from that $M_{0},M_{1},M_{2}$
are independent random variables, and~\eqref{eq:chain rule of mutual 1} and~\eqref{eq:chain rule of mutual 2} follow from the chain rule of   mutual information.
\end{IEEEproof}

\begin{rem}
If either $\Yc_{1}=\Yc_{2}$ or $\Mc_{1}=\Mc_{2}$, Proposition~\ref{prop:lem for improved outer bound 2} reduces to the well-known submodularity of entropy~\cite{onthecapacityindex,submodular}.
\end{rem}


\subsection{Proof of Theorem~\ref{thm:acyclic outer bound}}
\label{subset:acyclic outer bound}
In~\eqref{eq:defxw}, $\Nsf \Bsf x_{\Wc}$ represents the number of bits only cached by the users in $\Wc\subseteq[\Ksf]$.
For a demand vector $\dv\in [\Nsf]$ whose elements are distinct, a set $\Sc^\prime\subseteq [\Ksf]$, and a vector $\mathbf{v}=(v_{1},\ldots,v_{|\mathbf{v}|})$ where $v_{i}\in \Sc^\prime, \ \forall i\in [|\mathbf{v}|]$, we define
\begin{align}
&f(\dv,\Sc^\prime,\mathbf{v})  :=
\bigcup_{i\in[|\mathbf{v}|]}
\big\{
F_{d_{v_{i}},\Wc} : 
\Wc\subseteq\Sc^\prime \setminus \{v_{1},\ldots,v_{i}\}
\big\};
\label{eq:def of f}
\end{align} 
by~\cite[Lemma 1]{indexcodingcaching2020} the set $f(\dv,\Sc^\prime,\mathbf{v})$ forms an acyclic set in the directed graph $G_{\Tc_{\dv,\Zm}}$. 
Fix one   $\Qc\subseteq[\Hsf]$ with $|\Qc|\in [\rsf:\Hsf]$,  
and one permutation $\mathbf{p}(\Kc_{\Qc})$. For 
  each demand vector $\dv$ with distinct demands,
Proposition~\ref{prop:acyclic outer bound} with $\Jc=\Qc$ and  $\Sc = f\big(\dv,[\Ksf],\mathbf{p}(\Kc_{\Qc})\big)$  provides a converse bound on $\Rsf^{\star}_{\mathrm{u}}$.
By summing all the so obtained bounds for the  fixed $\Qc $ and $\mathbf{p}(\Kc_{\Qc})$,   we arrive at~\eqref{eq:outer bound 2 constrant 1}.

\subsection{First example on how to improve Theorem~\ref{thm:acyclic outer bound}}
\label{ex:example1}
Our first improvement to Theorem~\ref{thm:acyclic outer bound} is explained by  way of an example -- see also Remark~\ref{rem:why better 1}. 
Consider the combination network  with end-user-caches  in Fig.~\ref{fig: Combination_Networks}   where $\Hsf=4$, $\rsf=2$,  $\Ksf=\Nsf=6$, and 
  $\Msf=2$.   
Consider the demand vector $\dv=(1,\ldots,6)$.
Choose a set of relays $\Qc$ and divide $\Qc$ into several disjoint subsets, each of which has a length not less than $\rsf=2$. In this example, we let $\Qc=[\Hsf]=[4]$ and divide  $\Qc$ into 
$\Qc_{1}=\{1,2\}$ and $\Qc_{2}=\{3,4\}$;
so $\Kc_{\Qc_{1}}=\{1\}$ and $\Kc_{\Qc_{2}}=\{6\}$. 
We then consider the three permutations
$\mathbf{p}(\Kc_{\Qc_{1}}) = (1)$, 
$\mathbf{p}(\Kc_{\Qc_{2}}) = (6)$ and 
$\mathbf{p}\big(\Kc_{\Qc}\setminus (\Kc_{\Qc_{1}}\cup \Kc_{\Qc_{2}})\big) = (2,3,4,5)$.  
Recall the definition of $f(\cdot,\cdot,\cdot)$ given in~\eqref{eq:def of f} and let 
\begin{subequations}
\begin{align}
\Bc_{1}&=f\big(\dv,[\Ksf],\mathbf{p}(\Kc_{\Qc_{1}})\big) = \{F_{1,\Wc}:\Wc \subseteq [2:6]\}, \\
\Bc_{2}&=f\big(\dv,[\Ksf],\mathbf{p}(\Kc_{\Qc_{2}})\big) = \{F_{6,\Wc}:\Wc \subseteq [1:5]\}, \\ 
\Bc_{3}&=f\Big(\dv,[\Ksf]\setminus(\Kc_{\Qc_{1}}\cup \Kc_{\Qc_{2}}),\mathbf{p}\big(\Kc_{\Qc}\setminus (\Kc_{\Qc_{1}}\cup \Kc_{\Qc_{2}})\big)\Big) \nonumber\\& = \{F_{i,\Wc}:i\in[2:5],\Wc\subseteq [i+1:5]\}.
\end{align}
\end{subequations}
By using Proposition~\ref{prop:improved acyclic outer bound} with
$(\Jc,\Sc_{1},\Sc_{2})=(\Qc_{1},\Bc_{1},\Bc_{3})$ we get 
\begin{align}
&H(\Bc_{1}) \leq |\Qc_{1}|\Rsf^{\star}_{\mathrm{u}} \Bsf-H\big(X_{\Qc_{1}}|\Bc_{3}^{c}\big)+ 2\Bsf\varepsilon_{\Bsf},
\label{eq:example 1 B1}
\end{align}
and with
$(\Jc,\Sc_{1},\Sc_{2})=(\Qc_{2},\Bc_{2},\Bc_{3})$ we get
\begin{align}
&H(\Bc_{2}) \leq |\Qc_{2}|\Rsf^{\star}_{\mathrm{u}} \Bsf- H\big(X_{\Qc_{2}}|\Bc_{3}^{c}\big)+ 2\Bsf\varepsilon_{\Bsf}.
\label{eq:example 1 B2}
\end{align}
We sum~\eqref{eq:example 1 B1} and~\eqref{eq:example 1 B2} to obtain
\begin{subequations}
\begin{align}
H(\Bc_{1},\Bc_{2})
& \leq |\Qc|\Rsf^{\star}_{\mathrm{u}} \Bsf - 
\Big[ H\big(X_{\Qc_{1}}|\Bc_{3}^{c}\big)
    + H\big(X_{\Qc_{2}}|\Bc_{3}^{c}\big)\Big]+ 4\Bsf\varepsilon_{\Bsf}
 \\
&\leq |\Qc|\Rsf^{\star}_{\mathrm{u}} \Bsf-H(X_{\Qc}|\Bc_{3}^{c})+ 4\Bsf\varepsilon_{\Bsf} \\
&\leq |\Qc|\Rsf^{\star}_{\mathrm{u}} \Bsf-H(\Bc_{3})+ 5\Bsf\varepsilon_{\Bsf},
\label{eq:example B1 B2 line3}
\end{align}
\end{subequations}
where~\eqref{eq:example B1 B2 line3} follows from~\eqref{eq:acyclic outer bound1}.
%
With the above mentioned choice of permutations and $\Bsf \gg 1$, the bound in~\eqref{eq:example B1 B2 line3} becomes
\begin{align}
4 \Rsf^{\star}_{\mathrm{u}}  &\geq%
\sum_{\Wc\subseteq[6]\setminus\{1\}}%
|F_{1,\Wc}|/\Bsf+%
\sum_{\Wc\subseteq[6]\setminus\{6\}}%
|F_{6,\Wc}|/\Bsf \nonumber\\& + %
\sum_{i\in[2:5]}\sum_{\Wc\subseteq[i+1:5]}%
|F_{i,\Wc}|/\Bsf.
\label{eq:example1 outer bound based on fiW}
\end{align}
If we list all the inequalities in the form of~\eqref{eq:example1 outer bound based on fiW} for all the possible demands where users demand distinct files, and   sum them all together, then  we obtain (the definition of $x_{\Wc}$ is in~\eqref{eq:defxw}) 
\begin{align}
&4\Rsf^{\star}_{\mathrm{u}} \geq %
\sum_{\Wc\subseteq[2:6]} %
 x_{\Wc}+%
 \sum_{\Wc\subseteq[1:5]}%
  x_{\Wc}+%
  \sum_{i\in[2:5]}\sum_{\Wc\subseteq[i+1:5]}%
   x_{\Wc}.
\label{eq:example1 outer bound based on xw}
\end{align}

We then consider all the possible disjoint partitions of $\Qc$, and for each partition we consider all the possible combinations of permutations to write bounds as in the form of~\eqref{eq:example1 outer bound based on xw}. 
For $\Qc$ with  $|\Qc|\leq 3$, as $\Qc$ can not be divided into two sets each of which has length not less than $\rsf=2$, we directly use the bound in~\eqref{eq:outer bound 2 constrant 1}. 
With the file length and memory size constrains in~\eqref{eq:xfile size} and~\eqref{eq:memory size}, we can compute the converse bound by a linear programming with the above mentioned constraints and with variables $(\Rsf^{\star}_{\mathrm{u}} ,x_{\Wc}:\Wc\subseteq[\Ksf]=[6])$. 

By solving the linear programming numerically, the converse bound on $\Rsf^{\star}_{\mathrm{u}}$ given by the above method is $7/17\approx0.411$,
while Theorem~\ref{thm:acyclic outer bound} gives $9/23\approx0.391$.

\begin{rem}\label{rem:why better 1}
Notice that in~\eqref{eq:example B1 B2 line3}, $|\Qc|\Rsf^{\star}_{\mathrm{u}} \Bsf\geq H(\Bc_{1},\Bc_{2},\Bc_{3})$ where $\Bc_{1}\cup\Bc_{2}$ forms a directed circle. The technique   in this example provides  a tighter converse bound compared to Theorem~\ref{thm:acyclic outer bound} because it allows to deal with cycles in the directed graph that represents the equivalent index coding problem. 
\end{rem}

\subsection{Second example on how to improve Theorem~\ref{thm:acyclic outer bound}}
\label{ex:example2}
Our second improvement to Theorem~\ref{thm:acyclic outer bound} is explained by  way of an example -- see also Remark~\ref{rem:why better 2}. 

For a set  $\Sc$ and a vector $\mathbf{p}$, where each element of $\Sc$ is also an element in $\mathbf{p}$, we define $g(\Sc,\mathbf{p})$ as the vector obtained by removing the elements not in $\Sc$ from $\mathbf{p}$, e.g., $g(\{1,2,3\},(2,4,1,3))=(2,1,3)$. 

Consider the combination network  with end-user-caches  in Fig.~\ref{fig: Combination_Networks} where $\Hsf=4$, $\rsf=2$,  $\Ksf=\Nsf=6$, and 
  $\Msf=1/2$.  
Consider the demand vector $\dv=(1,\ldots,6)$. 
For an integer $b\in [\rsf:\Hsf]=[2:4]$, e.g., say $b=3$, 
consider each set of relays $\Qc$ with cardinality $b$. 
Consider a permutation $\mathbf{p}_{\Kc_{\Qc}}$ and a permutation $\mathbf{p}([\Ksf])$.  We apply Proposition~\ref{prop:improved acyclic outer bound} with 
$\Jc=\Qc$ 
so as to obtain
\begin{subequations}
\begin{align}
&|\Qc|  \Bsf \Rsf^{\star}_{\mathrm{u}} \geq 
H\big(X_{\Qc}\big)\geq
H\big(\Sc_{1}\big) 
+H\big(X_{\Qc}|\Sc_{2}^c\big)+ 2\Bsf\varepsilon_{\Bsf}, 
\label{eq:example2 by theorem 2}
\\&
\Sc_{1}=f\big(\dv,[\Ksf],\mathbf{p}(\Kc_{\Qc})\big),
\\&
\Sc_2=\Sc_{\Qc}:=f\Big(\dv,[\Ksf] \setminus \Kc_{\Qc},g\big([\Ksf] \setminus \Kc_{\Qc},\mathbf{p}([\Ksf])\big)\Big).
\label{eq:use this Sc2}
\end{align}
\end{subequations}
If $\Qc=\{1,2,3\}$ and thus $\Kc_{\Qc}=\{1,2,4\}$, we have 
\begin{align*}
 g\big([\Ksf] \setminus \Kc_{\Qc},\mathbf{p}([\Ksf])\big) &=g\big(\{3,5,6\},(1,\dots,6) \big)=(3,5,6),\\
 H\big(X_{\Qc}|\Sc_{\Qc}^c\big) &=H\Big(X_{\{1,2,3\}}\big| f\big(\dv,\{3,5,6\},(3,5,6)\big)\Big).
\end{align*}
We sum all the inequalities in the form of~\eqref{eq:example2 by theorem 2} for all the possible demands where the users request distinct files. With $\Bsf \gg1$, we have
\begin{subequations}
\begin{align}
&|\Qc|\Rsf^{\star}_{\mathrm{u}} \geq %
\sum_{j\in[|\Kc_{\Qc}|]}\sum_{\medspace\medspace\Wc\subseteq[\Ksf]\setminus\cup_{k\in[j]}\{ p_{k}(\Kc_{\Qc})\}}%
x_{\Wc}+ y_{\Qc,\mathbf{p}([\Ksf])},
\label{eq:sumxw}\\
&
y_{\Qc,\mathbf{p}([\Ksf])}:=\frac{1}{ \Bsf \Ksf!}    
\sum_{\dv: d_{i}\neq d_{j},  i\neq j} 
H\big(X_{\Qc}\big|
\Sc_{\Qc}^{c}\big), \ \text{with $\Sc_{\Qc}$ in~\eqref{eq:use this Sc2}.} 
\label{eq:def of yqp}
\end{align} 
\end{subequations}
For $\Qc=\{1,2,3\}$ and $\Qc=\{1,2,4\}$, we have
\begin{subequations}
\begin{align}
& H\big(X_{\{1,2,3\}}\big|\Ac^{c}\big)
+H\big(X_{\{1,2,4\}}\big|\Bc^{c}\big)
\nonumber\\ &\geq H\big(X_{\{1,2,3,4\}}|\{F_{6,\emptyset}\}^{c}\big)
    + H\big(X_{\{1,2\}}\big|
\Ac^{c} \cap 
\Bc^{c} \big)
\label{eq:example2 using lemma1 line1}
\\
&\geq H(F_{6,\emptyset})
    + H\Big(X_{\{1,2\}}\big|
\Ac^{c} \cap
\Bc^{c} \Big),
\label{eq:example2 using lemma1 line2}
\\&
\Ac:= f\big(\dv,\{3,5,6\},(3,5,6)\big) \nonumber\\& =\big\{ F_{3,\Wc}:\Wc\subseteq \{5,6\}\big\}\cup\{ F_{5,\emptyset}, F_{5,\{6\}}, F_{6,\emptyset}\},
 \\
&\Bc:= f\big(\dv,\{2,4,6\},(2,4,6)\big)\nonumber\\& =\big\{F_{2,\Wc}:\Wc\subseteq \{4,6\}\big\}\cup\{ F_{4,\emptyset}, F_{4,\{6\}}, F_{6,\emptyset}\}, 
\end{align} 
\end{subequations}
where to get~\eqref{eq:example2 using lemma1 line1} we used Proposition~\ref{prop:lem for improved outer bound 2} and the fact that $\Ac^{c}\cup \Bc^{c}=\{F_{6,\emptyset}\}^{c}$. 
Notice that without using Proposition~\ref{prop:lem for improved outer bound 2} we can not bound the sum of the two terms in the LHS (left hand side) of~\eqref{eq:example2 using lemma1 line1}. By using Proposition~\ref{prop:lem for improved outer bound 2}, we have the term $H\big(X_{\{1,2,3,4\}}|\{F_{6,\emptyset}\}^{c}\big)$ and all the relays connected to user $6$ demanding $F_{6,\emptyset}$ are in $\{1,2,3,4\}$, such that we can use Proposition~\ref{prop:acyclic outer bound} to bound this term by $H(F_{6,\emptyset})$.
Similarly,  
\begin{align}
\sum_{\Qc\subseteq[\Hsf]:|\Qc|=b} \negmedspace \negmedspace  y_{\Qc,\mathbf{p}([\Ksf])}\geq\frac{1}{\Ksf !\Bsf}\negmedspace \negmedspace \sum_{\dv:d_{i}\neq d_{j}, i\neq j}\negmedspace \negmedspace H(F_{1,\emptyset},\ldots,F_{6,\emptyset})=6 x_{\emptyset}.\label{eq:example sum of yqp}
\end{align}
We then consider each permutation $\mathbf{p}(\Kc_{\Qc})$ for $\Qc\subseteq[\Ksf]$ with $|\Qc|=b=3$ to write inequalities in the form of~\eqref{eq:sumxw}. With the constraints in~\eqref{eq:xfile size},~\eqref{eq:memory size} and~\eqref{eq:example sum of yqp} we can compute a converse bound on $\Rsf^{\star}_{\mathrm{u}}$ by solving a linear programming which gives $13/12$, while Theorem~\ref{thm:acyclic outer bound} gives $17/16$. 
Notice that in general we should consider each permutation $\mathbf{p}([\Ksf])$ to write  constraints in the form of~\eqref{eq:example sum of yqp}, but in this example it is enough to consider one permutation.
In order to reduce the number of variables, 
the constraint in~\eqref{eq:sumxw} is equivalent to the following
\begin{subequations}
\begin{align}
&|\Qc|\Rsf^{\star}_{\mathrm{u}} \geq %
\sum_{j\in[|\Kc_{\Qc}|]}\sum_{\medspace\medspace\Wc\subseteq[\Ksf]\setminus\cup_{k\in[j]}\{ p_{k}(\Kc_{\Qc})\}}%
x_{\Wc}+ y_{\Qc},\label{eq:prop 3 const 1}
\\&
y_{\Qc}:= \max_{\mathbf{p}([\Ksf])}y_{\Qc,\mathbf{p}([\Ksf])},\label{eq:def of yq}
\end{align}
\end{subequations}
satisfying for each permutation $\mathbf{p}([\Ksf])$,
\begin{align}
\sum_{\Qc\subseteq[\Hsf]:|\Qc|=b}
y_{\Qc}\geq \sum_{\Qc\subseteq[\Hsf]:|\Qc|=b}%
y_{\Qc,\mathbf{p}([\Ksf])},\label{eq:yq and yqp}
\end{align}
where $y_{\Qc,\mathbf{p}([\Ksf])}$ is defined in~\eqref{eq:def of yqp}. 

\begin{rem}\label{rem:why better 2}
In Theorem~\ref{thm:acyclic outer bound}, for each set $\Qc$ we have the constraint in~\eqref{eq:prop 3 const 1} but without $y_{\Qc}$. The above example shows that the sum of all the $y_{\Qc}$'s, where $|\Qc|=b$, is positive, 
thus the converse bound in~\eqref{eq:prop 3 const 1} is tighter than that in Theorem~\ref{thm:acyclic outer bound}.
\end{rem}

\subsection{Proof of Theorem~\ref{thm:improved outer bound 3}}
\label{subset:improved outer bound 3}
We now generalize and combine the two approaches illustrated in Sections~\ref{ex:example1} and~\ref{ex:example2} to improve the converse bound based on the acyclic index coding converse  in Theorem~\ref{thm:acyclic outer bound}.
 
We first focus on    a permutation of users $\mathbf{p}([\Ksf])$, a subset of relays $\Qc\subseteq [\Hsf]$ (assume that $|\Qc|=b$).
Consider  a demand vector where users demand different files, and 
 a partition 
$\Qc=\Qc_{1}\cup\cdots\cup\Qc_{a}$. Let $\mathbf{p}(\Kc_{\Qc_{1}}),\ldots,\mathbf{p}(\Kc_{\Qc_{a}}),\mathbf{p}\big(\Kc_{\Qc}\setminus (\Kc_{\Qc_{1}}\cup\cdots\cup\Kc_{\Qc_{a}})\big)$ be a combination of permutations.
We let $\Bc_{i}:=f\big(\dv,[\Ksf],\mathbf{p}(\Kc_{\Qc_{i}})\big)$ for $i\in [a]$;   $\Bc_{a+1}:=f\Big(\dv,[\Ksf]\setminus(\Kc_{\Qc_{1}}\cup \cdots \cup \Kc_{\Qc_{a}}),\mathbf{p}\big(\Kc_{\Qc}\setminus (\Kc_{\Qc_{1}}\cup \cdots \cup \Kc_{\Qc_{a}})\big)\Big)$; $\Sc_{\Qc}:=f\Big(\dv,[\Ksf]\setminus \Kc_{\Qc}, g\big([\Ksf]\setminus \Kc_{\Qc},\mathbf{p}([\Ksf])\big)\Big)$. For each $i\in [a]$, we write an inequality in the form  of~\eqref{eq:improved acyclic outer bound} by $\Jc=\Qc_{i}$, $\Sc_{1}=\Bc_{i}$, $\Sc_{2}=\Bc_{a+1}\cup \Sc_{\Qc}$. We then sum all of the $a$ inequalities to obtain
\begin{subequations}
 \begin{align}
&\sum_{i\in [a]}H(\Bc_{i})\leq |\Qc|\Rsf^{\star}_{\mathrm{u}} \Bsf- \sum_{i\in [a]}H\big(X_{\Qc_{i}}|(\Bc_{a+1}\cup \Sc_{\Qc})^{c}\big)+2a\Bsf\varepsilon_{\Bsf}\label{eq:combine proof line1}\\
& \leq |\Qc|\Rsf^{\star}_{\mathrm{u}} \Bsf-H\big(X_{\Qc}|(\Bc_{a+1}\cup \Sc_{\Qc})^{c}\big)+2a\Bsf\varepsilon_{\Bsf}\label{eq:combine proof line2}\\
& = |\Qc|\Rsf^{\star}_{\mathrm{u}} \Bsf-H\big(X_{\Qc}|\Sc_{\Qc}^{c}\big)-I\big(X_{\Qc};\Bc_{a+1}|(\Bc_{a+1}\cup\Sc_{\Qc})^{c}\big) \nonumber\\& +2a\Bsf\varepsilon_{\Bsf}\label{eq:combine proof line3}\\
& = \Qc|\Rsf^{\star}_{\mathrm{u}} \Bsf-H\big(X_{\Qc}|\Sc_{\Qc}^{c}\big)-H(\Bc_{a+1})\nonumber\\& +H\big(\Bc_{a+1}|(\Bc_{a+1}\cup\Sc_{\Qc})^{c},X_{\Qc}\big)+2a\Bsf\varepsilon_{\Bsf}\label{eq:combine proof line4}\\
& \leq |\Qc|\Rsf^{\star}_{\mathrm{u}} \Bsf -  H\big(X_{\Qc}|\Sc_{\Qc}^{c}\big) -  H(\Bc_{a+1})+(2a+1)\Bsf\varepsilon_{\Bsf},\label{eq:combine proof line5}
\end{align}
\end{subequations}
where from~\eqref{eq:combine proof line1} to~\eqref{eq:combine proof line2} the submodularity of entropy is used, and from~\eqref{eq:combine proof line4} to~\eqref{eq:combine proof line5} we use Fano's inequality  and the fact that  $\Bc_{a+1}$ is acyclic and $\Sc_{\Qc}$ does not include the side information of the user requiring $\Bc_{a+1}$.

We list all the inequalities in the form of~\eqref{eq:combine proof line5} for all the possible demands where users demand different files, and sum them to  obtain
\begin{align}
|\Qc|\Rsf^{\star}_{\mathrm{u}} &\geq  \sum_{i\in[a]}\sum_{\medspace\medspace j\in[|\Kc_{\Qc_{i}}|]}\sum_{\medspace\medspace\Wc\subseteq[\Ksf]\setminus\cup_{k\in[j]}\{ p_{k}(\Kc_{\Qc_{i}})\}}%
x_{\Wc} \nonumber\\
&+ \sum_{j\in\big[\big|\Kc_{\Qc}\setminus\Vc\big|\big]}\sum_{\medspace\medspace\Wc\subseteq([\Ksf]\setminus\Vc)\setminus\cup_{k\in[j]}\{p_{k}(\Kc_{\Qc}\setminus\Vc)\}}%
\negmedspace\negmedspace\negmedspace \negmedspace x_{\Wc} \nonumber\\
&+ \underbrace{\frac{1}{\Ksf !\Bsf}\sum_{\dv:d_{i}\neq d_{j}\textrm{\rm  for }i\neq j} %
  H(X_{\Qc}|S_{\Qc}^{c})}_{\text{$:= y_{\Qc,\mathbf{p}([\Ksf])}$ as defined in~\eqref{eq:def of yqp}}}
 .\label{eq:combined ypq}
\end{align} 

 From~\eqref{eq:combined ypq} and by the definition $y_{\Qc}:= \max_{\pv_{[\Ksf]}} y_{\Qc,\mathbf{p}([\Ksf])}$, we have 
 \begin{align}
 |\Qc|\Rsf^{\star}_{\mathrm{u}} &\geq  \sum_{i\in[a]}\sum_{\medspace\medspace j\in[|\Kc_{\Qc_{i}}|]}\sum_{\medspace\medspace\Wc\subseteq[\Ksf]\setminus\cup_{k\in[j]}\{ p_{k}(\Kc_{\Qc_{i}})\}}%
x_{\Wc} \nonumber\\
&+\sum_{j\in\big[\big|\Kc_{\Qc}\setminus\Vc\big|\big]}\sum_{\medspace\medspace\Wc\subseteq([\Ksf]\setminus\Vc)\setminus\cup_{k\in[j]}\{p_{k}(\Kc_{\Qc}\setminus\Vc)\}}%
\negmedspace\negmedspace\negmedspace \negmedspace x_{\Wc} + y_{\Qc}.\label{eq:combined yq}
\end{align}  

We then focus on the term   $y_{\Qc}$. 
 Choose an integer $b\in [\rsf:\Hsf]$ and a permutation $\mathbf{p}([\Ksf])$. 
In order to  prove~\eqref{eq:prop 3 const 2},  
  we aim to bound 
 \begin{align}
\sum_{\Qc\subseteq[\Hsf]:|\Qc|=b} %
 y_{\Qc}
 &\geq \sum_{\Qc\subseteq[\Hsf]:|\Qc|=b} %
 y_{\Qc,\mathbf{p}([\Ksf])}\nonumber\\
&=\frac{1}{\Ksf !\Bsf}\sum_{\dv:d_{i}\neq d_{j}\textrm{\rm  for }i\neq j}\sum_{\medspace\medspace\Qc\subseteq[\Hsf]:|\Qc|=b}%
  H(X_{\Qc}|S_{\Qc}^{c}).\label{eq:proof of th4 each permu} 
\end{align} 

Consider two sets of relays with cardinality $b$, $\Jc_{1}$ and $\Jc_{2}$. We can use Proposition~\ref{prop:lem for improved outer bound 2} to bound 
\begin{align}
&H\big(X_{\Jc_{1}}|\Sc_{\Jc_{1}}^{c}\big)+H\big(X_{\Jc_{2}}|\Sc_{\Jc_{2}}^{c}\big) \nonumber\\
&\geq H\big(X_{\Jc_{1}\cup \Jc_{2}}|(\Sc_{\Jc_{1}}\cap \Sc_{\Jc_{2}})^{c}\big)+H\big(X_{\Jc_{1}\cap \Jc_{2}}|(\Sc_{\Jc_{1}}\cup \Sc_{\Jc_{2}})^{c}\big).\label{eq:lemma 3 in J1 and J2}
\end{align}
 Each time we use Proposition~\ref{prop:lem for improved outer bound 2}, we call the first term in the RHS result as $\cup$-term and the second term as $\cap$-term. We then add $H\big(X_{\Jc_{3}}|\Sc_{\Jc_{3}}^{c}\big)$ to the RHS of~\eqref{eq:lemma 3 in J1 and J2}, where $\Jc_{3}$ is the third term in the sum $\sum_{\Qc\subseteq[\Hsf]:|\Qc|=b} H(X_{\Qc}|S_{\Qc}^{c})$. We use Proposition~\ref{prop:lem for improved outer bound 2} to bound the sum of the $\cup$-term of~\eqref{eq:lemma 3 in J1 and J2} and  $H(X_{\Jc_{3}}|\Sc_{\Jc_{3}}^{c})$. We put the $\cup$-term in the new RHS result at the first position, and use  Proposition~\ref{prop:lem for improved outer bound 2} again to bound the sum of the $\cap$-term in the RHS result of~\eqref{eq:lemma 3 in J1 and J2} and the $\cap$-term in the new RHS result. The $\cup$-term of the latest one is put at the second position while the $\cap$-term is at the third position. So after considering three sets, we now have three terms. Similarly, each time we consider the $j^{\textrm{\rm th}}$ set with cardinality $b$, we use Proposition~\ref{prop:lem for improved outer bound 2} to bound the sum of the term in the first position of the last iteration and $H\big(X_{\Jc_{j}}|\Sc_{\Jc_{j}}^{c}\big)$. The $\cup$-term of the result is put at the first position in this iteration. The $\cap$-term of the result should be added to the term at the second position at the last iteration. We perform this procedure until the term in the last position at the last iteration. We describe the above iterative procedure in Algorithm 2.

Notice that  when we apply Proposition~\ref{prop:lem for improved outer bound 2} to bound a sum of two terms, the $\cap$-term of the result may be $0$. When we use Proposition~\ref{prop:lem for improved outer bound 2} to bound the sum of $0$ and one term, the result is also the sum of this term (seen as the $\cup$-term) and $0$ (seen as the $\cap$-term). We should also notice that after each iteration, by assuming that the term at the $i_{1}^{\textrm{\rm th}}$ is $H\big(X_{\Gc_{i_{1}}}|\Ic_{i_{1}}^{c}\big)$ and the term at the $i_{2}^{\textrm{\rm th}}$ is $H(X_{\Gc_{i_{2}}}|\Ic_{i_{2}}^{c})$ where $i_{1}<i_{2}$, we can see that $\Gc_{i_{2}}\subseteq \Gc_{i_{1}}$ and $\Ic_{i_{2}}^{c}\subseteq \Ic_{i_{1}}^{c}$. \\
\rule{\textwidth/2}{0.8pt}

\vspace{-3bp}
\textbf{Algorithm 2 }Iterative Procedure by using Proposition~\ref{prop:lem for improved outer bound 2}
\vspace{-6bp}
\\
\rule{\textwidth/2}{0.8pt}
\begin{enumerate}
\item \textbf{input}: $H\big(X_{\Jc_{i}}|\Sc_{\Jc_{i}}^{c}\big)$ where $i\in [\binom{\Hsf}{b}]$  (each $\Jc_{i}$ is a distinct set of relays with cardinality $b$); \textbf{initialization}: $t=2$;
\item use Proposition~\ref{prop:lem for improved outer bound 2} to bound $H\big(X_{\Jc_{1}}|\Sc_{\Jc_{1}}^{c}\big)+H\big(X_{\Jc_{2}}|\Sc_{\Jc_{2}}^{c}\big)$; let $L_{t,1}$ be the $\cup$-term and $L_{t,2}$ be the $\cap$-term; 
\item use Proposition~\ref{prop:lem for improved outer bound 2} to bound $L_{t,1}+H\big(X_{\Jc_{t+1}}|\Sc_{\Jc_{t+1}}^{c}\big)$; let $L_{t+1,1}$ be the $\cup$-term and $T_{\textrm{\rm cap}}$ be the $\cap$-term;  
\item \textbf{for} $i=2,\ldots,t$, use Proposition~\ref{prop:lem for improved outer bound 2} to bound $L_{t,i}+T_{\textrm{\rm cap}}$ and let $L_{t+1,i}$ be the $\cup$-term and $T_{\textrm{\rm cap}}$ be the $\cap$-term;
 \item let $L_{t+1,t+1}=T_{\textrm{\rm cap}}$;
 \item \textbf{if} $t< \binom{\Hsf}{b}-1$, \textbf{then} $t=t+1$ and go to 3);
 \item \textbf{output}: $\sum_{i\in [\binom{\Hsf}{b}]}L_{\binom{\Hsf}{b},i}$.
\end{enumerate}
\rule{\textwidth/2}{0.8pt}

After considering all the sets of relays with cardinality $b$, we have a summation including $\binom{\Hsf}{b}$ terms.
In the end, for an acyclic set of subfiles $\Sc$, by using Proposition~\ref{prop:acyclic outer bound}, we have
\begin{align}
&H\big(X_{[\Hsf]}|\Sc^{c}\big)\geq H(\Sc).\label{eq:take out subfile}
\end{align}
Hence, we can bound this summation by a sum of the lengths of subfiles, then we obtain
\begin{align}
&\sum_{\Qc\subseteq[\Hsf]:|\Qc|=b} H(X_{\Qc}|S_{\Qc}^{c})\nonumber\\& \geq\sum_{i\in[\Ksf]}\sum_{\medspace\medspace\Wc\subseteq[\Ksf]\setminus\cup_{j\in[i]}\{p_{j}([\Ksf])\}}%
c\big(\{p_{i}([\Ksf])\}\cup\Wc,b\big)|F_{d_{p_{i}([\Ksf])},\Wc}|,\label{eq:sum of entropies}
\end{align}
where $c(\Wc_{1},l):=\binom{\Hsf-1}{l-1}-|\{\Qc\subseteq [\Hsf]:|\Qc|=l,\Kc_{\Qc}\nsubseteq [\Ksf]\setminus \Wc_{1}\}|$, which will be proved in the following. We focus on $|F_{d_{p_{i}([\Ksf])},\Wc}|$, where $\Wc\subseteq[\Ksf]\setminus\{p_{1}([\Ksf]),\ldots,p_{i}([\Ksf])\}$. For a set of relays $\Qc$ with cardinality $b$, in the term $H(X_{\Qc}|\Sc_{\Qc}^{c})$, we can see  that $F_{d_{p_{i}([\Ksf])},\Wc}\in \Sc_{\Qc}$ if and only if $\Kc_{\Qc}\cap (\{p_{i}([\Ksf])\}\cup\Wc)=\emptyset$ (in other words, $\Kc_{\Qc}\subseteq [\Ksf]\setminus (\{p_{i}([\Ksf])\}\cup\Wc)$). Focus on one relay $h\in [\Hsf]$.
When we apply Proposition~\ref{prop:lem for improved outer bound 2} to bound the sum of two terms by the sum of two new terms,  if among the two terms in the LHS of Proposition~\ref{prop:lem for improved outer bound 2}, one term includes $X_{h}$ not knowing $F_{d_{p_{i}([\Ksf])},\Wc}$ and the other does not include $X_{h}$ knowing $F_{d_{p_{i}([\Ksf])},\Wc}$, we can see that the number of terms in the RHS  of Proposition~\ref{prop:lem for improved outer bound 2} including $X_{h}$ not knowing $F_{d_{p_{i}([\Ksf])},\Wc}$ decreases  by $1$ compared to the LHS (the number of terms in the RHS not including $X_{h}$ but knowing $F_{d_{p_{i}([\Ksf])},\Wc}$ also decreases by $1$ compared to the LHS); otherwise, the number of terms in the RHS including $X_{h}$ not knowing $F_{d_{p_{i}([\Ksf])},\Wc}$  does not change compared to the LHS. In addition,  it can be checked that in any case when we use Proposition~\ref{prop:lem for improved outer bound 2}, the number of terms in the RHS not including $X_{h}$ but knowing $F_{d_{p_{i}([\Ksf])},\Wc}$ does not increase compared to the LHS. Hence, among all of the terms in the summation after the final iteration, the number of terms including $X_{h}$ not knowing $F_{d_{p_{i}([\Ksf])},\Wc}$ is not less than
\begin{align}
&\max\Big\{|\{\Qc \subseteq [\Hsf] :  |\Qc|=b,h\in\Qc,\Kc_{\Qc} \subseteq [\Ksf]\ \setminus (\{p_{i}([\Ksf])\}\cup\Wc)\}|\nonumber\\
&-|\{\Qc \subseteq [\Hsf]  :  |\Qc| = b,h \notin \Qc,\Kc_{\Qc}   [\Ksf]\ \setminus (\{p_{i}([\Ksf])\}\cup\Wc)\}|,0 \Big\}\nonumber\\
&= \max\left\{  \binom{\Hsf-1}{l-1} -|\{\Qc \subseteq [\Hsf] : |\Qc| 
 =  l,\Kc_{\Qc}\nsubseteq[\Ksf]\setminus\Wc_{1}\}|,0\right\} .\label{eq:one h}
\end{align}

In addition, after the final iteration, the term at the $i_{1}^{\textrm{\rm th}}$ position is $H(X_{\Gc_{i_{1}}}|\Ic_{i_{1}}^{c})$ and the term at the $i_{2}^{\textrm{\rm th}}$ position is $H(X_{\Gc_{i_{2}}}|\Ic_{i_{2}}^{c})$ where $i_{1}<i_{2}$, we can see that $\Gc_{i_{2}}\subseteq \Gc_{i_{1}}$ and $\Ic_{i_{2}}^{c}\subseteq \Ic_{i_{1}}^{c}$. So by~\eqref{eq:take out subfile} we have, $c(\Wc_{1},l)=\binom{\Hsf-1}{l-1}-|\{\Qc\subseteq [\Hsf]:|\Qc|=l,\Kc_{\Qc}\nsubseteq [\Ksf]\setminus \Wc_{1}\}|$ as defined in~\eqref{eq:def of c}.

Finally, from~\eqref{eq:proof of th4 each permu},~\eqref{eq:sum of entropies} and the value of $c(\Wc_{1},l)$, we obtain~\eqref{eq:prop 3 const 2}.
After proving~\eqref{eq:combined yq} and~\eqref{eq:prop 3 const 2}, the proof of Theorem~\ref{thm:improved outer bound 3} is completed.

\begin{rem}
\label{rem:N<K}
If $\Nsf<\Ksf$, we can not find a demand vector where each user has a distinct demand.
In this case, 
one should consider many subsystems with only $\min(\Nsf,\Ksf) = \Nsf$ users with distinct demands (as we did in~\cite{detailledisit} for shared-link model), which is not conceptually more difficult but requires a somewhat heavier notation.
\end{rem}

\begin{rem}
\label{rem:asymmetry}
It can be seen that the proposed strategies to tighten the acyclic index coding converse bound in combination networks do not rely on the symmetry of the network. Thus we can also use these strategies in general relay networks.
\end{rem}

\section{Conclusions}
\label{sec:conclusion and future work} 
In this paper we investigated the   combination networks with end-user-caches. For the achevability part, we proposed four delivery schemes with  the MAN cache placement to deliver the MAN multicast messages through the network. For the converse part, we   
  extended the acyclic index coding converse bound for the shared-link model to combination networks and improved it by leveraging network topology. The proposed achievable and converse bounds were shown  to be better than the state-of-the-art bounds. In addition, for the  combination networks with end-user-caches where $\Nsf\geq \Ksf$,  
  compared to the  existing order optimality results for the high memory size regime,  we additionally obtained the order optimality under the constraint of uncoded cache placement   for the small memory size regime, such that the remaining open case is when  $\frac{\rsf}{\Ksf} < \frac{\Msf}{\Nsf} < \frac{1}{2\rsf} $.

\section*{Acknowledgments}
We would like to thank    Dr.  Zhangchi Chen from Mathematical Department of University Paris-sud who proved Theorem~\ref{thm:circulant matrix rank}
and   Dr. Tang Liu (who was with the ECE Department of University of Illinois at Chicago when this work was done and  is now with the ECE Department of Princeton University) who inspired the derivation in Appendix~\ref{sec:discussion of full rank}.

\appendices

\section{Discussion of the Group Division of the Interference Elimination Scheme}
\label{sec:discussion of full rank}
To get the coefficients $[a_{g,1,j};\ldots;a_{g,\Hsf,j}]$ in equation~\eqref{eq:general equation}, $\mathbb{C}_{g}$ should be full-rank for each group $\Gc_{g}$. We should solve the following problem to ensure the feasibility where we introduce an integer $k=\rsf-1$ such that $2k+2=2\rsf$.

\textbf{Problem 1:} Let $k$ be a positive integer. We focus on all the $\binom{2k+1}{k}$ subsets of $[2k+2]$ with cardinality $k+1$ and
$2k+2$ is in each subset (because in~\eqref{eq:general interference} and~\eqref{eq:general useful} we only focus on the users connected to relay $\Hsf$). We want to divide these subsets into $\binom{2k+1}{k}/(2k+1)$ groups such that each group has $2k+1$ subsets. For each group $\Pc_{i}$, we create a $(2k+2)\times(2k+2)$ matrix. The first row is all $1$. For each subset in this group, we have one row of $0$ and $1$, where the $j^{\textrm{\rm th}}$ element is $1$ if and only if $j$ is in this subset. The condition that a  solution must satisfy is that each such matrix is full-rank.\footnote{\label{foot:solution}It can be seen that a solution for Problem 1 leads to a group division in the IES, where we partition the $\binom{2\rsf-1}{\rsf-1}$ sets in $\Vc_2$ into $\binom{2\rsf-1}{\rsf-1}/(2\rsf-1)$ groups, each containing $2\rsf-1$ sets. More precisely, assume that  $\Pc_i=\{u_1,\ldots,u_{2\rsf-1}\}$, where each of users in $\Pc_i$ is connected to relay $\Hsf$. We also assume that user $\bar{u}_j$ is connected to the relays in $[\Hsf]\setminus \Hc_{u_j}$ for each $j\in [2\rsf-1]$. Thus we have $\Gc_i=\big\{ \{u_1,\bar{u}_1\},\ldots, \{u_{2\rsf-1},\bar{u}_{2\rsf-1}\} \big\}.$}

In Appendix~\ref{sec:integer proof}, we prove that $\binom{2k+1}{k}/(2k+1)$ is an integer if $k$ is a positive integer.  We provide the following algorithm to construct such groups for Problem 1, which is shown by numerical evaluations to find such groups when $k\leq 12$. When $k>12$, this numerical simulation might be infeasible due to the complexity. 
\\
\rule{\textwidth/2}{0.8pt}

\vspace{-3bp}
\textbf{Algorithm 3 }Group division method for Problem 1
\vspace{-6bp}
\\
\rule{\textwidth/2}{0.8pt}
\begin{enumerate}
\item \textbf{input}: $k$, $\Pc=\{\Jc\subseteq [2k+2]:|\Jc|=k+1,(2k+2)\in \Jc\}$; \textbf{initialization}: $t_{1}=0$; $times=10;$ $\Pc_{i}=\emptyset$ for $i\in [\binom{2k+1}{k}/(2k+1)]$;
\item \textbf{for} $i\in [\binom{2k+1}{k}/(2k+1)]$,
\begin{enumerate}
\item $Test=0;$  randomly choose  $2k+1$ subsets in $\Pc$; create a $(2k+2)\times(2k+2)$ matrix denoted by $\mathbb{C}$ (the first row of $\mathbb{C}$ is all $1$ and for each chosen subset, there is one row of $0$ and $1$, where $j^{\textrm{\rm th}}$ element is $1$ if and only if $j\in [2k+2]$ is in this subset);
\item \textbf{if}  $\mathbb{C}$ is full-rank, \textbf{then} $Test=1$ and put the chosen subsets in $\Pc_{i}$;
\item \textbf{if} $Test=0$ and $t_{1}\leq times$, \textbf{then} $t_{1}=t_{1}+1$ and go to Step 2-a);
\end{enumerate}
\item \textbf{if} $\Pc_{i}\neq \emptyset$ for all $i\in [\binom{2k+1}{k}/(2k+1)]$, \textbf{then} \textbf{Output} $\Pc_{i}$ for all $i\in [\binom{2k+1}{k}/(2k+1)]$; \\
\textbf{else, then }go to Step 1);
\end{enumerate}
\rule{\textwidth/2}{0.8pt}

We can use Algorithm 3 to construct the groups.  
However, it is hard to prove the existence of the group division satisfying the full-rank condition for the general case. Instead of proving the existence of solution for Problem 1, we introduce Problem 2, the existence of whose solution is easier to analyse. As the number $2k+2$ appears in each subsets of Problem 1,  we do not consider the number $2k+2$ in Problem 2. In Appendix~\ref{sec:equivalence} we prove that the  we can add $2k+2$ into each subset in the solution of Problem 2 to get one solution of Problem 1.

\textbf{Problem 2:} Let $k$ be a positive integer. We focus on all the $\binom{2k+1}{k}$ subsets of $[2k+1]$ with cardinality $k$. We want to divide these subsets into $\binom{2k+1}{k}/(2k+1)$ groups such that each group has $2k+1$ subsets. In each group, the number of subsets containing each number in $[2k+1]$ is the same (equal to $k$). We create a $(2k+1)\times(2k+1)$ matrix, called {\it incident matrix}. There is one row of $0$ and $1$ in the incident matrix corresponding to  each subset in this group, where the $j^{\textrm{\rm th}}$ element in the row is $1$ if and only if $j$ is in this subset. The condition is that each incident matrix is full-rank.

Compared to Problem 1, Problem 2 has an additional constraint, which is that  in each group, the number of subsets containing each number in $[2k+1]$ is the same. In Example~\ref{ex:example of IE h6r3}, we have a group division satisfying Problem 1. In addition, if we take out the number $2k+2=6$ in each subset, it is a solution for Problem 2. To analyse the existence, we   review the following theorem given in~\cite[Theorem 1.1]{patecki2014cyclic}.
\begin{thm}[\cite{patecki2014cyclic}]
\label{thm:refcyclic}
Let $k^{\prime}$ and $n^{\prime}$ be positive integers, and let $\lambda$ be the smallest non-trivial divisor of $n^{\prime}$. Then all the $\binom{n^{\prime}}{k^{\prime}}$ subsets of $[n^{\prime}]$ with cardinality $k^{\prime}$ could be divided into $\binom{n^{\prime}}{k^{\prime}}/n^{\prime}$ non-overlapping groups, where each group includes $n^{\prime}$ subsets and its incident matrix is circulant, if and only if $n^{\prime}$ is relatively prime to $k^{\prime}$, $\lambda k^{\prime} > n^{\prime}$ and $n^{\prime}$ divides $\binom{n^{\prime}}{k^{\prime}}$.
\end{thm}
A circulant $n^{\prime}\times n^{\prime}$ matrix is uniquely determined by its first row $[c_0,c_1,\ldots,c_{n^{\prime}-1}]$ and its $i^{\textrm{\rm th}}$ row is obtained by shifting the first row rightwards by $i-1$ where $i\in [2:n^{\prime}]$. In Problem 2, $n^{\prime}=2k+1$ and $k^{\prime}=k$. It is easy to see that $2k+1$ and $k$ are relatively prime and that $\lambda \geq 3$, which leads   $\lambda k>2k+1$. In addition, in Appendix~\ref{sec:integer proof} we prove that $2k+1$ divides $\binom{2k+1}{k}$. Hence, if we choose $n^{\prime}=2k+1$ and $k^{\prime}=k$, the conditions in Theorem~\ref{thm:refcyclic} are satisfied. 
As the incident matrix of each group is circulant, we can see that in each group, the number of subsets containing each number in $[2k+1]$ is the same. Hence, it remains to analyse the rank of each incident matrix. It was shown in~\cite[Theorems 1.7 and 1.8]{circulantmatrices} that the following theorem holds.
\begin{thm}
\label{thm:circulant matrix rank}
Let $k$ be a positive integer. A $(2k+1)\times (2k+1)$ circulant matrix, where the number of $1$ in the first row is $k$ and the number of $0$ in the first row is $k+1$, is always invertible if $2k + 1=p^{v}$ or $pq$, where $p,q$ are different primes and $v$ is a positive integer.
\end{thm}

\section{Proof: $\binom{2k+1}{k}/(2k+1)$ is an integer.}
\label{sec:integer proof}
If $k$ is a positive integer,  we have that
\begin{align*}
 & \frac{1}{2k+1}\binom{2k+1}{k}=\frac{1}{k+1}\binom{2k}{k} =\frac{k+1-k}{k+1}\binom{2k}{k} \\&=\binom{2k}{k}-\frac{k}{k+1}\binom{2k}{k} 
  =\binom{2k}{k}-\binom{2k}{k+1}
\end{align*}
  is an integer.
  
\section{Proof: A solution of Problem 2 is a solution of Problem 1}
\label{sec:equivalence}
For a solution of Problem 2, we focus on any group $g$ and assume that the $(2k+1)\times (2k+1)$ incident matrix is $\mathbb{B}$. In the following, we will prove that the matrix $\mathbb{C}_{g}$ is also full-rank, where $\mathbb{C}_{g}=\left[\begin{array}{cc}
\mathbb{E}^{T} & 1\\
\mathbb{B} & \mathbb{E}
\end{array}\right]$ and $\mathbb{E}=[1;\ldots;1]$ with dimension $(2k+1)\times 1$. $\mathbb{E}^{T}$ is the transpose of $\mathbb{E}$. It can be seen that if  $\mathbb{C}_{g}$ with dimension $(2k+2)\times (2k+2)$ is full-rank for each group $g$, we can add $2k+2$ into each subset in the solution of Problem 2 to get one solution of Problem 1.

We prove it by contradiction, i.e., assume that $\mathbb{C}_{g}$ is not full-rank. There must exist   only one sequence  $[a_1,\dots,a_{2k+1}]$ such that $\sum_{i\in[2k+1]}a_{i}\mathbb{B}_{i}=\mathbb{E}^{T}$, where $\mathbb{B}_{i}$ represents the $i^{\textrm{\rm th}}$ row of $\mathbb{B}$. In other words, we have 
\begin{align*}
[ 
a_{1},
\cdots,
a_{2k+1}
] \mathbb{B} =\mathbb{E}^{T} \ \Longrightarrow \left[ 
a_{1},
\cdots,
a_{2k+1}
 \right]= \mathbb{E}^{T} \mathbb{B}^{-1}.
\end{align*}
For a solution of Problem 2,  in each group, the number of subsets containing each number in $[2k+1]$ is the same (equal to $k$). Hence, $[a_1,\dots,a_{2k+1}]=[1/k,\ldots,1/k]$. However, in this case, $\sum_{i\in[2k+1]}a_{i}\neq 1$, not satisfying the last column of $\mathbb{C}_{g}$. Hence, we prove that $\mathbb{C}_{g}$ is full-rank.

\section{Proof of Theorem~\ref{thm:exact optimallity}}
\label{sec:proof of thm exact opt}

\subsection{Proof of Theorem~\ref{thm:exact optimallity}  item~1}
We focus on each corner point with $\Msf=\frac{\Nsf t}{\Ksf}$ where $t \in \left[0:\left\lceil  \frac{\rsf}{\Hsf-\rsf}-1 \right\rceil \right]$.  In this case, we have  $\Hsf t/\rsf<t+1$.

Achievability.
We focus on one set of users $\Jc$ where $|\Jc|>t$. If each relay is connected to at most $t$ users of $\Jc$, then $\Jc$  contains at most $\Hsf t/\rsf$ users. So if     $\Hsf t/\rsf<t+1$, there must be one relay connected to $t+1$ users of $\Jc$. So if $\Hsf t/\rsf<t+1$, we have $\Vc_{2}=\emptyset$, i.e., for any set $\Jc$ of $t+1$ users, there exists at least one relay connected to all of these users. So the max-link load of the DIS  is $\frac{\Ksf(1-\Msf/\Nsf)}{\Hsf(1+\Ksf\Msf/\Nsf)}$.

Then we focus on the CICS  under this case. For each set $\Jc$ of $t+1$ users, we have $\max_{h\in [\Hsf]}|\Uc_{h} \cap \Jc|=t+1$. So the total link load to transmit  is $|W_{\Jc}|/(\Hsf\Bsf)$; thus the link load to transmit all the demanded subfiles of users is also $\frac{\Ksf(1-\Msf/\Nsf)}{\Hsf(1+\Ksf\Msf/\Nsf)}$.

Converse.
The converse bound in~\eqref{eq:cut-set outer bound combination} coincides with  the above load by taking with $x=\Hsf$.

\subsection{Proof of Theorem~\ref{thm:exact optimallity} item~2}
We focus on each corner point with  $\Msf=\frac{\Nsf t}{\Ksf}$ where $t \in [0:\Ksf]$.

Achievability.
By setting $\rsf=\Hsf-1$ in~\eqref{eq:def of v 2}, we have $\Vc_{2}=\emptyset$ when $t\leq \Ksf-2$, i.e., for any set $\Jc$ of $t+1$ users, there exists at least one relay connected to all of these users. So similar to the previous case, 
for each $t\in [0:\Ksf-2]$, the max-link load achieved by the DIS  or  the CICS   
is $\frac{\Ksf-t}{(t+1)\Hsf}$. 
For $t=\Ksf$, the link load is $0$. 
For $t\in [\Ksf-2:\Ksf]$, we use memory-sharing between the points $t=\Ksf-2$ and $t=0$.

Converse.
The converse bound in~\eqref{eq:cut-set outer bound combination} coincides with the lower convex envelop of the above loads by taking $x=\Hsf$ when $\Msf\leq \Nsf(\Ksf-2)/\Ksf$, and $x=\rsf$ when $\Msf\geq \Nsf(\Ksf-2)/\Ksf$.

\subsection{Proof of Theorem~\ref{thm:exact optimallity} item~3}
Achievability for $\Hsf<2\rsf$.
When $t=\Ksf\Msf/\Nsf=1$, we have $\Vc_{2}=\emptyset$ and thus   the load from the server to all the relays is $(\Ksf-t)/(t+1)=(\Ksf-1)/2$; due to the symmetry, the load from the server to each relay is the same, and thus the link load is $\frac{\Ksf-1}{2\Hsf}$. 
When $\Msf=0$, the load is $\Ksf/\Hsf$. 
By   memory-sharing between $\Msf=0$ and $\Msf=\Ksf\Msf/\Nsf$, we can achieve the load in~\eqref{eq:H<2r}. 

Converse for $\Hsf<2\rsf$.
The converse bound in~\eqref{eq:cut-set outer bound combination} coincides with the achievable bound by taking $x=\Hsf$ when $0\leq \Msf\leq \Nsf/\Ksf$.

Achievability for $\Hsf=2\rsf$.
When $0\leq \Msf\leq \Nsf/\Ksf$, from~\eqref{eq:achievable interference load} and by memory-sharing between $\Msf=0$ and $\Msf=\Nsf/\Ksf$, the load achieved by the  IES is $\frac{\Ksf(\Hsf-1)-(\frac{\Ksf\Hsf+\Hsf-\Ksf}{2}-1)\frac{\Ksf\Msf}{\Nsf}}{\Hsf(\Hsf-1)}$. 

Converse for $\Hsf=2\rsf$.
We use the converse bound in Theorem~\ref{thm:improved outer bound 3}. In Appendix~\ref{sub:Proof of eq:eliminate x0 and x1} we prove that 
\begin{align}
\Hsf(\Hsf-1)\Rsf_\text{u}^{\star}&\geq  \Ksf(\Hsf-1)-\left(\frac{\Ksf\Hsf+\Hsf-\Ksf}{2}-1\right)\frac{\Ksf\Msf}{\Nsf}\nonumber\\& +\sum_{\Wc\subseteq [\Ksf]:|\mathcal{W}|>1}z_{\Wc}x_{\Wc},
\label{eq:eliminate x0 and x1}
\end{align} 
where $z_{\Wc} \geq 0 $ is the coefficient of $x_{\Wc}$ { in~\eqref{eq:defxw}}.
Hence, the bound in~\eqref{eq:eliminate x0 and x1}
coincides with the achievable bound when $0\leq \Msf\leq\Nsf/\Ksf$.

\subsection{Proof of~\eqref{eq:eliminate x0 and x1} }
\label{sub:Proof of eq:eliminate x0 and x1}
When $\Hsf=2\rsf$, we compute the converse bound in Theorem~\ref{thm:improved outer bound 3}. 
By choosing $a=1$, the converse bound in~\eqref{eq:combine outer bounds xw} becomes 
\begin{align}
|\Qc|\Rsf^{\star}_{\mathrm{u}} \geq %
\sum_{j\in[|\Kc_{\Qc}|]}\sum_{\medspace\medspace\Wc\subseteq[\Ksf]\setminus\cup_{k\in[j]}\{ p_{k}(\Kc_{\Qc})\}}%
x_{\Wc}+ y_{\Qc}.\label{eq:prop 3 const 111}
\end{align}

We sum all the inequalities as~\eqref{eq:prop 3 const 111} for all the sets of relays $\Qc$ where $|\Qc|=\Hsf-1$ and for all the permutations $\mathbf{p}(\Kc_{\Qc})$. In~\eqref{eq:prop 3 const 1}, there are $|\Kc_{\Qc}|=\Ksf/2$ terms of $x_{\emptyset}$ and $(\Ksf-1)+\ldots+(\Ksf-|\Kc_{\Qc}|)=(3\Ksf/2-1)\Ksf/4$ terms of $x_{\Wc}$ where $|\Wc|=1$. Because of the symmetry, from the sum we obtain
\begin{align}
 \Hsf(\Hsf-1)\Rsf^{\star}_{\mathrm{u}} &\geq \frac{\Ksf}{2}\Hsf x_{\emptyset}+\frac{\Hsf}{4}(\frac{3}{2}\Ksf-1)x_{1}\nonumber\\& +\sum_{\Wc\subseteq [\Ksf]:|\Wc|>1}   v_{\Wc}x_{\Wc}+\sum_{\Qc:|\Qc|=b}y_{\Qc},\label{eq:sum all Q}
\end{align}
where $x_{1}=\sum_{\Wc\subseteq [\Ksf]:|\Wc|=1}x_{\Wc}$ and $v_{\Wc}\geq 0$ represents the coefficient of $x_{\Wc}$. Then, we focus on~\eqref{eq:prop 3 const 2}. Notice that $c_{\Wc_{1},\Hsf-1}=\binom{\Hsf-1}{\Hsf-2}-\rsf=\rsf-1$ where $|\Wc_{1}|=1$. So in~\eqref{eq:prop 3 const 2}, the total coefficient of $x_{\emptyset}$ is $\Ksf(\rsf-1)$. Then we focus on a set of user $\Wc_{1}\subseteq [\Ksf]$ where $|\Wc_{1}|=2$ under the assumption that the two users in $\Wc_{1}$ are $k_{1}$ and $k_{2}$. In~\eqref{eq:prop 3 const 2}, there is only one term with coefficient $c_{\Wc_{1},\Hsf-1}$  for each $\Wc_{1}\subseteq [\Ksf]$ where $|\Wc_{1}|=2$. Now we want to compute $c_{\Wc_{1},\Hsf-1}$ for each $\Wc_{1}\subseteq [\Ksf]$ where $|\Wc_{1}|=2$. If $\Hc_{k_1}\cap \Hc_{k_2}=\emptyset$, we have $c_{\Wc_{1},\Hsf-1}=\max \{\Hsf-1-\Hsf,0\}=0$. In addition, there are $\frac{\Ksf}{2}\binom{\rsf}{0}\binom{\rsf}{\rsf}$ such sets. If $|\Hc_{k_1}\cap \Hc_{k_2}|=1$, we have $c_{\Wc_{1},\Hsf-1}=\max \{\Hsf-1-\Hsf-1,0\}=0$. There are $\frac{\Ksf}{2}\binom{\rsf}{1}\binom{\rsf}{\rsf-1}$ such sets. If $|\Hc_{k_1}\cap \Hc_{k_2}|=i\in [2:\rsf-1]$, we have $c_{\Wc_{1},\Hsf-1}=\max \{\Hsf-1-(\Hsf-i),0\}=i-1$. There are $\frac{\Ksf}{2}\binom{\rsf}{i}\binom{\rsf}{\rsf-i}$ such sets.  Hence, we have 
\begin{align*}
&\sum_{\Wc_{1} \subseteq [\Ksf]:|\Wc_{1}|=2}c_{\Wc_{1},\Hsf-1}=\sum_{i\in[2:\rsf-1]}(i-1)\frac{\Ksf}{2}\binom{\rsf}{i}\binom{\rsf}{\rsf-i}\\& =\frac{\Ksf}{2}\big\{\sum_{i\in[0:\rsf]}(i-1)\binom{\rsf}{i}\binom{\rsf}{\rsf-i}+1-(\rsf-1)\big\}\\
&=\frac{\Ksf}{2}\big\{\sum_{i\in[0:\rsf]}i\binom{\rsf}{i}\binom{\rsf}{\rsf-i}-\binom{2\rsf}{\rsf}-\rsf+2\big\}\\& =\frac{\Ksf}{2}\big\{\rsf\sum_{i\in[0:\rsf]}\frac{i}{\rsf}\binom{\rsf}{i}\binom{\rsf}{\rsf-i}-\Ksf-\rsf+2\big\}\\
&=\frac{\Ksf}{2}\big\{\rsf\sum_{i\in[1:\rsf]}\binom{\rsf-1}{i-1}\binom{\rsf}{\rsf-i}-\Ksf-\rsf+2\big\}\\& =\frac{\Ksf}{2}\big\{\rsf\binom{2\rsf-1}{\rsf}-\Ksf-\rsf+2\big\}=\frac{\Ksf}{2}(\rsf-2)\Big(\frac{\Ksf}{2}-1\Big).
\end{align*}
Hence, we can sum all the inequalities as~\eqref{eq:prop 3 const 2} for all the permutations $\mathbf{p}([\Ksf])$ to obtain,
\begin{align}
\sum_{\Qc:|\Qc|=b}y_{\Qc}\geq \Ksf(\rsf-1)x_{\emptyset} +\frac{1}{2}(\rsf-2)\Big(\frac{\Ksf}{2}-1\Big)x_{1}.\label{eq:sum yq}
\end{align}
From~\eqref{eq:sum all Q} and~\eqref{eq:sum yq}, we have
\begin{align}
&\Hsf(\Hsf-1)\Rsf^{\star}_{\mathrm{u}}\geq \frac{\Ksf}{2}\Hsf x_{\emptyset}+\frac{\Hsf}{4}(\frac{3}{2}\Ksf-1)x_{1}\nonumber\\& +\sum_{\Wc\subseteq [\Ksf]:|\Wc|>1}v_{\Wc}x_{\Wc}+\Ksf(\rsf-1)x_{\emptyset} +\frac{1}{2}(\rsf-2)(\frac{\Ksf}{2}-1)x_{1}\nonumber\\
&=\Big(\frac{\Ksf}{2}\Hsf+\Ksf(\rsf-1) \Big)x_{\emptyset}+\frac{\Hsf(\Ksf-1)-(\Ksf-2)}{2}x_{1} \nonumber\\& + \sum_{\Wc\subseteq [\Ksf]:|\Wc|>1}v_{\Wc}x_{\Wc}.\label{eq:sum q +sum yq}
\end{align}
Then, we want to eliminate $x_{\emptyset}$ and $x_{1}$ with the help of~\eqref{eq:xfile size} and~\eqref{eq:memory size}. From~\eqref{eq:xfile size}, we have
\begin{align}
&\Big(\frac{\Ksf}{2}\Hsf + \Ksf(\rsf -  1) \Big) (x_{\emptyset} +  x_{1}) \nonumber\\& = \Big(\frac{\Ksf}{2}\Hsf + \Ksf(\rsf -  1) \Big)(1 -     \sum_{\Wc\subseteq [\Ksf]:|\Wc|>1}      x_{\Wc}).\label{eq:from file size}
\end{align}
From~\eqref{eq:memory size}, we have
\begin{align}
&\Big(\frac{\Ksf - \Hsf - \Ksf\Hsf}{2} +  1\Big)x_{1} \nonumber\\ & = \Big(\frac{\Ksf - \Hsf - \Ksf\Hsf}{2} +  1\Big)\big(\frac{\Ksf\Msf}{\Nsf} -     \sum_{\Wc\subseteq [\Ksf]:|\Wc|>1}      x_{\Wc}\big).\label{eq:from memory size}
\end{align}
We take~\eqref{eq:from file size} and~\eqref{eq:from memory size} into~\eqref{eq:sum q +sum yq} to obtain,
\begin{align}
\Hsf(\Hsf-1)\Rsf^{\star}_{\mathrm{u}}&\geq \Ksf(\Hsf-1)-\Big(\frac{-\Ksf + \Hsf + \Ksf\Hsf}{2} -  1\Big)\frac{\Ksf\Msf}{\Nsf} \nonumber\\& +\sum_{\Wc\subseteq [\Ksf]:|\Wc|>1} z_{\Wc}x_{\Wc},\label{eq:proof eliminate x0 and x1}
\end{align}
where $z_{\Wc}=v_{\Wc}+\big(\frac{-\Ksf+\Hsf+\Ksf\Hsf}{2}- 1\big)\geq 0$. So we prove~\eqref{eq:eliminate x0 and x1}.

\section{Proof of Theorem~\ref{thm:order optimality}}
\label{sec:proof of thm order opt}
By taking $x=\Hsf$ and comparing the cut-set bound under the constraint of uncoded placement and the achievable bounds in~\eqref{eq:Mingyue inner bound}, we can straightforward obtain Theorem~\ref{thm:order optimality}-1).

\subsection{Proof of Theorem~\ref{thm:order optimality} item~2  and item~3}
We first focus on $\Msf=\frac{\Nsf t}{\Ksf}$, where $t\in [0:\Ksf]$.

It can be seen that for each $\Jc\subseteq [\Ksf]$ where $|\Jc|=t+1$, $\min_{h\in[\Hsf]}|\Jc\setminus \Uc_h|\leq |\Jc|-1=t$. Hence, from~\eqref{eq:load of CICS},
\begin{align}
\Rsf_{\mathrm{CICS}}\leq \binom{\Ksf}{t+1}\frac{1+t/\rsf}{\Hsf\binom{\Ksf}{t}}.
\label{eq:outer bound of CICS}
\end{align}
By the cut-set converse bound in~\eqref{eq:cut-set outer bound combination}, $\Rsf^{\star}_{\mathrm{u}}$ is lower  bounded by the lower convex envelop of $\Big(\frac{\Nsf t}{\Ksf},\frac{\binom{\Ksf}{t+1}}{\Hsf\binom{\Ksf}{t}}\Big)$ for $t\in [0:\Ksf]$. Hence, the multiplicative gap between the CICS  and the cut-set converse bound in~\eqref{eq:cut-set outer bound combination} is   within     $1+t/\rsf$.

Next, for any $t \in [0,\Ksf]$, it is obvious that the multiplicative gap between the CICS  and the cut-set converse bound in~\eqref{eq:cut-set outer bound combination} is   within     $1+\left\lceil t \right\rceil /\rsf$.

So we can sequentially prove Theorem~\ref{thm:order optimality}-2).

\subsection{Proof of Theorem~\ref{thm:order optimality} item~4}
To encode each message $W_{\Jc}$ where $\Jc \in \Vc_{1}$,  the total link load of the IES  is $|W_{\Jc}|/\Bsf$. To encode each message $W_{\Jc}$ where $\Jc \in \Vc_{2}$, we transmit one linear combination including $W_{\Jc}$ and other $2\rsf-2$ messages  to $2\rsf$ relays. So the total link load to transmit $W_{\Jc}$ is $2\rsf/(2\rsf-1)$. 
Compared to the converse bound in~\eqref{eq:cut-set outer bound combination} with $x=\Hsf$ when $0\leq \Msf\leq \Nsf/\Ksf$, the IES is order optimal within a factor of $\frac{2\rsf}{2\rsf-1}\leq \frac{4}{3}$ under the constraint of uncoded cache placement.

\bibliographystyle{IEEEtran}
\bibliography{IEEEabrv,IEEEexample}

\begin{IEEEbiographynophoto}
						{Kai Wan} (S '15 -- M '18)
						received  the B.E. degree in    Optoelectronics from  Huazhong University of Science and Technology, China, in 2012, the   M.Sc. and Ph.D. degrees in Communications from Universit{\'e}  Paris-Saclay, France, in 2014 and 2018.  He is currently a post-doctoral    researcher with the Communications and Information Theory Chair   (CommIT) at Technische Universit\"at Berlin, Berlin, Germany. His   research interests include information theory, coding techniques, and   their applications on coded caching,  index coding, distributed storage,  distributed computing, wireless communications,   privacy and security. He has served as an Associate Editor of IEEE Communications Letters from Aug. 2021.
					\end{IEEEbiographynophoto}
					
						\begin{IEEEbiographynophoto}{Daniela Tuninetti}  (M '98 -- SM '13 -- F '21)
 is currently a Professor within the Department of Electrical
and Computer Engineering at the University of Illinois at Chicago (UIC),
which she joined in 2005. Dr. Tuninetti got her Ph.D. in Electrical Engineering
in 2002 from ENST/T{\'e}l{\'e}com ParisTech (Paris, France, with work done at the
Eurecom Institute in Sophia Antipolis, France), and she was a postdoctoral
research associate at the School of Communication and Computer Science
at the Swiss Federal Institute of Technology in Lausanne (EPFL, Lausanne,
Switzerland) from 2002 to 2004. Dr. Tuninetti is a recipient of a best paper
award at the European Wireless Conference in 2002, of an NSF CAREER
award in 2007, and named University of Illinois Scholar in 2015. Dr. Tuninetti
was the editor-in-chief of the IEEE Information Theory Society Newsletter
from 2006 to 2008, an editor for IEEE COMMUNICATION LETTERS from
2006 to 2009, for IEEE TRANSACTIONS ON WIRELESS COMMUNICATIONS
from 2011 to 2014; and for IEEE TRANSACTIONS ON INFORMATION
THEORY from 2014 to 2017. She is currently a distinguished lecturer for the
Information Theory society. She is also currently an editor for IEEE Transactions on Communications.
 Dr. Tuninetti's research interests are in the
ultimate performance limits of wireless interference networks (with special
emphasis on cognition and user cooperation), coexistence between radar and
communication systems, multi-relay networks, content-type coding, cache-aided
systems and distributed private coded computing.

					\end{IEEEbiographynophoto}
					
										\begin{IEEEbiographynophoto}{Mingyue Ji}
(S '09 -- M '15) received the B.E. in Communication Engineering from Beijing University of Posts and Telecommunications (China), in 2006, the M.Sc. degrees in Electrical Engineering from Royal Institute of Technology (Sweden) and from University of California, Santa Cruz, in 2008 and 2010, respectively, and the PhD from the Ming Hsieh Department of Electrical Engineering at University of Southern California in 2015. He subsequently was a Staff II System Design Scientist with Broadcom Corporation (Broadcom Limited) in 2015-2016. He is now an Assistant Professor of Electrical and Computer Engineering Department and an Adjunct Assistant Professor of School of Computing at the University of Utah. He received the IEEE Communications Society Leonard G. Abraham Prize for the best IEEE JSAC paper in 2019, the best paper award in IEEE ICC 2015 conference, the best student paper award in IEEE European Wireless 2010 Conference and USC Annenberg Fellowship from 2010 to 2014.  He has served as an Associate Editor of IEEE Transactions on Communications from 2020. He is interested the broad area of information theory, coding theory, concentration of measure and statistics with the applications of caching networks, wireless communications, distributed storage and computing systems, distributed machine learning, and (statistical) signal processing.
					\end{IEEEbiographynophoto}
					
					\begin{IEEEbiographynophoto}{Pablo Piantanida} (Senior Member, IEEE) received the  B.Sc.  degree in electrical engineering and the M.Sc. degree from the University of Buenos Aires, Argentina,   in   2003,   and the   Ph.D.   degree   from Universit{\'e} Paris-Sud, Orsay, France, in 2007. He is currently Full Professor with the Laboratoire des Signaux et Syst\`emes (L2S), CentraleSup{\'e}lec together with CNRS and Universit{\'e} Paris-Saclay. He is also an associate member of Com\`ete - Inria research team (Lix - Ecole Polytechnique). His research interests include information theory, machine learning, security of learning systems and the secure processing of information and applications to computer vision, health, natural language processing, among others. He has served as the General Co-Chair for the 2019 IEEE  International  Symposium on  Information  Theory  (ISIT).  He served as an Associate  Editor for the  IEEE  TRANSACTIONS  ON  INFORMATION FORENSICS AND SECURITY and Editorial Board of Section "Information Theory, Probability and Statistics" for Entropy. He is member of the IEEE Information Theory Society Conference Committee. 
\end{IEEEbiographynophoto}

\end{document}